\pdfoutput=1% forces arxiv to use pdflatex
\def\Title{Using Cadabra for tensor computations in\\ General Relativity.}
\def\Author{Leo Brewin}
\documentclass[a4paper,12pt]{article}
\usepackage{cdblatex}
\usepackage{hyperref}
\usepackage{geometry}
\usepackage{brewin}

\hypersetup{colorlinks=true,
            citecolor=blue,
            linkcolor=red,
            pdfauthor= \Author,
            pdftitle = \Title}

\numberwithin{equation}{section}% requires amsmath

\begin{document}

\input{./cadabra/dgeom.cdbtex}

\CdbSetup{action=verbatim}% we won't be processing any Cadabra code,
                          % just listing code from other sources

\lstset{numbers=left,gobble=2,basicstyle=\ttfamily\small,escapechar=|}

% ============================================================================================
\title{\Title}
\author{\Author\\[10pt]%
School of Mathematical Sciences\\%
Monash University, 3800\\%
Australia}
% \date{20-Jun-2018}% 1st version started
% \date{10-Sep-2019}% 1st version finished
% \date{28-Oct-2019}% 2nd version finished
\date{10-Dec-2019}% 1st arxiv version finished
% \reference{Preprint: arXiv:0903.5365\\[5pt]
%            Journal: {\it Phys.Rev.D.} {\bf 80} 084030 (2009)}
\reference{Preprint}

\maketitle

% ============================================================================================
\begin{abstract}
\noindent
Cadabra is an open access program ideally suited to complex tensor commutations in General
Relativity. Tensor expressions are written in LaTeX while an enhanced version of Python is
used to control the computations. This tutorial assumes no prior knowledge of Cadabra. It
consists of a series of examples covering a range of topics from basic syntax such as
declarations, functions, program control, component computations, input and output through
to complete computations including a derivation of two of the BSSN equations from the ADM
equations. Numerous exercises are included along with complete solutions. All of the source
code for the examples, exercises and solutions are available on GitHub.
\end{abstract}

% ============================================================================================
% \bgroup
% \parskip=0pt plus 0pt minus 0pt
%
% \tableofcontents
%
% \vfill
% \egroup
%
% \clearpage

% ============================================================================================
\section*{Introduction}
\label{sec:intro}

The main goal in writing this tutorial was to provide the reader with sufficient knowledge
so that they can use Cadabra \cite{peeters:2017-01,peeters:2017-02} to do meaningful
computations in general relativity. It was written for readers with no prior knowledge of
Cadabra and is presented as a series of examples using familiar computations (such as
verifying that the Levi-Civita connection is a metric connection) as vehicles to present the
various elements of Cadabra.

The tutorial contains many exercises (with complete solutions) that allow the reader to test
their understanding as well to explore some of the side issues raised in the main thread of
the tutorial.

The LaTeX and Cadabra sources for the tutorial can be found on the
\href{https://github.com/leo-brewin/cadabra-tutorial}{author's GitHib} site
(see Part 4 for the relevant URL).

This tutorial is a significantly extended version of a similar tutorial written in 2009
\cite{brewin:2009-02}. Though the 2009 tutorial has been updated to comply with the
version 2.0 syntax it does not contain any of the extensive additions introduced in version
2.0. It should be noted that version 1.0 of Cadabra is no longer supported and all users are
encouraged to migrate to version 2.0. Note also the version 2.0 syntax is not backward
compatible with that of version 1.0.

\clearpage

% ============================================================================================
\section*{Computations in General Relativity}

Here are three examples of the kinds of computation that are often required in General
Relativity.

\begin{description}
\item[Numerical computation.]\hfil\break\vskip-15pt
Use a numerical method to evolve the time symmetric initial data for a geodesic slicing of a
Schwarzschild spacetime in an isotropic gauge.

\item[Algebraic computation.]\hfil\break\vskip-15pt
Compute the Riemann tensor for the metric $ds^2=\Phi(r)^2\left(dr^2+r^2d\Omega^2\right)$.

\item[Tensor computations.]\hfil\break\vskip-15pt
Verify that $0=g_{ab;c}$ given
$\Gamma^{a}{}_{bc} = \frac{1}{2}g^{ad}\left(g_{dc,b} + g_{bd,c} - g_{bc,d}\right)$.
\end{description}

What tools are available to perform these computations? For the first example, it is hard to
envisage \emph{not} using a computer to do the job. The second example is one which could
easily be done by hand or on a computer (using, for example, GRTensorIII
\cite{grtensorIII:2017-01}, Maxima \cite{maxima:2014-01} or Cadabra, the subject of this
tutorial). The third examples is so simple that most people would use traditional pencil and
paper methods. However, there are many other tensor computations in General Relativity that
are particular tedious to push through by hand (e.g., developing higher order Riemann normal
expansions of the metric or performing perturbation expansions of the vacuum field
equations). So there is very good reason to seek help by way of a computer program designed
specially to manipulate tensor expressions. This tutorial will provide a brief introduction
to one such program, Cadabra, and how it can be used in General Relativity.

Besides Cadabra, there are a number of other programs that, to varying degrees,
can manipulate tensor expressions, including
GRTensorIII \cite{grtensorIII:2017-01},
Maxima \cite{maxima:2014-01},
Canon \cite{manssur:2004-01},
Riemann \cite{portugal:1997-01} and
xAct \cite{xact:2008-01}.
No attempt will be made here to provide even a cursory review of the above programs
(however, see the recent review by MacCallum \cite{maccallum:2018-01}). Instead, the
intention in this tutorial will be to show how Cadabra can be used to do useful work in
General Relativity.

Given that Cadabra is just one of a number of programs that can manipulate tensor
expressions, the obvious question would be -- why chose Cadabra?

One of Cadabra's main selling points is its elegant and simple syntax. This is based on a
subset of LaTeX to express tensor expressions, Python to coordinate the computations and some
unique Cadabra syntax to describe properties of various objects (e.g., index sets,
symmetries, commutation rules etc.). This leads to a shallow learning curve and codes that
are clear and easy to read. The core program of Cadabra is written in {\tt C++} including
highly optimised procedures for simplifying complex tensor expressions. It has a strong user
base, active discussion forums and is under active development.

Another strong point of Cadabra is its use of LaTeX for tensor expressions for not only
\emph{input to} Cadabra but also for \emph{output from} Cadabra. This means that output from
one Cadabra code can be easily used in other Cadabra codes or even in separate LaTeX
documents. Indeed this document is a case in point -- all of the results appearing later in
this document were computed in separate Cadabra codes and included, without change, from the
corresponding Cadabra output.

The following examples were deliberately constructed so as to require little mathematical
development (for the current audience) while being of sufficient complexity to allow
Cadabra's features to be properly showcased. For the majority of this tutorial no
assumptions will be made about the dimensionality of the space other than in Example
\ref{sec:ex-10} (4 dimensions) and Example \ref{sec:ex-13} (3 dimensions). The
Misner-Thorne-Wheeler (MTW \cite{mtw:1973-01}) conventions will be used for the metric
signature and the Riemann tensor. The connection will be assumed to be metric compatible
(i.e., the Levi-Civita connection). Abstract index notation will be used but on the odd
occasion where an explicit component based equation is given, the components will be given
in a coordinate basis.

% ============================================================================================
\section*{The Cadabra software}

The full source for Cadabra can be found on the GitHub page \cite{peeters:2017-02} while
binaries for popular versions of Linux and Windows can be found in the downloads section of
the Cadabra home page \cite{peeters:2017-01}. There are no binaries for macOS but it is a
simple matter to compile from the source using Homebrew. Complete instructions are available
on the Cadabra GitHub page \cite{peeters:2017-02}.

There are two main ways to run Cadabra, either through the command line or through a GUI
interface similar to the notebook interfaces used by Jupyter and Mathematica. The command
line version of Cadabra works with plain text files, such as {\tts foo.cdb}. These files can
be created using any text editor and contain Cadabra statements. To run Cadabra on the file
{\tts foo.cdb} you need only type

\bgroup
\lstset{numbers=none}
\begin{lstlisting}
   cadabra2 foo.cdb
\end{lstlisting}
\egroup

on the command line. In contrast, files like {\tts foo.cnb} are Cadabra notebooks and can
contain not only Cadabra statements but also Cadabra output as well as LaTeX markup. These
files are not intended to be edited in a plain text editor but rather are created, edited
and executed entirely from within the Cadabra GUI. To initiate the GUI and load the notebook
{\tts foo.cnb} type

\bgroup
\lstset{numbers=none}
\begin{lstlisting}
   cadabra2-gtk foo.cnb
\end{lstlisting}
\egroup

on the command line. Once the GUI has started you can edit or execute the current notebook or
use the {\tts File} menu item to navigate to other notebooks. The menu in the GUI contains
the usual set of entries that should need little explanation. However, if it is not obvious
what a particular menu item does, then just click on that item and note what happens --
though do chose to work on a test file. There are not many menu items so this click and
observe method should reveal most of the menu actions in a short time.

\clearpage

% ============================================================================================
\section*{The tutorial sources}

The complete set of LaTeX and Cadabra sources can be obtained by cloning from the GitHub
site.

\bgroup
\lstset{numbers=none}
\begin{lstlisting}
   git clone https://github.com/leo-brewin/cadabra-tutorial
\end{lstlisting}
\egroup

This will create a directory {\tts cadabra-tutorial} containing all of the sources. Here is a brief description of the main directories and their contents.

\def\atMYn{\vrule height  0pt depth  5pt width 0pt}
\def\atMYm{\vrule height 15pt depth  5pt width 0pt}
\def\atMYN{\vrule height 15pt depth  0pt width 0pt}
\def\Box#1{\parbox[t]{0.60\textwidth}{\sloppy \atMYN#1\atMYn}}

\begin{tabular}{ll}
   {\tts pdf/}&\Box{This document as well as the {\tts .pdf} files for the
   exercises and examples.}\\
   {\tts source/tex/}&\Box{The LaTeX source for this document.}\\
   {\tts source/cadabra/example*.tex}&%
   \Box{The LaTeX/Cadabra source for each of the examples in
   this document. These are written in the hybrid-latex format in which the Cadabra
   code is embeded in a LaTeX document. The tools to process these files are provided
   in the {\tts hybrid-latex} directory.}\\
   {\tts source/cadabra/cdb/}&%
   \Box{The raw Cadabra sources extracted from the hybrid-latex
   files in {\tts source/cadabra/example*.tex}. These are in {\tts .cdb} format and are
   provided for readers who like to copy-paste the Cadabra code into a {\tts cadabra2-gtk}
   window.}\\
   {\tts source/cadabra/exercises/}&%
   \Box{This directory contains the worked solutions for all of the exercises.}\\
   {\tts source/cadabra/fragments/}&%
   \Box{Some of the exercises asks the reader to use specific fragments of code. Those
   fragments can be found in this directory -- saving the reader from the tedium of
   writing the code by hand.}\\
   {\tts source/cadabra/tests/}&%
   \Box{This directory is used only when running tests (quelle surprise). To check that
   everything is working correctly just run {\tts make tests} from the {\tts source/cadabra}
   directory. See the main {\tts README.md} file for more details.}\\
   {\tts hybrid-latex/}&%
   \Box{This directory contains all the tools needed to process the hybrid-latex files.
   See the file {\tts hybrid-latex/INSTALL.txt} for instructions on where to copy
   theses files.}
\end{tabular}

\clearpage

% ============================================================================================
\section*{Notation}

\input{./cadabra/example-05.cdbtex}

The following examples will contain lines of Cadabra code as well as the
corresponding output. The question here is -- how is that correspondence conveyed to the
reader? The device used here will be to match the output against the line number of the code.

Here is a small fragment of a larger Cadabra code (drawn from Example \ref{sec:ex-05}).
\begin{cadabra}
   expr := A_{a} v^{a} + B_{a} v^{a} + C_{a} v^{a};    |\lstlabel{ex-05.660}|
   zoom       (expr, $B_{a} Q??$)                      |\lstlabel{ex-05.661}|
   substitute (expr, $v^{a} -> w^{a}$);                |\lstlabel{ex-05.662}|
   unzoom     (expr)                                   |\lstlabel{ex-05.663}|
\end{cadabra}
Try to ignore the code (for the moment) and focus instead on the small line numbers in the
left hand margin. These numbers are not part of the Cadabra syntax but have been added here
so that individual lines of code can be identified. They are also used as tags to match
against the Cadabra output which, in this case, just happens to be (have faith)
\begin{align*}
   \cdb{ex-05.100} &= \cdb{ex-05.101}\tag*{\lstref{ex-05.661}}\\
                   &= \cdb{ex-05.102}\tag*{\lstref{ex-05.662}}\\
                   &= \cdb{ex-05.103}\tag*{\lstref{ex-05.663}}
\end{align*}
The weird looking equation numbers on the right hand side are matched to the line numbers in
the Cadabra code. Thus \lstref{ex-05.661} is the output generated by line
\lstref*{ex-05.661} of the Cadabra code, likewise the output for line \lstref*{ex-05.662} is
given by \lstref{ex-05.662}.

\clearpage

% ============================================================================================
% Part 1. Getting started. With whimsical section headings.
% ============================================================================================
\hrule height 0pt
\vskip 4cm
{\Huge\bf Part 1 Essential elements}
\vskip 2cm

This first part of the tutorial consists of a set of examples that are intended for readers
with little or no experience with Cadabra. Each section is built around a simple example
based on some familiar elements of general relativity. These context based examples are used
to introduce the essential elements of Cadabra required for routine tensor computations.

The second part of this tutorial switches the focus from introducing Cadabra to applying
Cadabra to more substantial questions (which once again cover well known topics in general
relativity). The examples are a hotchpotch reflecting the research interest of the author.

The examples are supported by many exercises with full solutions. The exercises are not
essential for progression through the later examples but they do help the reader to test
their understanding of basic aspects of Cadabra. They also explore aspects of Cadabra not
otherwise covered in the main thread of this tutorial.

\clearpage

% ============================================================================================
\section{Hello metric connection}
\label{sec:ex-01}
\ResetCounters

\input{./cadabra/example-01.cdbtex}

How might Cadabra be used to verify that $0=\nabla_{c}g_{ab}$ given $2\Gamma^{a}{}_{bc} =
g^{ad}\left(\partial_{b}g_{dc} + \partial_{c} g_{bd} - \partial_{d} g_{bc}\right)$?

This may seem an odd way to start but here is the full Cadabra code.

\begin{cadabra}
   # Define some properties                                                  |\lstlabel{ex-01.c1}|

   {a,b,c,d,e,f,h,i,j,k,l,m,n,o,p,q,r,s,t,u#}::Indices.                      |\lstlabel{ex-01.01}|

   g_{a b}::Metric.                                                          |\lstlabel{ex-01.02}|
   g_{a}^{b}::KroneckerDelta.                                                |\lstlabel{ex-01.03}|

   \nabla{#}::Derivative.                                                    |\lstlabel{ex-01.04}|
   \partial{#}::PartialDerivative.                                           |\lstlabel{ex-01.05}|

   # Define rules for covariant derivative and the Christoffel symbol        |\lstlabel{ex-01.c2}|

   nabla := \nabla_{c}{g_{a b}} -> \partial_{c}{g_{a b}} - g_{a d}\Gamma^{d}_{b c}    |\lstlabel{ex-01.06}|
                                                         - g_{d b}\Gamma^{d}_{a c};   |\lstlabel{ex-01.07}|

   Gamma := \Gamma^{a}_{b c} -> (1/2) g^{a d} (  \partial_{b}{g_{d c}}       |\lstlabel{ex-01.08}|
                                               + \partial_{c}{g_{b d}}       |\lstlabel{ex-01.09}|
                                               - \partial_{d}{g_{b c}} );    |\lstlabel{ex-01.10}|

   # Start with a simple expression                                          |\lstlabel{ex-01.c3}|

   cderiv := \nabla_{c}{g_{a b}};                                            |\lstlabel{ex-01.11}|

   # Do the computations                                                     |\lstlabel{ex-01.c4}|

   substitute          (cderiv, nabla);                                      |\lstlabel{ex-01.12}|
   substitute          (cderiv, Gamma);                                      |\lstlabel{ex-01.13}|
   distribute          (cderiv);                                             |\lstlabel{ex-01.14}|
   eliminate_metric    (cderiv);                                             |\lstlabel{ex-01.15}|
   eliminate_kronecker (cderiv);                                             |\lstlabel{ex-01.16}|
   canonicalise        (cderiv);                                             |\lstlabel{ex-01.17}|
\end{cadabra}

The output from the above code is

\begin{dgroup*}[compact,spread={2pt}]
   \Dmath*{\cdb{nabla.100}\hfill\lstref{ex-01.06}}
   \Dmath*{\cdb{Gamma.100}\hfill\lstref{ex-01.08}}
\end{dgroup*}

\begin{dgroup*}[compact,spread={2pt}]
   % \Dmath*{\cdb{nabla.100}\hfill\lstref{ex-01.06}}
   % \Dmath*{\cdb{Gamma.100}\hfill\lstref{ex-01.08}}
   \Dmath*{\cdb{ex-01.100} = \cdb{ex-01.101}\hfill\lstref{ex-01.12}
                           = \cdb{ex-01.102}\hfill\lstref{ex-01.13}
                           = \cdb{ex-01.103}\hfill\lstref{ex-01.14}
                           = \cdb{ex-01.104}\hfill\lstref{ex-01.15}
                           = \cdb{ex-01.105}\hskip11cm\hfill\V{12pt}\lstref{ex-01.16}
                           = \cdb{ex-01.106}\hfill\lstref{ex-01.17}}
\end{dgroup*}

Each of these line shows selected stages of processing by Cadabra. The zero in the final
line shows that $\nabla_{c} g_{ab}$ is indeed zero for the given choice of $\Gamma^{a}{}_{bc}$.

Note that for each line of output shown above, Cadabra wrote only the part between the equals
sign and the (apparent) equation number on the far right. Everything else was
added by the author to put the Cadabra output into context. The number on the far right
matches the line number in the source while the text to the left of the equals sign
identifies the object associated with the Cadabra output. So though the above output is not
exactly what would be seen in the GUI it is important to note that the Cadabra output has
not been modified in any way other than to be sandwiched between the equals sign on the left
and the line number on the right.

Looking back at the above code, the obvious question is -- what does each line do? For some
lines the answer is clear but for others there are elements of the syntax that do require
further explanation. Thus at this point it is useful to spend a bit of time working through
the above Cadabra code in some detail.

Statements in the Cadabra grammar fall into a number of distinct categories:
\emph{comments}, \emph{properties}, \emph{expressions}, \emph{algorithms} and a broad
category that consists of any valid Python statement. Comments in Cadabra are single lines
that begin with one or more spaces (or tabs) followed by the \verb|#| character. Any text
after the \verb|#| will be treated as a comment. There are four comments in the above
example (lines \lstref*{ex-01.c1}, \lstref*{ex-01.c2}, \lstref*{ex-01.c3} and
\lstref*{ex-01.c4}). The statements in lines \lstref*{ex-01.01} to \lstref*{ex-01.05} assign
\emph{properties} to some symbols, while those in lines \lstref*{ex-01.06} to
\lstref*{ex-01.11} define three \emph{expressions} named \verb|nabla|, \verb|Gamma| and
\verb|cderiv|. The remaining statements apply \emph{algorithms} to the expressions (i.e.,
they perform the computations). Note that \emph{algorithms} are, in the eyes of Python,
ordinary Python functions. Python functions can also be applied to Cadabra objects and thus
could also be described as algorithms. But as this may lead to some confusion the convention
adopted in this tutorial is that the term \emph{algorithm} will be reserved exclusively for
Cadabra's own functions.

Cadabra statements can consist of one or more lines of text. Thus Cadabra sets clear rules
about how a statement can be constructed from a series of lines. It will read its input,
line by line, while also looking for a clear marker to indicate the end of the current
statement. For \emph{properties} and \emph{expressions} the statement will be terminated by
either a dot {\tts .} or a semi-colon {\tts ;}. The situation is slightly different for
\emph{algorithms} -- they are terminated either by a dot, a semi-colon or by the closing
right parenthesis of the algorithm. In all cases, Cadabra will generate output only for
those statements that end with a semi-colon. Python statements are terminated in the normal
Python manner.

Once Cadabra has digested the source it will pass a slightly modified copy onto its own
internal version of Python (enhanced to support Cadabra's algorithms). Thus the original
Cadabra source must conform to Python's strict (but simple) indentation rules.

What do these statements actually mean? The first statement
\bgroup
\lstset{firstnumber=3}
\begin{cadabra}
   {a,b,c,d,e,f,h,i,j,k,l,m,n,o,p,q,r,s,t,u#}::Indices.
\end{cadabra}
\egroup
simply declares a set of symbols that may be used as indices. The last symbol \verb|u#|
informs Cadabra that an infinite set of indices of the form \verb|u1,u2,u3|{$\>\cdots$} is
allowed. If you prefer to work with Greek indices then you could declare
\begin{cadabra}[numbers=none]
   {\alpha,\beta,\gamma,\mu,\nu,\theta,\phi#}::Indices.
\end{cadabra}
Note that all of the usual LaTeX Greek symbols are understood by Cadabra. They can be used as
indices or symbols (e.g., \verb|\Gamma| can be used to denote a Christoffel symbol). However,
they can not be used as identifiers (i.e., they can not appear on the left hand side of an
assignment). Thus the following statement will raise a syntax error
\bgroup
\begin{cadabra}[numbers=none]
   \Gamma := (1/2) g^{a d} (  \partial_{b}{g_{d c}}
                            + \partial_{c}{g_{b d}}
                            - \partial_{d}{g_{b c}} );

\end{cadabra}
\egroup

The next pair of statements
\bgroup
\lstset{firstnumber=5}
\begin{cadabra}
   g_{a b}::Metric.
   g_{a}^{b}::KroneckerDelta.
\end{cadabra}
\egroup
declares that \verb|g_{a b}| represents a (symmetric) metric and that $g_a{}^b$ is the usual
Kronecker delta (i.e., $g_a{}^b = \delta_a^b$).

The following pair of statements
\bgroup
\lstset{firstnumber=8}
\begin{cadabra}
   \nabla{#}::Derivative.
   \partial{#}::PartialDerivative.
\end{cadabra}
\egroup
assigns a derivative property to the symbols \verb|\nabla| and \verb|\partial|. The
distinction between the \verb|::Derivative| and \verb|::PartialDerivative| properties is that
only those derivative operators declared as \verb|::PartialDerivative| will be taken as
self-commuting ($\partial_a \partial_b = \partial_b \partial_a$). Note that the \verb|#| in
each declaration signifies that any number of indices (up or down) are allowed. That is both
\verb|\partial{a}| and \verb|\partial{a b c d}| will be seen by Cadabra as derivative
operators. This interpretation of \verb|{#}| carries over to other declarations, for example
\verb|\delta{#}::KroneckerDelta| declares \verb|\delta| to be a Kronecker delta with any
number of upper or lower indices (and in any order).

The next three statements define three expressions, \verb|nabla|, \verb|Gamma| and
\verb|cderiv|.
\bgroup
\lstset{firstnumber=11}
\begin{cadabra}
   # Define rules for covariant derivative and the Christoffel symbol

   nabla := \nabla_{c}{g_{a b}} -> \partial_{c}{g_{a b}} - g_{a d}\Gamma^{d}_{b c}
                                                         - g_{d b}\Gamma^{d}_{a c};

   Gamma := \Gamma^{a}_{b c} -> (1/2) g^{a d} (  \partial_{b}{g_{d c}}
                                               + \partial_{c}{g_{b d}}
                                               - \partial_{d}{g_{b c}} );

   # Start with a simple expression

   cderiv := \nabla_{c}{g_{a b}};
\end{cadabra}
\egroup
The name of the expression appears to the left of the `\verb|:=|' characters while the
corresponding tensor expression appears on the right using a familiar LaTeX syntax. Tensor
indices such as \verb|a,b,c...| should always be separated by one or more spaces (unlike the
case in LaTeX). This ensures that Cadabra knows exactly how many indices belong to an object
(e.g., \verb|g_{ab}| would be interpreted as an object with \emph{one} covariant index
\verb|ab|). This rule can be relaxed when the index set contains its own delimiter such as
the slash that appears when indices are written using LaTeX names. Thus an object like
\verb|g_{\alpha\beta}| clearly contains just two indices.

Note carefully the braces around the metric term in \verb|\partial_{c}{g_{a b}}|. This is
essential -- the symbol \verb|\partial| is an operator and thus needs an argument to act on,
namely, the argument contained inside the pair of braces.

There is one very important operational difference between the expressions for \verb|cderiv|
and those for \verb|nabla| and \verb|Gamma|. The expression \verb|cderiv| defines a Cadabra
object that will be manipulated in stages towards the final result (in line
\lstref*{ex-01.17}). These changes are obtained by applying Cadabra's algorithms (lines
\lstref*{ex-01.12} to \lstref*{ex-01.17}) to \verb|cderiv|. The other expressions,
\verb|nabla| and \verb|Gamma|, are known as substitution rules and each are of the form
\begin{cadabra}[numbers=none]
   rule := target -> replacement;
\end{cadabra}
They provide Cadabra with all the information needed to replace any instances of $\nabla_{c}
g_{ab}$ and $\Gamma^{a}{}_{bc}$ with the appropriate combination of the metric and its
derivatives. The application of these rules can be seen in lines \lstref*{ex-01.12} and
\lstref*{ex-01.13} with each call to \verb|substitute| applying a rule to the expression
\verb|cderiv|.

After Cadabra has executed the calls to \verb|substitute|, the object \verb|cderiv| will
consist solely of terms built from the metric and its derivatives. Though this may look
simple there is a very important and subtle detail that must be noted. The substitution rule
\verb|Gamma| as given above was for $\Gamma^{a}{}_{bc}$ yet the expression for \verb|cderiv|
requires $\Gamma^{d}{}_{bc}$ and $\Gamma^{d}{}_{ac}$. Cadabra handles this index
manipulation by relabelling dummy indices in such a way as to avoid index clashes. This
feature also exists in xAct. % (but not so in GRTensorIII).

% In this simple example separate statements were used to create and later execute the
% rules. This ability to defer the application of substitution rules at the users
% discretion is one of Cadabra's main features. In other algebraic software, such as
% Maple, assignments are propagated forward from the point at which they are declared.
% They will also be imposed indirectly on existing expressions. Cadabra, in contrast,
% allows the user to choose when a rule should be applied as well as on which
% expressions the rule will be applied to. This gives the user considerable freedom in
% developing a strategy to achieve a desired computational goal.

The remaining few statements

\bgroup
\lstset{firstnumber=28}
\begin{cadabra}
   distribute          (cderiv);
   eliminate_metric    (cderiv);
   eliminate_kronecker (cderiv);
   canonicalise        (cderiv);
\end{cadabra}
\egroup

serve only to massage the expression towards the expected result -- zero. Each of the
statements applies an algorithm to the expression \verb|cderiv| with the result replacing
the original value of \verb|cderiv|. That is, Cadabra's algorithms makes in-place changes to
Cadabra objects. The algorithm \verb|distribute| is used to expand products, it will expand
\verb|a (b+c)| into \verb|a b + a c|. In line \lstref*{ex-01.02} of the code the property
\verb|::Metric| was given to \verb|g_{a b}|. This is used by the \verb|eliminate_metric|
algorithm to convert combinations such as \verb|g_{a c} g^{c b}| into a Kronecker-delta
$\delta^b{}_a$ which (not surprisingly) is eliminated by the \verb|eliminate_kronecker|
algorithm. The \verb|canonicalise| algorithm is one of Cadabra's most useful algorithms (on
a par with \verb|substitute|) as it can apply a wide range of simplifications and general
housekeeping. In this case it makes use of the symmetric property of the metric to complete
the final step of the calculation. The result in line \lstref{ex-01.17} is zero as expected.

% --------------------------------------------------------------------------------------------
\subsection{Cadabra syntax summary}

The above discussion has introduced some key elements of the Cadabra syntax. Other elements
will be discussed later as the need arises. Though this does present a shallow learning curve
(consider the alternative where mastery of the full syntax tree is required before seeing
any examples) it does mean that important information is scattered throughout the tutorial.
This of course makes it harder to find key information after the first reading. To mitigate
that problem, here is a short summary of the Cadabra syntax that will be seen in later
examples and exercises.

This summary will only cover the very basics needed to work through this tutorial. Many
elements of the Cadabra syntax will not be discussed here. For a complete and definitive
reference please see the Cadabra web pages \url{https://cadabra.science/help.html}.

The first point to emphasise is that Cadabra is built upon Python and LaTeX and thus Cadabra
codes must adhere to their respective syntaxes.

% --------------------------------------------------------------------------------------------
{\bf Parsing}\\[5pt]
Parsing a Cadabra program serves two purposes. First, it checks for correctness of the code.
Second, it converts any statements unique to Cadabra (such as \verb|{a,b,c}::Indices|) into
statements that can be understood by Python. The result is a new program written entirely in
Python (with the Cadabra elements implemented as function calls to an external library).
This preprocess step can be seen in action using the command line tool
{\tts cadabra2python}. To create the Python code for the file {\tts foo.cdb} you need only
type
\bgroup
\lstset{numbers=none}
\begin{lstlisting}
   cadabra2python foo.cdb foo.py
\end{lstlisting}
\egroup

% --------------------------------------------------------------------------------------------
{\bf Statement termination}\\[5pt]
Since statements can be composed of one or more lines of text there must be some rule for
deciding when a series of lines constitutes a single statement. Python statements are
terminated according to Python's rules. Here are some examples of valid and invalid Python
statements.
\begin{cadabra}[numbers=none]
   foo = bah                 # valid
   foo = simplify (bah)      # valid
   bah := derive (foo)       # invalid, use = not := for Python assignment
\end{cadabra}
A Cadabra statement can be terminated using either a dot \verb|.|, a semi-colon \verb|;|
or the closing right parenthesis \verb|)| for functions and algorithms. Using a semi-colon
to terminate a statement will force Cadabra to print the output generated by the statement.
Here are two Cadabra statements, only the first is valid.
\begin{cadabra}[numbers=none]
   foo := A_{a} B_{b};       # valid
   bah  = B_{a} A_{b}.       # invalid, use := not = for Cadabra assignment
\end{cadabra}

% --------------------------------------------------------------------------------------------
{\bf Continuation}\\[5pt]
Python statements can be split across lines in a number of ways including line breaks
between items in a list. A slash at the end of line also signifies a continuation. This is
standard Python. For Cadabra the rules are not so simple. Properties (e.g.,
\verb|::Indices|) can \emph{not} be split across multiple lines. However, multiple instances
are allowed and will be stored as a sequence of property lists. Examples of this will be
seen later in Exercise \exref{ex:1.06} and Example \ref{sec:ex-12}. In contrast, Cadabra
expressions such as \verb|foo := A_{a} B_{b}.| \emph{can} be split across more than one line
by including line breaks as needed and with proper termination (e.g., a dot or a
semi-colon). Note that Python's indentation rules apply only to the first line of a group --
the remaining lines can be indented to suit. See also the discussion on very long lines in
the Miscellaneous section of \hyperlink{link2part3}{Part 3}.

% --------------------------------------------------------------------------------------------
{\bf Identifiers}\\[5pt]
Identifiers can be built using standard alphanumeric characters (excluding the special
characters like \verb|!@#$%^$| etc.). Python allows underscore characters but as they are
also used by LaTeX to introduce subscripts it is best to \emph{not} use an underscore in a
Cadabra identifier (it is allowed but it can cause confusion for the reader). In
this tutorial all identifiers will be built from the alphanumeric characters (a to z, A to Z
and 0 to 9) and occasionally LaTeX symbol names.

% --------------------------------------------------------------------------------------------
{\bf Assignment}\\[5pt]
Assignments in Python are made using \verb|=| as in \verb|foo = "abc"| while in Cadabra they
are made (mostly) using \verb|:=|. One reason for this small difference is the simple fact
that Python does not understand assignments made from LaTeX expressions. For example,
\verb|foo = A_{a} B_{b}| would make no sense in pure Python. Thus \verb|:=| is used to
signal that the assignment \verb|foo := A_{a} B_{b}.| must be made by Cadabra rather than
Python.

The same assignment can also be made using Cadabra's \verb|Ex| function using
\verb|foo = Ex(r"A_{a} B_{b}")|. This function takes a (raw) string, converts it into a
Python compliant datastructure (an \verb|Exnode|) and assigns the result to the left hand
side (i.e., to \verb|foo|). Since this statement is handled by Cadabra's own enhanced
version of Python (to include \verb|Ex|) the assignment uses \verb|=| rather than \verb|:=|.
Note also the absence of an explicit termination character (no dot or colon) and also the use
of the raw string \verb|r"..."|. The raw string is not needed in this example but would be
required if the string contained any slashes (e.g., a LaTeX symbol like \verb|\Gamma|). The
function \verb|Ex| is very useful when building expressions from smaller pieces (see for
example the function \verb|truncate| in Example \ref{sec:ex-04}).

It must be noted that identifiers carry no residual information about their origins
(\verb|=| or \verb|:=|). Thus any statement like \verb|bah = foo| will have the usual Python
meaning, namely, that \verb|bah| and \verb|foo| share one copy of the data pointed to by
\verb|foo|. There are many occasions were a second distinct copy of the data is required.
Copies of pure Cadabra objects (i.e., created using \verb|:=| or \verb|Ex|) can be made
using statements like \verb|bah := @(foo);|. The \verb|@(...)| is simply a function that
returns a copy of the given Cadabra object. This construction will be used many times in
this tutorial (the first instance can be seen in Exercise \exref{ex:1.07}).

% --------------------------------------------------------------------------------------------
{\bf Comment character}\\[5pt]
The hash character \verb|#| is used in Python to start a comment. All text on the line
following and including the hash will be ignored by Python. But in Cadabra the hash
character is used in many property declarations. Here are some examples (put aside for the
moment what these mean, just accept that they are valid Cadabra statements)
\begin{cadabra}[numbers=none]
   {a,b,c,d,e#}::Indices.
   \delta{#}::KroneckerDelta.
   D{#}::LaTeXForm{"\nabla"}.
\end{cadabra}
This dual use of the hash character forces a compromise to be made -- comments are not
allowed as trailing text on a pure Cadabra line (e.g., on the end of a property
declaration). Despite this restriction, there are a few occasions in this tutorial were
trailing comments are used for convenience and to save space. These trailing comments would
need to be removed before passing the code to Cadabra\footnote{This task is handled
automatically by the \cdbverb{hybrid-latex} scripts, see
\url{https://github.com/leo-brewin/hybrid-latex}}. Other comments, for example lines that
begin with a hash or as trailing text on a pure Python line, are allowed.

% --------------------------------------------------------------------------------------------
{\bf Indentation}\\[5pt]
All Cadabra programs must conform to Python's indentation rules. These rules may,
at first sight, seem strange for people not familiar with Python but they are not
too hard to understand. The basic idea is that code blocks that might in other
languages be wrapped in \verb|{}| or \verb|begin/end| pairs are indented by at least
one space (usually four spaces) from the surrounding code. This applies to
if-then-else blocks, for-loops, function definitions and nested blocks (and more).
Here are a few examples
\begin{cadabra}[numbers=none]
   foo = 123
   if foo == 123:
      bah = 456
      print ("in True")
   else:
      bah = 789
      print ("in False")
   print (bah)

   def swap (my_string):
       first_char   = my_string[0]
       second_char  = my_string[1]
       my_string[0] = second_char
       my_string[1] = first_char
       return my_string
\end{cadabra}
Similar indenting is often used in other languages as a way to improve the
readability of the code. In Python this use of indentation is mandatory.

% --------------------------------------------------------------------------------------------
{\bf CamelCase and snake\_case}\\[5pt]
Though Cadabra is case sensitive it does not stipulate which case to use for various
constructions. However, the common practice is to use \verb|CamelCase| for properties (e.g.,
\verb|::Indices|, \verb|::Derivative|) and \verb|snake_case| for algorithms and
functions (e.g., and \verb|sort_product|, \verb|product_rule|). Two obvious (trivial)
exceptions are the function \verb|Ex| and the use of uppercase LaTeX names for identifiers
such as \verb|\Gamma|.

% --------------------------------------------------------------------------------------------
{\bf Substitution rules}\\[5pt]
Cadabra's \verb|substitute| algorithm works its magic on an expression under the control
of a substitution rule (or a list of rules, see below). The rules can be specified either
as a named rule, for example,
\begin{cadabra}[numbers=none]
   foo := A^{a b} -> B^{a} C^{b}.
   substitute (bah, foo)
\end{cadabra}
or as an anonymous rule, for example,
\goodbreak
\begin{cadabra}[numbers=none]
   substitute (bah, $A^{a b} -> B^{a} C^{b}$)
\end{cadabra}
Both of these examples do the same job -- replacing $A^{ab}$ in \verb|bah| with
$B^{a} C^{b}$.

Rules can also be built using an equal sign rather than the forward arrow \verb|->|.
Thus the above examples could be written as
\begin{cadabra}[numbers=none]
   foo := A^{a b} = B^{a} C^{b}.
   substitute (bah, foo)
\end{cadabra}
and
\begin{cadabra}[numbers=none]
   substitute (bah, $A^{a b} = B^{a} C^{b}$)
\end{cadabra}
The end result will be exactly as before (replacing $A^{ab}$ with $B^{a} C^{b}$). There is
one important difference between these two constructions. Rules built using the equal sign
must have identical free indices on either side of the equal sign. In contrast, rules built
using \verb|->| are are not bound by this rule. This gives the programmer great flexibility
in manipulating the index structure of an expression -- it also opens the door for making a
complete mess of the expression (with great power comes great responsibility :). See items
\itref{it:SubRuleChoices} and \itref{it:SubRuleDanger} in Part 3 for more details on this
point.

Anonymous rules are convenient for one-off substitutions (and are used extensively in the
\hyperlink{GaussEqtn}{Gauss equation} example). They also provide data locality -- the rule
is in plain sight making clear the changes that are about to be applied. In contrast, a
named rule may be defined far away from its use thus forcing the reader to hunt through the
code for the definition. Named rules are useful when that rule is likely to be used many
times (for example a rule for a covariant derivative) or when the rule has an important
meaning (e.g., a rule for the Riemann tensor). Named rules, unlike anonymous rules, can also
be split across one or more lines, for example
\begin{cadabra}[numbers=none]
   RFromGamma  := R -> g^{a b} g^{c d} (    \partial_{c}{\Gamma_{a b d}}
                                          - \partial_{d}{\Gamma_{a b c}}
                                          + \Gamma_{e a d} \Gamma^{e}_{b c}
                                          - \Gamma_{e a c} \Gamma^{e}_{b d} ).
\end{cadabra}

You can also create a \emph{list} of rules using
\begin{cadabra}[numbers=none]
   RFromGamma := { R -> R_{a b} g^{a b},
                   R_{a b} -> R_{a c b d} g^{c d},
                   R_{a b c d} ->   \partial_{c}{\Gamma_{a b d}}
                                  - \partial_{d}{\Gamma_{a b c}}
                                  + \Gamma_{e a d} \Gamma^{e}_{b c}
                                  - \Gamma_{e a c} \Gamma^{e}_{b d} }.
\end{cadabra}
This rule contains three simple rules, delimited by commas and wrapped in braces (so it is a
Python list). To properly apply this rule you would need to use
\begin{cadabra}[numbers=none]
   RicciScalar := R.
   substitute (RicciScalar, RFromGamma)
   substitute (RicciScalar, RFromGamma)
   substitute (RicciScalar, RFromGamma)
\end{cadabra}
Now you might well ask -- Why are there three calls to \verb|substitute|? In the first call
only the first rule \verb|R -> R_{a b} g^{a b}| will be matched. The second and third calls
are needed to match the terms introduced in the previous calls. Having to call
\verb|substitute| many times is a bit tedious but the good news is that Cadabra provides a
short cut by allowing you to write
\begin{cadabra}[numbers=none]
   RicciScalar := R.
   substitute (RicciScalar, RFromGamma, repeat=True)
\end{cadabra}
The \verb|repeat=True| clause will force Cadabra to keep applying rules until the expression
settles down (i.e., no changes from one substitution to the next).

Lists of rules can be concatenated with other rules using the addition operator. Thus you
can also write
\begin{cadabra}[numbers=none]
   RFromRab := R -> R_{a b} g^{a b}.
   RabFromGamma := { R_{a b} -> R_{a c b d} g^{c d},
                     R_{a b c d} ->   \partial_{c}{\Gamma_{a b d}}
                                    - \partial_{d}{\Gamma_{a b c}}
                                    + \Gamma_{e a d} \Gamma^{e}_{b c}
                                    - \Gamma_{e a c} \Gamma^{e}_{b d} }.
   RFromGamma = RFromRab + RabFromGamma
\end{cadabra}

There is one caveat that must be noted -- the version of Cadabra\footnote{\CdbVersion} used
in this tutorial does not allow the addition of rules that \emph{each} contains just a
\emph{simple} rule (as simple rules are not \emph{lists}). A simple hack is to convert each
simple rule to a list by adding a trivial rule, (e.g., replace \verb|foo->bah| with
\verb|{foo->bah,x->x}|).

% --------------------------------------------------------------------------------------------
{\bf Line splitting}\\[5pt]
Cadabra allows \emph{expressions} to be split across one or more lines such as
\begin{cadabra}[numbers=none]
   Rabcd := R_{a b c d} ->   \partial_{c}{\Gamma_{a b d}}
                           - \partial_{d}{\Gamma_{a b c}}
                           + \Gamma_{e a d} \Gamma^{e}_{b c}
                           - \Gamma_{e a c} \Gamma^{e}_{b d}.
\end{cadabra}
However, it does \emph{not} allow property lists or anonymous rules (i.e., \verb|$...$|) to
be split. Thus each of the following statements will raise an error.
\begin{cadabra}[numbers=none]
   {a, b, c, d,
    e, f, g, h, i, j}::Indices.

   {R_{a b c d},
    \partial_{e}{R_{a b c d}}}::SortOrder.

    substitute (foo, $R -> R_{a b} g^{a b},
                      R_{a b} -> R_{a c b d} g^{c d}$)
\end{cadabra}
For property lists the preferred solution is to use one line (no matter how long it might
be). Thus you would use
\begin{cadabra}[numbers=none]
   {a, b, c, d, e, f, g, h, i, j}::Indices.

   {R_{a b c d},\partial_{e}{R_{a b c d}}}::SortOrder.
\end{cadabra}
The easiest solution for anonymous rules is to replace them with a named rule.

See the discussion on line splitting in \hyperlink{link2part3}{Part 3} below for an
alternative solutions for property lists and anonymous rules.

% \clearpage

% --------------------------------------------------------------------------------------------
\begin{Exercises}

   \begin{Exercise}
      \exlabel{ex:1.01}%
      Given that
      \begin{align*}
         \cdb{libdg.Gamma.000} = \cdb{libdg.Gamma.001}
      \end{align*}
      use Cadabra to verify that
      \begin{align*}
         \Gamma^{a}{}_{bc} = \Gamma^{a}{}_{cb}
      \end{align*}
      {\bf Hint:} Define a rule for $\Gamma^{a}{}_{bc}$ based on the above definition.
      Then apply that rule to the expression $\Gamma^{a}{}_{bc} - \Gamma^{a}{}_{cb}$ and
      finally use suitable Cadabra algorithms to simplify the result.
   \end{Exercise}

   % this pushes to the footnote to the bottom of the page
   % \hrule height 0pt% probably not needed
   % \vfill% this does the trick

   \begin{Exercise}
      \exlabel{ex:1.02}%
      Define $\Gamma_{abc}$ (the Christoffel symbols of the first kind) by
      \begin{align*}
         \Gamma_{abc} = g_{ad} \Gamma^{d}{}_{bc}
      \end{align*}
      Use Cadabra to verify that
      \begin{align*}
         \Gamma_{abc} + \Gamma_{bac} = \partial_{c} g_{ab}
      \end{align*}
      {\bf Hint:} Define two rules, one for $\Gamma^{a}{}_{bc}$ as per the previous exercise
      and one for $\Gamma_{abc}$ as per the above definition. Apply both rules to the
      expression $\Gamma_{abc} + \Gamma_{bac} - \partial_{c} g_{ab}$ then use suitable
      Cadabra algorithms to simplify the result.
   \end{Exercise}

   \begin{Exercise}
      \exlabel{ex:1.03}%
      Modify your Cadabra code from the previous example to apply just \emph{one} rule to
      $\Gamma_{abc} + \Gamma_{bac} - \partial_{c} g_{ab}$.

      {\bf Hint:} Cadabra allows rules to act not only on expressions but also on other
      rules. Use this feature to construct a single rule from the original pair.

      {\bf Note.} To avoid a Cadabra runtime error you may need to replace \verb|::Indices.|
      with \verb|::Indices(position=independent).| This point will be discussed in more
      detail in the following example (on covariant differentiation).
   \end{Exercise}

   \begin{Exercise}
      \exlabel{ex:1.04}%
      This exercise is a brief experiment with Cadabra's \verb|sort_product| algorithm.
      Apply \verb|sort_product| to each of the following expressions and carefully note the
      result. You should be able to glean from these examples the default sort order used by
      Cadabra.
      \begin{align*}
         (1)\hskip 1cm & C^{f} w^{e} B^{d} v^{c} A^{b} u^{a}\\
         (2)\hskip 1cm & \Omega_{f} \gamma_{e} \Pi_{d} \beta_{c} \Gamma_{b} \alpha_{a}\\
         (3)\hskip 1cm & C^{f} w^{e} B^{d} v^{c} A^{b} u^{a} \Omega_{f}
                         \gamma_{e} \Pi_{d} \beta_{c}
                         \Gamma_{b} \alpha_{a}\\
         (4)\hskip 1cm & \partial_{f}{C^{f}} w^{l}
                         \partial_{d}{B^{d}} v^{k} \partial_{b}{A^{b}} u^{j}
                         \Omega_{i} \partial^{e}{\gamma_{e}} \Pi_{h} \partial^{c}{\beta_{c}}
                         \Gamma_{g} \partial^{a}{\alpha_{a}}\\
         (5)\hskip 1cm & \partial{C} w \partial{B} v \partial{A} u \Omega
                         \partial{\gamma} \Pi
                         \partial{\beta} \Gamma \partial{\alpha}\\
         (6)\hskip 1cm & A_{b} A_{a} A_{c d e} A_{f g}\\
         (7)\hskip 1cm & A_{a} A^{a}+A^{a} A_{a}
      \end{align*}
      The results of the first four examples shows that Cadabra's default sort can be
      summarised as
      \begin{center}
         UPPERCASE $<$ SLASH-UPPERCASE $<$ slash-lowercase $<$ lowercase
      \end{center}
      The fifth example shows that Cadabra's default sort ordering usually ignores indices.
      The exception, as shown in the final pair of examples, is when object names are
      repeated. In such cases Cadabra will sort the terms based on their indices.

      Cadabra does allow some control over the sort order by explicitly listing the order
      in a \verb|::SortOrder| property. Each of the following are valid instances of
      a sort order list
      \begin{cadabra}
         {F,E,D,C,B,A}::SortOrder.
         {R_{a b}, R_{a b c d}, R^{a b c d}}::SortOrder.
         {\partial_{a}{g_{b c}}, \partial{a b}{R}}::SortOrder.
      \end{cadabra}
   \end{Exercise}

   \begin{Exercise}
      \exlabel{ex:1.05}%
      Look back at the last example in the previous exercise. Cadabra returned $A_{a} A^{a}
      + A^{a} A_{a}$ which, assuming $A$ is self-commuting, can be simplified to $2A_{a}
      A^{a}$. If the original expression had been $A_{a} Z^{a} + Z^{a} A_{a}$ then the
      result (after \verb|sort_product|) would have been $2 A_{a} Z^{a}$. This should give
      you a clue as to how the first expression (involving just $A$) can be sorted to give
      $2A_{a} A^{a}$. Write a code that does the job. An extension of this idea will
      be developed later in Exercise \exref{ex:4.06}.
   \end{Exercise}

   \begin{Exercise}
      \exlabel{ex:1.06}%
      Cadabra does allow multiple instances of the \verb|::SortOrder| property. Run the
      following code through Cadabra and observe the result.
      \begin{cadabra}
         {D,C,B,A}::SortOrder.

         foo := A B C D;
         sort_product (foo);

         {V,U}::SortOrder.

         foo := U V A B C D;
         sort_product (foo);

         {A,B,C,D}::SortOrder.

         foo := U V D C B A;
         sort_product (foo);
      \end{cadabra}
      The results may seem surprising. The final results for \verb|foo| is
      \verb|foo = D C B A V U|.
      But looking at third instance of \verb|SortOrder| it is reasonable to expect
      \verb|foo = A B C D V U|. How can this be? The answer lies in how Cadabra handles
      multiple instances of the \verb|SortOrder|. The logic is a bit tricky but it goes as
      follows. The sorting is done using Bubble Sort. This works by sorting a list one pair
      at a time. Suppose \verb|P| and \verb|Q| define a pair \verb|PQ|. The correct order
      might require the pair to be swapped. That decision, to swap or not, is made by first
      searching for the first list that contains \verb|P|. If that list also contains
      \verb|Q| then that list will be used to determine if \verb|P|and \verb|Q| should be
      swapped. In all other cases (i.e., when a suitable \verb|SortOrder| list can not be
      found) the correct order for \verb|PQ| will be found from Cadabra's default sort order.

      The upshot is that repeat entries in \verb|SortOrder|, either in a list or across
      lists, play no part in setting the order. The repeat entries, such as the entire third
      list above, will in effect by ignored.

      An alternative to using \verb|SortOrder| will be presented later in Exercise
      \exref{ex:4.06}.
   \end{Exercise}

   \begin{Exercise}
      \exlabel{ex:1.07}%
      This exercise explores the differences between \verb|foo = bah| and
      \verb|foo := @(bah)|.
      The first ensures that both \verb|foo| and \verb|bah| share the same data.
      Any changes to either \verb|foo| or \verb|bah| will be shared by its partner. Using
      \verb|foo := @(bah)| creates a fresh copy of \verb|bah| and assigns \verb|foo| to that
      copy. Any subsequent changes to \verb|foo| will not be reflected in \verb|bah| and
      vice-versa.

      The following code demonstrates this behaviour by using the \verb|id| function to
      reveal the location in the computer's memory where the object resides (i.e., a memory
      address). Careful inspection of the source and the corresponding output should
      convince you that the above description is correct.

      \begin{cadabra}
         {a,b,c,d,e,f,h#}::Indices.

         foo := B_{b} A_{a}.
         bah := A_{a} C_{c}.

         print("foo = "+str(foo))
         print("bah = "+str(bah)+"\n")

         print("type foo = "+str(type(foo)))
         print("type bah = "+str(type(bah))+"\n")

         print("id foo = "+str(id(foo)))
         print("id bah = "+str(id(bah))+"\n")

         bah = foo

         print("foo = "+str(foo))
         print("bah = "+str(bah)+"\n")

         sort_product (foo)

         print("bah = "+str(bah)+"\n")

         print("id foo = "+str(id(foo)))
         print("id bah = "+str(id(bah))+"\n")

         bah := @(foo).

         print("id foo = "+str(id(foo)))
         print("id bah = "+str(id(bah))+"\n")
      \end{cadabra}

   \end{Exercise}

   \begin{Exercise}
      \exlabel{ex:1.08}%
      The following code contains a number of syntax errors. Identify and correct the errors
      then test the corrected code by running it through Cadabra.
      \begin{cadabra}
         {a,b,c,d,e,f#}::Indices.
         C{#}::Symmetric.

         foo := A_{a} B_{b} + C_{ab}.
         bah := B_{b} A_{a} + C_{ba}.
         meh := @(foo) - @(bah)

         if meh == 0:
            print ("meh is zero, and all is good")
               success = True.
         else:
            print ("meh is not zero, oops")
               success = False.

         canonicalise (meh).
         sort_product (meh);

         {\alpha\beta\gamma}::Indices.

         foo := Ex ("A_{ab} - A_{a b}");
         bah := Ex ("A_{\alpha\beta} - A_{\alpha \beta}");
      \end{cadabra}

   \end{Exercise}

   \begin{Exercise}
      \exlabel{ex:1.09}%
      This and the following two exercises deal with simple index manipulations. Consider a
      pair of tensors $A_a$ and $B_b$ defined by
      \begin{align*}
         A_{a} = A_{a c} C^{c}\qquad\text{and}\qquad B_{b} = B_{b c} C^{c}
      \end{align*}
      The two tensors have distinct free indices but share a common dummy index $c$. How
      does Cadabra handle the possible index clash when constructing a product of $A_{a}$
      with $B_{b}$? The answer can be found by running this simple code
      \begin{cadabra}
         {a,b,c,d,e,f,h#}::Indices.

         foo := A_{a c} C^{c}.
         bah := B_{b c} C^{c}.

         foobah := @(foo) @(bah).
      \end{cadabra}
      Run the above code and look closely at the result. You should notice that Cadabra has
      automatically adjusted the dummy indices to avoid a clash.
   \end{Exercise}

   \begin{Exercise}
      \exlabel{ex:1.10}%
      Another common index operation is to relabel the free indices. Write a Cadabra code
      that relabels $A_{a b c}$ to $A_{u v w}$. You can do this by contracting
      $A_{a b c}$ with suitably chosen Kronecker deltas.
   \end{Exercise}

   \begin{Exercise}
      \exlabel{ex:1.11}%
      Suppose now that you need to cycle the free indices, say from $A_{a b c}$ to
      $A_{b c a}$. This can be done using two rounds of Kronecker deltas. But there is an
      elegant and simpler solution using Cadabra's substitution rules. The idea is to
      create a rule that replaces a temporary object like $T_{a b c}$ with $A_{a b c}$.
      Then apply that rule (using \verb|susbtitute|) to $T_{b c a}$. Note the cycled
      indices on $T$. Write a Cadabra code that implements this neat trick.
   \end{Exercise}

\end{Exercises}
% --------------------------------------------------------------------------------------------

\clearpage

% ============================================================================================
\section{Covariant differentiation}% LCB: something better please ?
\label{sec:ex-02}
\ResetCounters

\input{./cadabra/example-02.cdbtex}

Cadabra does not have native algorithms to compute covariant derivatives, Riemann
tensors, Ricci tensors and so on. One of its strengths is that it provides a rich set of
simple tools by which such objects can be constructed. This second example will show how
Cadabra can be trained to compute covariant derivatives.

For a simple vector such as $v^a$ the standard textbook definition of the covariant
derivative $\nabla_{b} v^{a}$ is
\begin{align*}
   \nabla_{b} v^{a} = \partial_{b} v^{a} + \Gamma^{a}{}_{cb} v^{c}
\end{align*}
A simple way to implement this in Cadabra would be to first define symbols to represent the
derivative operators
\begin{cadabra}[numbers=none]
   \nabla{#}::Derivative.
   \partial{#}::PartialDerivative.
\end{cadabra}
and then define a rule for the actual covariant derivative
\begin{cadabra}[numbers=none]
   deriv := \nabla_{a}{v^{b}} -> \partial_{a}{v^{b}} + \Gamma^{b}_{c a} v^{c}.
\end{cadabra}
This rule could then be used to replace any instances of $\nabla_{b} v^a$ in a Cadabra
expression such as \verb|foo| with the appropriate partial derivatives and Christoffel
symbols using
\begin{cadabra}[numbers=none]
   substitute (foo,deriv)
\end{cadabra}
From here it is a simple matter to construct a working code -- just add a definition for the
indices and some lines to simplify the output. This leads to the following minimal working
code.
\begin{cadabra}
   {a,b,c,d,e,f,g,h,i,j,k,l,m,n,o,p,q,r,s,t,u#}::Indices.

   \nabla{#}::Derivative.
   \partial{#}::PartialDerivative.

   # rule for covariant derivative of v^{a}

   deriv := \nabla_{a}{v^{b}} -> \partial_{a}{v^{b}} + \Gamma^{b}_{c a} v^{c}. |\lstlabel{ex-02.001}|

   # create an expression

   foo := \nabla_{a}{v^{b}}.

   # apply the rule, then simplify

   substitute    (foo,deriv)    |\lstlabel{ex-02.102}|
   canonicalise  (foo)          |\lstlabel{ex-02.103}|
\end{cadabra}
The corresponding output is
\begin{dgroup*}[spread={3pt}]
   \Dmath*{\cdb{ex-02.101} = \cdb{ex-02.102}\hfill\lstref{ex-02.102}
                           = \cdb{ex-02.103}\hfill\lstref{ex-02.103}}
\end{dgroup*}
The first line in the output is as expected -- it simply repeats the definition given above.
However, the second line is not exactly as expected -- note how the second index on the
Christoffel symbol has been raised (while the corresponding index on $v$ has been lowered).
Though this is mathematically correct, it is not standard practice and it would be better if
Cadabra could be persuaded to not do such index gymnastics. The solution is to inform Cadabra
that the upper and lower indices are to be left as is by adding the qualifier
\verb|position=independent| to the \verb|::Indices| property. That is
\bgroup
\lstset{firstnumber=1}
\begin{cadabra}
   {a,b,c,d,e,f,g,h,i,j,k,l,m,n,o,p,q,r,s,t,u#}::Indices(position=independent).
\end{cadabra}
\egroup
The corresponding output is now
\begin{dgroup*}[spread={3pt}]
   \Dmath*{\cdb{ex-02.201} = \cdb{ex-02.202}\hfill\lstref{ex-02.102}
                           = \cdb{ex-02.203}\hfill\lstref{ex-02.103}}
\end{dgroup*}
In this instance the changes brought about by specifying \verb|(position=independent)| are
simply cosmetic. There are, however, cases where strict control must be maintained over the
raising and lowering of indices (usually by explicit use of the metric). This is particularly
true for expressions that involve derivative operators. Without the
\verb|(position=independent)| qualifier the \verb|canonicalise| algorithm might
(incorrectly) raise or lower an index inside the derivative, such as $b$ in $\partial_{a}
V^{b}$. Of course, if the derivative operator is compatible with the metric (e.g., $\nabla
g=0$) then there is no issue and the indices can be declared without the
\verb|(position=independent)| qualifier (though the aesthetics of the output might not be
ideal).

The above discussion also explains why \verb|(position=independent)| was required in
Exercise \exref{ex:1.03}. Without it Cadabra will treat \verb|\Gamma_{a b c}| and
\verb|\Gamma^{a}_{b c}| as one and the same. Thus any attempt to apply a substitution on
\verb|\Gamma^{a}_{b c}| in the rule
\Break
\verb|\Gamma_{a b c} -> g_{a d}\Gamma^{d}_{b c}|
will actually be applied to both \verb|\Gamma| terms. This removes all trace of
\verb|\Gamma| from the rule (you can verify this by making small changes to your code from
Exercise \exref{ex:1.03}). See also Exercise \exref{ex:2.08} for more adventures with
indices.

There remains one minor problem with the above code -- the rule in line \lstref*{ex-02.001}
was designed explicitly for covariant derivatives of $v^a$ and thus is not applicable to
other objects such as $u^a$ or expressions like $u^a+v^a$. The solution lies in defining a
rule that is applicable to a wider class of objects. Cadabra has a simple syntax that uses a
single post-fix question mark to define a generic object. Thus \verb|A?| will match objects
such as \verb|P|, \verb|Q|, \verb|PQ| etc. The upshot is that the original rule for the
covariant derivative can be generalised to
\bgroup
\lstset{firstnumber=6}
\begin{cadabra}
   # template for covariant derivative of a vector

   deriv := \nabla_{a}{A?^{b}} -> \partial_{a}{A?^{b}} + \Gamma^{b}_{c a} A?^{c}.
\end{cadabra}
\egroup
This rule will work as expected when applied to $\nabla_a u^b$, $\nabla_a v^b$ and
$\nabla_a u^b + \nabla_a v^b$.

% \clearpage

% --------------------------------------------------------------------------------------------
\begin{Exercises}

   \begin{Exercise}
      \exlabel{ex:2.01}%
      Use the definitions
      \begin{align*}
         \nabla_{a} u^b &= \partial_{a} u^b + \Gamma^{b}{}_{ca} u^c\\
         \noalign{and}
         \nabla_{a} v_b &= \partial_{a} v_b - \Gamma^{c}{}_{ab} v_c
      \end{align*}
      to verify that
      \begin{align*}
         \nabla_{a}\left(v_b u^b\right) = \partial_{a}\left(v_b u^b\right)
      \end{align*}
      {\bf Hint:} Begin by applying the product rule to $\nabla_{a}\left(v_b
      u^b\right)-\partial_{a}\left(v_b u^b\right)$. You can do so using either Cadabra's
      \verb|product_rule| algorithm or you can create two rules, one for each of the
      derivative operators, $\nabla$ and $\partial$. You might also need \verb|product_sort|
      and \verb|rename_dummies| for housekeeping.

   \end{Exercise}

   \begin{Exercise}
      \exlabel{ex:2.02}%
      Given $A_a$ and $B_b$, define $v_{ab}$ by $v_{ab} = A_a B_b$. Adapt your Cadabra
      codes for $\nabla_{a} v_b$ to verify that
      \begin{align*}
         \nabla_{a}{v_{bc}} = \partial_{a}{v_{bc}}
                            - \Gamma^{d}{}_{ba} v_{dc}
                            - \Gamma^{d}{}_{ca} v_{bd}
      \end{align*}
   \end{Exercise}

   \begin{Exercise}
      \exlabel{ex:2.03}%
      In a similar vein, given $v^{a}{}_{b} = A^{a} B_{b}$ show that
      \begin{align*}
         \nabla_{a}{v^{b}{}_{c}} = \partial_{a}{v^{b}{}_{c}}
                                 + \Gamma^{b}{}_{da} v^{d}{}_{c}
                                 - \Gamma^{d}{}_{ca} v^{b}{}_{d}
      \end{align*}
      This and the previous exercise show how easy it is to use Cadabra to verify standard
      textbook definitions for covariant derivatives. Setting $v^{a}{}_{b} = A^{a} B_{b}$
      might appear to limit the validity of the above result. However, since any tensor can
      be built as a linear combination of products of vectors and dual-vectors and as the
      above is linear in $v^{a}{}_{b}$ it follows that the result does hold for any choice
      of $v^{a}{}_{b}$ (as expected). This same trick could be used to discover equations
      for covariant derivatives of any tensor, however, it is much easier to just code up
      the textbook definition as shown in the following example.
   \end{Exercise}

   \begin{Exercise}
      \exlabel{ex:2.04}%
      The objective in this and the following exercise is to build a single rule that
      expresses $\nabla_a \nabla_b v^c$ in terms of $v$, $\Gamma$ and their partial
      derivatives. As a start, use the following fragment to build a Cadabra code.
      Observe the result of the call to \verb|substitute|.
      \begin{cadabra}
         deriv1 := \nabla_{a}{v^{b}}             -> \partial_{a}{v^{b}}
                                                  + \Gamma^{b}_{d a} v^{d}.

         deriv2 := \nabla_{a}{\nabla_{b}{v^{c}}} -> \partial_{a}{\nabla_{b}{v^{c}}}
                                                  + \Gamma^{c}_{d a} \nabla_{b}{v^{d}}
                                                  - \Gamma^{d}_{b a} \nabla_{d}{v^{c}}.

         substitute (deriv2,deriv1)
      \end{cadabra}

   \end{Exercise}

   \begin{Exercise}
      \exlabel{ex:2.05}%
      The previous exercise showed that calls to \verb|substitute| will be applied to
      all terms in an expression, in this case to both sides of the rule \verb|deriv2|.
      One way to avoid this problem is to ensure that the left hand side of the expression
      does not contain the target of the rule being applied. Using the same rules as above
      for \verb|deriv1| and \verb|deriv2| build a new code using
      \begin{cadabra}
         expr := v^{c}_{b a} -> \nabla_{a}{\nabla_{b}{v^{c}}}.

         substitute (expr,deriv2)
         substitute (expr,deriv1)
      \end{cadabra}
      You might like to tidy the final result by substituting $\nabla_{a}{\nabla_{b}{v^{c}}}$
      for $v^{c}{}_{b a}$. A variation on this code will be presented in the following
      section on the Riemann tensor.

   \end{Exercise}

   \begin{Exercise}
      \exlabel{ex:2.06}%
      Use Cadabra to verify that for any scalar function $\phi$
      \begin{align*}
         (\nabla_{a} \nabla_{b} -  \nabla_{b} \nabla_{a}) \phi
         = (\Gamma^{c}{}_{ab}-\Gamma^{c}{}_{ba}) \partial_{c} \phi
      \end{align*}
   \end{Exercise}

   \begin{Exercise}
      \exlabel{ex:2.07}%
      A popular strategy in proving various theorems in differential geometry is to first
      assume that coordinates have been chosen so that the metric connection
      vanishes at some (arbitrarily) chosen point. This step kills a whole raft of
      terms and from there the theorem becomes almost trivial to prove. Suppose that this
      step, of setting $\Gamma = 0$, is to be applied to the following expression
      (this is not part of any deep theorem it was invented just to set the scene)
      \begin{align*}
         \Gamma^{a}{}_{bc}(x) = \Gamma^{a}{}_{b c}
                            + x^{d} \partial_{d}{\Gamma^{a}{}_{b c}}
      \end{align*}
      Write a Cadabra code that uses a substitution rule to set $\Gamma = 0$ while
      retaining the partial derivative.
   \end{Exercise}

   \begin{Exercise}
      \exlabel{ex:2.08}%
      Cadabra actually has three choices for the \verb|position| keyword, namely
      \verb|position=free|, \verb|position=fixed| and \verb|position=independent|. with
      \verb|position=free| as the default. The difference between the three choices is the
      degree of freedom given to Cadabra in raising and lowering indices. As already seen,
      \verb|position=free| allows Cadabra to freely raise and lower indices while
      \verb|position=indpendent| instructs Cadabra to leave index raising and lowering to the
      user. The choice \verb|position=fixed| lies between these two extremes. It will allow
      \verb|canonicalise| to raise and lower matching dummy indices. These three cases are
      demonstrated in the following code. Run the code and look closely the output. You
      should see the behaviour just described.
      \begin{cadabra}
         {a,b,c}::Indices(position=free).

         foo := A_{a b} + A^{a b}.

         substitute (foo, $A_{a b} -> B_{a b}$)

         {p,q,r}::Indices(position=fixed).

         foo := A_{p q} B^{p q} + A^{p q} B_{p q}.

         canonicalise (foo)

         {u,v,w}::Indices(position=independent).

         foo := A_{u v} B^{u v} + A^{u v} B_{u v}.

         canonicalise (foo)
      \end{cadabra}
      Note that mixed indices as in $A_{ab}+A^{ab}$ should never occur in general
      relativity. Cadabra will flag such cases as an error when using \verb|position=fixed|
      or \verb|position=independent|.

   \end{Exercise}

\end{Exercises}
% --------------------------------------------------------------------------------------------

These exercises show that it is not too hard to create rules for each covariant derivative
of interest though it might be tedious listing all possible cases (even when using
constructions like \verb|A?| etc.). Unfortunately, Cadabra's pattern matching repertoire,
such as \verb|A?|, does not extend to arbitrary tensors. Thus it is not possible to write a
single rule that covers every possible form of covariant derivative. However, with Cadabra's
native interface to Python, it is possible to write a function that will return the full
covariant derivative for an arbitrary tensor. Unfortunately, the inner workings of this
function draw upon many aspects of Cadabra's core syntax that are beyond the scope of this
tutorial. For full details see \url{https://cadabra.science/notebooks/ref_programming.html}

\clearpage

% ============================================================================================
\section{To Riemann and beyond}
\label{sec:ex-03}
\ResetCounters

\input{./cadabra/example-03.cdbtex}

The Riemann tensor for a symmetric connection can be computed (in a coordinate basis) using
\Dmath*{\cdb{libdg.Rabcd.000} = \cdb{libdg.Rabcd.001}}
A standard computation in differential geometry then shows that
\Dmath*{\cdb{ex-03.104} = - R^{a}{}_{dbc} V^d}
where the symbol \verb|;| denotes covariant differentiation for the connection
$\Gamma^{a}{}_{bc}$. The purpose of this example is to show how Cadabra can be used to
recover the above definition of $R^{a}{}_{bcd}$ by direct computation of the left hand side
of the previous equation.

One way to expand $\cdb{ex-03.104}$ is combine two expressions, one for $V^{a}{}_{;b}$ and
one for $V^{a}{}_{b;c}$ with $V^{a}{}_{b}$ equal to $V^{a}{}_{;b}$. This suggest the
following Cadabra fragment
\begin{cadabra}[numbers=none]
   # rules for the first two covariant derivs of V^a

   deriv1 := V^{a}_{; b}      -> \partial_{b}{V^{a}}
                               + \Gamma^{a}_{c b} V^{c}.

   deriv2 := V^{a}_{; b ; c}  -> \partial_{c}{V^{a}_{; b}}
                               + \Gamma^{a}_{d c} V^{d}_{; b}
                               - \Gamma^{d}_{b c} V^{a}_{; d}.
\end{cadabra}
Though this is a faithful transcription of the underlying mathematics this fragment is taking
a small liberty with the syntax -- Cadabra might treat the \verb|;| as a tensor index despite
not being declared in the list of valid indices (i.e., the \verb|::Indices|). It turns out
that Cadabra is smart enough to not make this mistake, either by good design or by good
fortune. However, any ambiguity (on Cadabra's part) can be avoided by using
\begin{cadabra}[numbers=none]
   # force ; to not be seen as a tensor index

   ;::Symbol;
\end{cadabra}
There remains one issue (before looking at the complete code) -- How can Cadabra be informed
that the connection is symmetric? Cadabra does support the \verb|::Symmetric| and
\verb|::AntiSymmetric| properties but these apply to \emph{all} of the indices of
the attached objects. For the case of $\Gamma^{a}{}_{bc}$, which is symmetric only on the
lower pair of indices, Cadabra provides a more sophisticated property as follows
\begin{cadabra}[numbers=none]
   \Gamma^{a}_{b c}::TableauSymmetry(shape={2}, indices={1,2});
\end{cadabra}
This does look a bit cryptic so a brief explanation of the syntax would be helpful. But
doing so at this stage will take the discussion to far from the current objective -- to
compute the Riemann tensor. Thus a deeper explanation will be deferred until after the main
results have been presented. Here now is the complete Cadabra code.
\begin{cadabra}
   {a,b,c,d,e,f,g,h,i,j,k,l,m,n,o,p,q,r,s,t,u#}::Indices(position=independent).

   \partial{#}::PartialDerivative.

   \Gamma^{a}_{b c}::TableauSymmetry(shape={2}, indices={1,2});

   # force ; to not be seen as a tensor index

   ;::Symbol;

   # rules for the first two covariant derivs of V^a

   deriv1 := V^{a}_{; b}      -> \partial_{b}{V^{a}}            |\lstlabel{ex-03.101}|
                               + \Gamma^{a}_{c b} V^{c}.

   deriv2 := V^{a}_{; b ; c}  -> \partial_{c}{V^{a}_{; b}}      |\lstlabel{ex-03.102}|
                               + \Gamma^{a}_{d c} V^{d}_{; b}
                               - \Gamma^{d}_{b c} V^{a}_{; d}.

   substitute (deriv2,deriv1)       |\lstlabel{ex-03.103}|

   # commute the second covariant derivatives

   Vabc := V^{a}_{; b ; c} - V^{a}_{; c ; b}.   |\lstlabel{ex-03.104}|

   substitute (Vabc,deriv2)         |\lstlabel{ex-03.105}|

   distribute     (Vabc)            |\lstlabel{ex-03.106}|
   product_rule   (Vabc)            |\lstlabel{ex-03.107}|

   # tidy up

   sort_product   (Vabc)            |\lstlabel{ex-03.108}|
   rename_dummies (Vabc)            |\lstlabel{ex-03.109}|
   canonicalise   (Vabc)            |\lstlabel{ex-03.110}|
   sort_sum       (Vabc)            |\lstlabel{ex-03.111}|
   factor_out     (Vabc,$V^{a?}$)   |\lstlabel{ex-03.112}|
\end{cadabra}

The three rules used in the above code are reported by Cadabra as follows
\begin{dgroup*}[spread={3pt}]
   \Dmath*{\cdb{ex-03.101}\hfill\lstref{ex-03.101}}
   \Dmath*{\cdb{ex-03.102}\hfill\lstref{ex-03.102}}
   \Dmath*{\cdb{ex-03.103}\hfill\lstref{ex-03.103}}
\end{dgroup*}
The last of these (obtained by substituting the first rule into the second) can be used to
expand $\cdb{ex-03.104}$. This leads to
\begin{dgroup*}[spread={3pt}]
   \Dmath*{\cdb{ex-03.104} = \cdb{ex-03.105}\hfill\lstref{ex-03.105}}
\end{dgroup*}
Now the simplifications begin. First the brackets are expanded
\begin{dgroup*}[spread={3pt}]
   \Dmath*{\cdb{ex-03.104} = \cdb{ex-03.106}\hfill\lstref{ex-03.106}}
\end{dgroup*}
followed by the product rule
\begin{dgroup*}[spread={3pt}]
   \Dmath*{\cdb{ex-03.104} = \cdb{ex-03.107}\hfill\lstref{ex-03.107}}
\end{dgroup*}
Notice that some obvious cancelations have not been made (e.g., the $\partial^2_{bc} V^a$
terms could be cancelled). These cancellations (and other minor aesthetic improvements) will
be handled by the \verb|canonicalise| algorithm. In order to allow \verb|canonicalise| to
catch as many simplifications as possible it is common to do some basic housekeeping on the
expression before calling \verb|canonicalise|. In most cases it is sufficient to sort the
products then rename the dummy indices. This leads to
\begin{dgroup*}[spread={3pt}]
   \Dmath*{\cdb{ex-03.104} = \cdb{ex-03.108}\hfill\lstref{ex-03.108}
                           = \cdb{ex-03.109}\hfill\lstref{ex-03.109}
                           = \cdb{ex-03.110}\hfill\lstref{ex-03.110}}
\end{dgroup*}
The final pair of lines in the above code massages the Cadabra output into a familiar form
\begin{dgroup*}[spread={3pt}]
   \Dmath*{\cdb{ex-03.104} = \cdb{ex-03.111}\hfill\lstref{ex-03.111}
                           = \cdb{ex-03.112}\hfill\lstref{ex-03.112}}
\end{dgroup*}

% --------------------------------------------------------------------------------------------
\subsection{Symmetry and Young diagrams}
\label{sec:young-diagrams}

As noted above, the syntax involving the \verb|::TableauSymmetry| does require some (limited)
explanation. Cadabra uses sophisticated algorithms to handle tensor symmetries based on
the Littlewood-Richardson algorithm for finding a basis of the irreducible representations of
totally symmetric groups. The algorithm uses Young diagrams which consist of a set of cells
arranged as series of rows which in turn are described by the \verb|::TableauSymmetry|
property. In short, the index symmetries of a tensor are encoded in these diagrams. The
\verb|shape={...}| parameter describes the shape of a Young diagram, in this case it consists
of one row with two cells. The \verb|indices={...}| parameter describes how the tensor's
indices are assigned to the cells. For this purpose, the indices on the tensor are counted
from left to right starting with zero. So in the above example the lower two indices $b$ and
$c$ are counted as 1 and 2 and they are assigned to the two cells of the Young diagram. More
details on using tableaux as a way to describe tensor symmetries can be found in the Cadabra
manual.

% --------------------------------------------------------------------------------------------
\subsection{A cheap hack for a symmetric connection}
\label{sec:cheap-hack}

If Young diagrams and tableaux seem a bit too cryptic then there is a (less than ideal)
alternative. One way to obtain a symmetric connection is to temporarily put
$\Gamma^{a}{}_{bc} = G^{a} G_{bc}$ where $G_{bc}=G_{cb}$, ask Cadabra to make its
simplifications and then return the $\Gamma^{a}{}_{bc}$ to the result. This is not a
mathematical operation, it is just a trick to help Cadabra spot what symmetries are
available. Here is a fragment of code that does the job (in the absence of any
\verb|::TableauSymmetry|)

\begin{cadabra}[numbers=none]
   ...
   # trick to impose zero torsion (symmetric connection)

   G_{a b}::Symmetric.

   substitute     (Vabc,$\Gamma^{a}_{b c} ->  G^{a} G_{b c}$)
   sort_product   (Vabc)
   rename_dummies (Vabc)
   canonicalise   (Vabc)
   substitute     (Vabc,$G^{a} G_{b c} -> \Gamma^{a}_{b c}$,repeat=True)

   # tidy up and display the results
   ...
\end{cadabra}

The problem with this approach is that if the pair of terms $G^{a}$ and $G_{bc}$ ever get
separated (e.g., from a product rule) then it may not be possible to complete the last step
of this trick, that is, to eliminate the $G^{a}$ and $G_{ab}$ in favour of
$\Gamma^{a}{}_{bc}$. Another road to danger lies in playing this trick when products of
connections are involved. For example, using this trick on
$\Gamma^{a}{}_{bc}\Gamma^{d}{}_{ef} - \Gamma^{d}{}_{bc}\Gamma^{a}{}_{ef}$ would cause all
terms to cancel giving zero as the result. This is clearly wrong. But if it can be shown
that such problems can not arise (e.g., there are no derivatives or the equations are linear
in the connection) then this method is rather easy to apply. It also provides a quick way to
implement more complicated symmetries (e.g., if $A_{abcde}$ is symmetric in the first two
and last three indices put $A_{abcde} = G_{ab}G_{cde}$ with both $G_{ab}$ and $G_{abc}$
declared as \verb|::Symmetric|).

Note the use of \verb|repeat=True| in the call to \verb|substitute| in the above code.
It ensures that every product $G^{a} G_{b c}$ is replaced with $\Gamma^{a}{}_{b c}$. This
point is explored further in Exercise \exref{ex:3.10} below.

% \clearpage

% --------------------------------------------------------------------------------------------
\begin{Exercises}

   \begin{Exercise}
      \exlabel{ex:3.01}%
      Write one or more Cadabra codes to verify the following symmetries of $R^{a}{}_{bcd}$
      \begin{align*}
          0 &= R^{a}{}_{bcd} + R^{a}{}_{bdc}\\
          0 &= R^{a}{}_{bcd} + R^{a}{}_{dbc} + R^{a}{}_{cdb}\\
          0 &= R^{a}{}_{bcd;e} + R^{a}{}_{bec;d} + R^{a}{}_{bde;c}
      \end{align*}
   \end{Exercise}

   \begin{Exercise}
      \exlabel{ex:3.02}%
      Rewrite the code given in the above text for $R^{a}{}_{bcd}$ to use $\nabla$ as the
      derivative operator rather than the symbol $;$.

      {\bf Hint:} You may want to look back at Exercise \exref{ex:2.05}.
   \end{Exercise}

   \begin{Exercise}
      \exlabel{ex:3.03}%
      Using
      \begin{align*}
         \cdb{libdg.dgab.000}  &= \cdb{libdg.dgab.001}\\
         \cdb{libdg.Rabcd.000} &= \cdb{libdg.Rabcd.001}
      \end{align*}
      write a Cadabra code to express $R_{abcd} = g_{ae} R^{e}{}_{bcd}$
      in terms of $\Gamma^{a}{}_{b c}$, $\Gamma_{a b c}$ and their partial derivatives.
   \end{Exercise}

   \begin{Exercise}
      \exlabel{ex:3.04}%
      Use the result of the previous exercise to verify that
      \begin{align*}
         R_{abcd} &= - R_{bacd}\\
         R_{abcd} &=   R_{cdab}
      \end{align*}
      {\bf Hint:} Rewrite the equations in the form $Q=0$ (for some suitable choice of $Q$)
      then use Cadabra to evaluate and simplify the left hand side.
   \end{Exercise}

   \begin{Exercise}
      \exlabel{ex:3.05}%
      Use Cadabra to verify that
      \begin{align*}
         \left(\nabla_{d}\nabla_{c} - \nabla_{c}\nabla_{d}\right)\left(A_{a} B_{b}\right)
         = B_{b}\left(\nabla_{d}\nabla_{c} - \nabla_{c}\nabla_{d}\right) A_{a}
         + A_{a}\left(\nabla_{d}\nabla_{c} - \nabla_{c}\nabla_{d}\right) B_{b}
      \end{align*}
      This exercise involves little more than successive applications of the product rule.
      You do not need to express $\nabla$ in terms of the connection.

      The above result leads to a nice simplification. Define a new derivative operator
      $D_{ab}$ by
      \begin{align*}
         D_{ab} = \nabla_{b}\nabla_{a} - \nabla_{a}\nabla_{b}
      \end{align*}
      then the above result can be written as
      \begin{align*}
         D_{cd}\left(A_{a}B_{b}\right)
         = D_{cd}\left(A_{a}\right) B_{b}
         + A_{a} D_{cd}\left(B_{b}\right)
      \end{align*}
      This is easy to remember -- it has the form of a product rule for a typical
      derivative operator.
   \end{Exercise}

   \begin{Exercise}
      \exlabel{ex:3.06}%
      Suppose $R_{abcd} = A_{a} B_{b} C_{c} D_{d}$. Use the $D$ operator introduced in
      the previous exercise to verify that
      \begin{align*}
         \left(\nabla_{f}\nabla_{e}
              -\nabla_{e}\nabla_{f}\right) R_{abcd}
         = R_{g b c d} R^{g}{}_{a e f}
         + R_{a g c d} R^{g}{}_{b e f}
         + R_{a b g d} R^{g}{}_{c e f}
         + R_{a b c g} R^{g}{}_{d e f}
      \end{align*}
      You may need the following equation
      \begin{align*}
         \left(\nabla_{c}\nabla_{b}
              -\nabla_{b}\nabla_{c}\right) V_{a}
         = D_{b c} \left(V_{a}\right)
         = R^{d}{}_{a b c} V_{d}
      \end{align*}
   \end{Exercise}

   \begin{Exercise}
      \exlabel{ex:3.07}%
      This exercise is a variation of the previous exercise -- it is the full computation
      made without any tricks or assumptions on the form of $R_{abcd}$.

      You will find it easier to use the symbol \verb|;| as the derivative operator rather
      than $\nabla$ (as per Example \ref{sec:ex-03}). You will also need to create rules for
      both the first and second covariant derivatives of $R_{abcd}$ and, for the final step,
      a rule to recover $R^{a}{}_{bcd}$ from terms involving the connection and its partial
      derivatives.

      This exercise requires a lot more work than the previous exercise. Do not try to
      write a complete code from scratch. Start with a trivial code. Then extend that code
      one line at a time looking closely at Cadabra's output before choosing the next
      Cadabra statement.
   \end{Exercise}

   \begin{Exercise}
      \exlabel{ex:3.08}%
      Another standard result in differential geometry is that the Ricci tensor $R_{ab}$
      for a symmetric connection is itself symmetric. That is, given
      \begin{align*}
          0 &= R^{a}{}_{bcd} + R^{a}{}_{dbc} + R^{a}{}_{cdb}\\
          0 &= R_{abcd} + R_{bacd}\\
          0 &= R_{abcd} + R_{abdc}
      \end{align*}
      then
      \begin{align*}
         R_{ab} = R^{c}{}_{acb} = R_{ba}
      \end{align*}
      The same result can also be obtained directly using
      \begin{align*}
         \partial_{c} g^{a b}  &= - g^{a e} g^{b f} \partial_{c} g_{e f}\\
         \cdb{libdg.Gamma.000} &= \cdb{libdg.Gamma.001}\\
         \cdb{libdg.Rabcd.000} &= \cdb{libdg.Rabcd.001}
      \end{align*}
      Use this last set of equations as a basis for a Cadabra code to verify that $R_{ab} =
      R_{ba}$.
   \end{Exercise}

   \begin{Exercise}
      \exlabel{ex:3.09}%
      Adapt your Cadabra code from the previous exercise to express $R_{ab}$ solely in terms
      of $g_{ab}$, its first and second partial derivatives and $g^{ab}$. Your answer should
      not contain any partial derivatives of $g^{ab}$.
   \end{Exercise}

   \begin{Exercise}
      \exlabel{ex:3.10}%
      The code given in Example \ref{sec:cheap-hack} included the following line
      \begin{cadabra}
         substitute (Vabc,$G^{a} G_{b c} -> \Gamma^{a}_{b c}$,repeat=True)
      \end{cadabra}
      Is the \verb|repeat=True| argument really necessary? Modify the source for Example
      \ref{sec:ex-03} by removing this argument, run the new code and observe the result.
      You should see that the result differs from the original result. This behaviour can be
      understood using the following simplified code.
      \begin{cadabra}
         foo := A B + A B A B + A B A B A B + A B A B A B A B .
         bah := A B + A B A B + A B A B A B + A B A B A B A B .

         substitute (foo,$A B -> A$)
         substitute (bah,$A B -> A$,repeat=True)
      \end{cadabra}
      Before the two calls to \verb|substitute| both \verb|foo| and \verb|bah| will equal
      $A B + A B A B + A B A B A B + A B A B A B A B$ and after the two calls to
      \verb|substitute| their values will be
      \begin{align*}
         {\tt foo} &= A + A A B + A A B A B + A A B A B A B\\
         {\tt bah} &= A + A A + A A A + A A A A
      \end{align*}
      By inspection it is easy to infer the action of \verb|substitute| with and without the
      \verb|repeat=True| argument. Without the \verb|repeat=True| argument only the first
      occurrence of the target in each product term will be substituted. In contrast, when
      \verb|repeat=True| argument is used Cadabra will repeat the process until all possible
      matches have been made.
   \end{Exercise}

\end{Exercises}
% --------------------------------------------------------------------------------------------

\clearpage

% ============================================================================================
\section{Feel the Function}
\label{sec:ex-04}
\ResetCounters

\input{./cadabra/example-04.cdbtex}

Since Cadabra's core language is built on Python (and implemented in \verb|C++| for
efficiency) it inherits all of the functionality of Python including the use of functions.
Here is a simple example of a function in Cadabra
\begin{cadabra}
   def tidy (expr):
      sort_product   (expr)
      rename_dummies (expr)
      canonicalise   (expr)
      return expr
\end{cadabra}
This function takes a single argument \verb|expr|, applies a sequence of Cadabra algorithms
to \verb|expr| and finally returns the updated version of \verb|expr|. Since \verb|tidy| is
a Python function, it must conform to all of Python's rules governing functions in
particular the use of a consistent indentation in the body of the function. The function can
be called using
\begin{cadabra}[numbers=none]
   foo = tidy (bah)
\end{cadabra}
This applies \verb|tidy| to \verb|bah| and saves the result in \verb|foo|. Note that this is
a pure Python statement and thus the assignment is made using \verb|=| rather than
\verb|:=|. This also explains the absence of a Cadabra statement terminator (such as
\verb|.|) -- it is a Python statement and thus it should conform to the Python's rules for
statement termination.

As a more involved example consider the situation where you are asked to extract the cubic
terms from a polynomial such as
\begin{cadabra}[numbers=none]
   Quartic := c^{a}
            + c^{a}_{b} x^b
            + c^{a}_{b c} x^b x^c
            + c^{a}_{b c d} x^b x^c x^d
            + c^{a}_{b c d e} x^b x^c x^d x^e.
\end{cadabra}
One approach (there are others, e.g., emulating a truncated Taylor series) is to use
Cadabra's \verb|::Weight| property and the \verb|keep_weight| algorithm. The idea is to
assign (invisible) weights to the terms of the polynomial (through the \verb|::Weight|
property) and then extract terms matching a chosen weight (using the \verb|keep_weight|
algorithm).

Here is a Cadabra function that does the job.
\begin{cadabra}
   def truncate (poly,n):

       # assign a weight to x^{a} and give it a label
       x^{a}::Weight(label=\epsilon).             |\lstlabel{ex-04.001}|

       # start with an empty expression
       ans = Ex("0")                              |\lstlabel{ex-04.002}|

       # loop over selected terms in the source
       for i in range (0,n+1):                    |\lstlabel{ex-04.003}|

          foo := @(poly).                         |\lstlabel{ex-04.004}|
          bah  = Ex("\epsilon = " + str(i))       |\lstlabel{ex-04.005}|

          # extract a single term
          keep_weight (foo, bah)                  |\lstlabel{ex-04.006}|

          # update the running sum
          ans = ans + foo                         |\lstlabel{ex-04.007}|

       # all done, return final answer
       return ans                                 |\lstlabel{ex-04.008}|
\end{cadabra}
Though this function follows a fairly standard idiom -- start with an empty sum and loop over
all terms while updating the rolling sum -- some elements of the syntax have not been
described so far and thus a few lines of explanation are warranted.

Line \lstref*{ex-04.001} identifies $x^a$ as the target to carry the weights (and is given
the label \verb|\epsilon| to distinguish it from other targets declared by other instances
of \verb|::Weight|). Cadabra now sees the polynomial as if it had been written as
\begin{cadabra}[numbers=none]
   Quartic := c^{a}
            + c^{a}_{b} x^b \eps
            + c^{a}_{b c} x^b x^c \eps^2
            + c^{a}_{b c d} x^b x^c x^d \eps^3
            + c^{a}_{b c d e} x^b x^c x^d x^e \eps^4.
\end{cadabra}
The function \verb|Ex| (used in lines \lstref*{ex-04.002} and \lstref*{ex-04.005}) is a
Cadabra function that takes a string (or zero) and returns a Cadabra expression for that
string. Thus line \lstref*{ex-04.002} sets the rolling sum \verb|ans| to zero while line
\lstref*{ex-04.005} sets the target \verb|bah| for the next term to extract from the
polynomial. The syntax \verb|foo := @(poly)| is Cadabra's way of creating a fresh copy of
\verb|poly| and saving it in \verb|foo|. Line \lstref*{ex-04.006} extracts the desired term
from \verb|foo| and overwrites \verb|foo| with the result (as per most Cadabra algorithms).
The Python \cdbverb{for-loop} starts with \verb|i=0| and continues for \verb|n+1| iterations
thus covering the range \verb|i=0,1,2,...n|.

The function could be called as follows
\begin{cadabra}[numbers=none]
   Cubic = truncate (Quartic,3)
\end{cadabra}
with the final result exactly as expected -- the leading cubic part of the original quartic polynomial.

% \clearpage

% --------------------------------------------------------------------------------------------
\begin{Exercises}

   In each of the following exercises you can assume that each polynomial is of the form
   \begin{align*}
      p(x) = c^{a}
           + c^{a}{}_{b} x^b
           + c^{a}{}_{b c} x^b x^c
           + c^{a}{}_{b c d} x^b x^c x^d
           + c^{a}{}_{b c d e} x^b x^c x^d x^e
   \end{align*}
   where the coefficients $c^{a}$, $c^{a}{}_{b}$, $c^{a}{}_{b c}$ etc. may vary from one
   polynomial to another. The restriction to quartics is just to avoid the complexities that
   might otherwise arise with highr order polynomials.

   \begin{Exercise}
      \exlabel{ex:4.01}%
      Write a function that returns the first partial derivative of a polynomial, that is
      $\partial_{b}(p(x))$. For a quadratic such as
      \begin{align*}
         p(x) = c^{a}
              + c^{a}{}_{b} x^b
              + c^{a}{}_{b c} x^b x^c
      \end{align*}
      the function should return
      \begin{align*}
        \partial_{b}(p(x)) = c^{a}{}_{b}
                           + c^{a}{}_{b c} x^c
                           + c^{a}{}_{c b} x^c
      \end{align*}
      {\bf Hint:} Begin by making substitutions like $x^{a} \rightarrow x^{a} + \delta^{a}$
      then expand in powers of $\delta^{a}$. At a later stage in your function you will need
      to make a second substitution $\delta^{a} \rightarrow 1$. This last step is not
      without its risks as you will discover in the following exercise.

      {\bf Note.} There are other (better) ways to differentiate expressions. One
                  such method can be found following the solution for this exercise.
   \end{Exercise}

   \begin{Exercise}
      \exlabel{ex:4.02}%
      The solution for the previous exercise contains the following three lines
      \begin{cadabra}
         sort_product   (foo)
         rename_dummies (foo)
         factor_out     (foo,$\delta^{a?}$)
      \end{cadabra}
      Comment out those lines and then re-run the code. Look carefully at the output. Are you
      worried? You should be! Look at the free indices -- $(a,b)$ for the first and third
      terms and $(a,c)$ on the second term. The source of this problem is the substitution
      \verb|\delta^{a} -> 1|. This changes the index structure and is thus a very risky
      operation. It should only ever be used when it is clear that problems such as that
      just seen can not arise. One way to prepare an expression for rules like
      \verb|\delta^{a} -> 1| is to use \verb|sort_product|, \verb|rename_dummies|,
      \verb|canonicalise| and \verb|factor_out| to ensure that the expression contains just
      one instance of $\delta^{a}$.
   \end{Exercise}

   \begin{Exercise}
      \exlabel{ex:4.03}%
      Write a function that accepts two polynomials and returns their product truncated to a
      given order. The easiest solution is to multiply both polynomials, expand and then
      truncate at the desired order. The problem with this solution is that it requires the
      full product to be computed which wastes both time and memory. The better solution is
      to construct the product term by term, starting from 0-th order and stopping at the
      required order. You could start by writing a function that returns a term of a given
      order from a polynomial (you can use the \verb|truncate| function from the main
      example as a starting point). This function could then be embedded in a loop to build
      a single term of the product. A further loop can be used to construct all of the
      required terms.

      {\bf Note.} When testing your function do ensure that the free indices on the two
      polynomials do not clash.
   \end{Exercise}

   \begin{Exercise}
      \exlabel{ex:4.04}%
      Here is a simple expression that is crying out for some TLC.
      \begin{align*}
         p(x) =   \frac{1}{3} A_{a b} x^{a} x^{b}
                + \frac{1}{9} B_{e c} x^{c} x^{e}
                - \frac{1}{5} C_{p c} B_{d q} g^{c d} x^{p} x^{q}
      \end{align*}
      The formatting could definitely be improved by factoring out the $x^{a}$ and by
      clearing the fractions. Write a function with two arguments -- the above polynomial
      and a scale factor. The scale factor should be used to clear the numerators. Your
      function should return the following expression
      \begin{align*}
         p(x) = \frac{1}{45} x^{a} x^{b} \left(15 A_{ab}+5B_{ab}-9B_{ca}C_{bd}g^{dc}\right)
      \end{align*}
   \end{Exercise}

   \begin{Exercise}
      \exlabel{ex:4.05}%
      This is a simple extension of the previous exercise. This time the messy polynomial is
      \begin{align*}
         p(x) &=  \frac{1}{7} A_{e} x^{e}
                - \frac{1}{3} B_{f} x^{f}\\
              & \quad
                + \frac{1}{3} A_{a b} x^{a} x^{b}
                + \frac{1}{9} B_{e c} x^{c} x^{e}
                - \frac{1}{5} C_{p c} B_{d q} g^{c d} x^{p} x^{q}\\
              & \qquad
                + \frac{3}{7} A_{a b c} x^{a} x^{b} x^{c}
                - \frac{1}{5} B_{a b} C_{c d e} g^{c d} x^{a} x^{b} x^{e}
                + \frac{7}{11} B_{a b} B_{c d} C_{e f g} g^{b c} g^{d f} x^{a} x^{e} x^{g}
      \end{align*}
      This expression contains 1st, 2nd and 3rd order terms in $x^{a}$. Write a function
      that first extracts the 1st, 2nd and 3rd order terms then tidies each using a function
      based on that from the previous exercise. Finally, rebuild the expression using the
      three (tidy) terms. You should obtain
      \begin{align*}
         p(x) &= \frac{1}{21} x^{a} \left(3A_{a}-7B_{a}\right)\\
              &  \quad  + \frac{1}{45} x^{a} x^{b}
                          \left(15 A_{a b} + 5 B_{a b} - 9 B_{c a} C_{b d} g^{d c}\right)\\
              &  \qquad + \frac{1}{385} x^{a} x^{b} x^{c}
                          \left(  165 A_{a b c}
                                -  77 B_{a b} C_{d e c} g^{d e}
                                + 245 B_{a d} B_{e f} C_{b g c} g^{d e} g^{f g}\right)
      \end{align*}
   \end{Exercise}

   \begin{Exercise}
      \exlabel{ex:4.06}%
      As noted in an earlier Exercise \exref{ex:1.06}, successive instances of
      \verb|SortOrder| might not produce the desired result (e.g., using
      \verb|{A,B}::SortOrder| as an attempt to undo a previous \verb|{B,A}::SortOrder|
      will fail). How can such problems be avoided? If the expression that needs to be
      sorted is composed solely of items with names like \verb|AAA01|, \verb|AAA02|,
      \verb|AAA03| etc. then the sorting can be done using Cadabra's default sort order
      (i.e., no need to declare \verb|SortOrder|).

      Write a function that uses the \verb|substitute| algorithm to replace targeted objects
      with names like \verb|AAA01|, \verb|AAA02|, \verb|AAA03| etc. Use this as a basis to
      sort the expression. Then undo the substitutions and return the now sorted expression.

      Test your function by sorting the following expression to place all of the $x^{a}$ to
      the left of all other terms
      \begin{cadabra}
         expr := g_{a b} x^{a} x^{b} + \Gamma_{a b c} x^{a} x^{b} x^{c}
      \end{cadabra}
      The value of this approach is that it allows you to create bespoke sort functions that
      will work as intended every time. The coding is certainly more tedious than using
      \verb|::SortOrder| though the certainty of the result probably justifies the effort.
   \end{Exercise}

   \begin{Exercise}
      \exlabel{ex:4.07}%
      Since Cadabra's functions like \verb|sort_product|, \verb|canoniclaise| etc.
      can alter their argument in place it is possible to write functions like
      \begin{cadabra}
         def tidy (obj):
             sort_product   (obj)
             rename_dummies (obj)
             canonicalise   (obj)

         foo := C^{f} B^{a} A_{f a}.
         tidy (foo)
      \end{cadabra}
      Notice the absence of a line like \verb|return obj|. This function will work
      as expected but it is not standard Python practice. However, a function like
      \begin{cadabra}
         def tidy (obj):
             bah := @(obj)
             sort_product   (bah)
             rename_dummies (bah)
             canonicalise   (bah)
             obj := @(bah)

         foo := C^{f} B^{a} A_{f a}.
         tidy (foo)
      \end{cadabra}
      will \emph{not} return the correct result. Verify these claims by running each of
      the above codes and observing the result for \verb|foo|. A good working practice is to
      always use a Python \verb|return| statement to return the final result of the function.
   \end{Exercise}

\end{Exercises}
% --------------------------------------------------------------------------------------------

\clearpage

% ============================================================================================
\section{Stay focused}
\label{sec:ex-05}
\ResetCounters

\input{./cadabra/example-05.cdbtex}

When massaging an expression towards a desired form it is often the case that some terms in
the expression need special attention while others can be left as they stand. One way to
implement this workflow in Cadabra is to manually pull apart the expression then allow
Cadabra to do its magic on the separate pieces. This is not ideal and it would be better if
the expression could be left intact while restricting Cadabra's actions to targeted parts of
the expression. Cadabra provides two algorithms \verb|zoom| and \verb|unzoom| designed to
focus Cadabra's attention to specific targets in an expression.

As an example, consider the task of replacing the second \verb|v^{a}| in the following
expression with \verb|w^{a}|
\begin{cadabra}[numbers=none]
   expr := A_{a} v^{a} + B_{a} v^{a} + C_{a} v^{a};
\end{cadabra}
Using \verb|substitute (expr,$v^{a}->w^{a}$)| would replace \emph{each} instance of
\verb|v^{a}| with \verb|w^{a}|. Thus some further information must be given to Cadabra to
restrict its attention to just the middle term -- this is where the \verb|zoom| and
\verb|unzoom| algorithms enter the scene. Here is a short Cadabra fragment that uses
\verb|zoom| and \verb|unzoom| to do the job properly.
\begin{cadabra}
   expr := A_{a} v^{a} + B_{a} v^{a} + C_{a} v^{a};    |\lstlabel{ex-05.100}|
   zoom       (expr, $B_{a} Q??$)                      |\lstlabel{ex-05.101}|
   substitute (expr, $v^{a} -> w^{a}$);                |\lstlabel{ex-05.102}|
   unzoom     (expr)                                   |\lstlabel{ex-05.103}|
\end{cadabra}
The corresponding output is as follows.
\begin{align*}
   \cdb{ex-05.100} &= \cdb{ex-05.101}\tag*{\lstref{ex-05.101}}\\
                   &= \cdb{ex-05.102}\tag*{\lstref{ex-05.102}}\\
                   &= \cdb{ex-05.103}\tag*{\lstref{ex-05.103}}
\end{align*}
The \verb|zoom| algorithm is designed to zoom in on selected parts of an expression. When
\verb|zoom| is in play Cadabra will use ellipses $\dots$ to denote those parts of the
expression currently hidden from view. This can be seen in lines \lstref{ex-05.101} and
\lstref{ex-05.102} of the above output. The original view is recovered with the call to
\verb|unzoom| in line \lstref*{ex-05.103}. Note that for the duration of a
\verb|zoom/unzoom| pair, Cadabra retains the full expression even though it only displays
the parts selected by \verb|zoom|.

A close look at the call to \verb|zoom| in line \lstref*{ex-05.101} above reveals that
\verb|zoom| takes two arguments, the first is the expression that will be \verb|zoom|'ed and
the second is a pattern that describes the target. In this case the pattern is
\verb|B_{a}Q??|. The first part of this pattern \verb|B_{a}| is easy to understand while the
second part \verb|Q??| is suggestive of a pattern matching rule. This is the second of
Cadabra's pattern matching rules\footnote{Cadabra also supports conditional and regular
expression patterns. See \url{https://cadabra.science/notebooks/ref_patterns.html} for more
details.} -- the first pattern, such as \verb|A?_{a}|, matches any tensor with a single
downstairs index, the second pattern, such as \verb|Q??|, matches an arbitrary expression
composed of sums and products of arbitrary tensors. Thus \verb|Q??| will match each of the
following expressions $A^{a}$, $A^{a} B_{a}$, $V_{a} + W_{a b c} P^{b c}$. The pattern used
in the above example was \verb|B_{a} Q??| and thus will match only the middle term of the
original expression. The upshot is that the call to \verb|substitute| will only alter the
middle term. This can be seen clearly in lines \lstref{ex-05.101} and \lstref{ex-05.102} of
the above output. The final line of the output is the result of the call to \verb|unzoom| in
line \lstref*{ex-05.103}. This shows that first and third terms where indeed left untouched
by \verb|substitute|.

Note that the choice of \verb|Q| in the pattern \verb|Q??| is not ordained by Cadabra -- any
symbol could be used. Note also that these double question mark patterns can be used in
substitution rules. For example, the rule \verb|A_{a} W?? -> B_{a} W??| would replace
\verb|A_{a}| with \verb|B_{a}| in any expression that begins with \verb|A_{a}|.

Finally note that nesting of calls to \verb|zoom| and \verb|unzoom| is allowed and this can
be used for greater control in selecting targets within an expression.

% --------------------------------------------------------------------------------------------
\subsection{Tags}

Suppose that $V_{ab}$ is an anti-symmetric tensor, i.e., $V_{ab} = - V_{ba}$. Then it is
clear that an expression like $2V_{ab} - 3V_{ba}$ can be reduced to just $5V_{ab}$. This
reduction could easily be implemented in Cadabra using something like the following
\begin{cadabra}[numbers=none]
   V_{a b}::AntiSymmetric.
   expr := 2 V_{a b} - 3 V_{b a}.
   canonicalise (expr)
\end{cadabra}
Now suppose that you wished to achieve the same result but \emph{without} assigning the
\verb|AntiSymmetric| property to \verb|V_{a b}|. Clearly the call to \verb|canoncialise|
will no longer swap the indices on the second term and thus the expression will remain in its
original form. The challenge is to persuade Cadabra to swap the indices on the second term.
This suggests a substitution like the following
\begin{cadabra}[numbers=none]
   expr := 2 V_{a b} - 3 V_{b a}.
   substitute (expr, $3 V_{a b} -> - 3 V_{a b})
\end{cadabra}
Alas, this too will fail as Cadabra will report a runtime error -- numerical factors on
the left of a rule, such as 3 in the above code, are not allowed. A similar problem arises
when trying to use \verb|zoom| to shift the focus. Thus the following code
\begin{cadabra}[numbers=none]
   expr := 2 V_{a b} - 3 V_{b a}.
   zoom       (expr, $3 V_{b a})
   substitute (expr, $V_{a b} -> - V_{a b})
\end{cadabra}
will produce a similar runtime error.

One solution to this problem is to modify the expression by adding unique tags to each term.
These tags can then be used as the targets for \verb|zoom|. The tagged expression can then be
manipulated to achieve the desired result after which the tags are removed. For the current
example, reducing $2V_{ab}-3V_{ba}$ to $5V_{ab}$, this is certainly a case of using a jet
plane to cross a street but the general method is applicable to much more challenging
problems (as shown in Example \ref{sec:ex-12} in part 3).

The process of adding and clearing tags can be achieved with calls to the following pair of
functions
\begin{cadabra}
   def add_tags (obj,tag):                           |\lstlabel{ex-05.110}|
      n = 0                                          |\lstlabel{ex-05.111}|
      ans = Ex('0')                                  |\lstlabel{ex-05.112}|
      for i in obj.top().terms():                    |\lstlabel{ex-05.113}|
         foo = obj[i]                                |\lstlabel{ex-05.114}|
         bah = Ex(tag+'_{'+str(n)+'}')               |\lstlabel{ex-05.115}|
         ans := @(ans) + @(bah) @(foo).              |\lstlabel{ex-05.116}|
         n = n + 1                                   |\lstlabel{ex-05.117}|
      return ans                                     |\lstlabel{ex-05.118}|

   def clear_tags (obj,tag):                         |\lstlabel{ex-05.120}|
      ans := @(obj).                                 |\lstlabel{ex-05.121}|
      foo  = Ex(tag+'_{a?} -> 1')                    |\lstlabel{ex-05.122}|
      substitute (ans,foo)                           |\lstlabel{ex-05.123}|
      return ans                                     |\lstlabel{ex-05.124}|
\end{cadabra}
The operation of each function involves a simple mix of Python and Cadabra constructs. Both
functions take two arguments, the first is the expression to be tagged and the second is a
string that describes the base of the tag. The tag base, for example $\mu$, is used to
generate a sequence of tags such as $\mu_0,\mu_1,\mu_2,\dots$. The \verb|add_tags| function
uses a \verb|for-loop| to select each term in the expression (line \lstref*{ex-05.114}),
multiplies that term by the tag and then updates a rolling sum (line \lstref*{ex-05.116}).
The \verb|clear_tags| function does its job by simply replacing all tags with the number 1.

Here is a short code fragment that demonstrates how these functions can be used to reduce
$2V_{ab}-3V_{ba}$ to $5V_{ab}$.
\begin{cadabra}
   expr := 2 V_{p q} - 3 V_{q p}.                    |\lstlabel{ex-05.200}|

   expr = add_tags (expr,'\\mu')                     |\lstlabel{ex-05.201}|

   zoom       (expr, $\mu_{1} Q??$)                  |\lstlabel{ex-05.202}|
   substitute (expr, $V_{a b} -> - V_{b a}$)         |\lstlabel{ex-05.203}|
   unzoom     (expr)                                 |\lstlabel{ex-05.204}|

   expr = clear_tags (expr,'\\mu')                   |\lstlabel{ex-05.205}|
\end{cadabra}
The corresponding output is as follows.
\begin{align*}
   \cdb{ex-05.200} &= \cdb{ex-05.201}\tag*{\lstref{ex-05.201}}\\
                   &= \cdb{ex-05.202}\tag*{\lstref{ex-05.202}}\\
                   &= \cdb{ex-05.203}\tag*{\lstref{ex-05.203}}\\
                   &= \cdb{ex-05.204}\tag*{\lstref{ex-05.204}}\\
                   &= \cdb{ex-05.205}\tag*{\lstref{ex-05.205}}
\end{align*}
The main objection to this method is that it requires explicit knowledge of the left to
right order of the terms in an expression. Consider for example an expression that happens
to have say 10 terms and assume that the tag function have been used to target the $\mu_7$
term. If some changes are made to the code preceding that expression then it is possible
that the term of interest may no longer be matched to $\mu_7$. The re-ordering of the terms
might now find the target term matched with $\mu_4$. This would require corresponding
changes to the calls to \verb|zoom|. It is also possible that this same problem could arise,
not through any change of the users code, but by changes made by the Cadabra team to the
internal workings of its own functions. The bottom line is that the user must take care
when using these functions -- careful scrutiny of the output should be standard practice!

% \clearpage

% --------------------------------------------------------------------------------------------
\begin{Exercises}

   \begin{Exercise}
      \exlabel{ex:5.01}%
      Verify that the following substitution rule
      \begin{cadabra}
         expr := A_{a} (P^{b}+Q^{b}) + C_{a} V^{b}.
         swap := A_{a} B?? + C_{a} D?? -> A_{a} D?? + C_{a} B??
      \end{cadabra}
      can be used to swap the expressions attached to \verb|A_{a}| and \verb|C_{a}|.
   \end{Exercise}

   \begin{Exercise}
      \exlabel{ex:5.02}%
      Verify the claim that Cadabra will report a runtime error when attempting the
      following substitution
      \begin{cadabra}
         expr := 2 V_{a b} - 3 V_{b a}.
         substitute (expr, $3 V_{b a} -> - 3 V_{a b})
      \end{cadabra}
   \end{Exercise}

   \begin{Exercise}
      \exlabel{ex:5.03}%
      Use a suitable substitution pattern to delete the second term in the
      following polynomial
      \begin{align*}
         p(x) = A_{a b} B^{a b} + A_{a b} A_{c d} B^{a b} B^{c d} - C_{a b} B^{a b}
      \end{align*}
      Do not use a tagged expression -- that approach will be left for the next exercise.
   \end{Exercise}

   \begin{Exercise}
      \exlabel{ex:5.04}%
      Repeat the previous exercise but this time making use of the \verb|add_tags|
      and \verb|clear_tags| functions.

      {\bf Hint:} A simple way to delete a term is to multiply it by zero.
   \end{Exercise}

   \begin{Exercise}
      \exlabel{ex:5.05}%
      A common method of introducing a Riemann tensor into a computation is to make
      use of the simple commutation rule for covariant derivatives, namely
      \begin{align*}
         V^{a}{}_{;b;c} = V^{a}{}_{;c;b} - R^{a}{}_{dbc} V^{d}
      \end{align*}
      Use this equation as the basis of a Cadabra code to simplify the
      expression $V^{a}{}_{;b;c} - V^{a}{}_{;c;b}$ to the expected result, namely
      $-R^{a}{}_{dbc} V^{d}$.

      {\bf Hint:} You will need to work with a tagged expression.
   \end{Exercise}

\end{Exercises}
% --------------------------------------------------------------------------------------------

\clearpage

% ============================================================================================
\section{Full disclosure}
\label{sec:ex-06}
\ResetCounters

\input{./cadabra/example-06.cdbtex}

Previous examples in this tutorial have shown that Cadabra is no slouch when it comes to
complex tensor algebra. The purpose of this example is to show that it is also a dab hand at
component computations, that is, given a set of coordinates, compute the components of a
tensor in those coordinates.

Here is a rather simple first example. Given the components of $V_{a}$, compute the
components of a new tensor $dV$ defined by $dV_{a b} = \partial_{b}{V_{a}} -
\partial_{a}{V_{b}}$. This computation entails a few basic steps -- choose a set of
coordinates, express the components of $V$ in those coordinates then evaluate $dV$. The
information required by Cadabra is much the same as just described -- a set of coordinates,
the components of $V_{a}$ and a way to compute $dV_{a b}$ from $V_{a}$. Here is a short
Cadabra code that does the job.
\begin{cadabra}
   {\theta, \varphi}::Coordinate.
   {a,b,c,d,e,f,g,h#}::Indices(values={\theta, \varphi}, position=independent).

   \partial{#}::PartialDerivative.

   V  := { V_{\theta} = \varphi, V_{\varphi} = \sin(\theta) }. |\lstlabel{ex-06.100}|
   dV := \partial_{b}{V_{a}} - \partial_{a}{V_{b}}.            |\lstlabel{ex-06.101}|

   evaluate (dV, V)
\end{cadabra}
The first line declares a pair of symbols for the coordinates while the second line attaches
those coordinates to the indices. Note the use of Greek letters to denote the coordinates in
contrast to the Latin characters used for the tensor indices. This is an aesthetic choice
commonly used in research articles in General Relativity to make clear the distinction
between an abstract expression for a tensor and its components in a given frame (known as
the Penrose abstract index notation). Note also that only two coordinates were declared in
the first line -- this implies that the tensors live in a two dimensional space. The
components of $V_{a}$ are described in line \lstref*{ex-06.100} as an explicit list of
values. Each entry in the list is of the form \verb|foo = bah| where \verb|foo| is one of
the components of a tensor (in this case $V_{a}$) while \verb|bah| is a scalar expression
(i.e., an expression that does not contain any free indices, there are other restrictions as
noted below). Line \lstref*{ex-06.101} informs Cadabra how to compute $dV_{a b}$ from $V_{a}$
while the final line completes the job -- it evaluates each of the components of $dV_{a b}$.
The output from the above code is
\begin{align*}
             V_{a} &= \cdb{ex-06.100}\tag*{\lstref{ex-06.100}}\\[10pt]
   \cdb{ex-06.101} &= \cdb{ex-06.102}\tag*{\lstref{ex-06.101}}
\end{align*}
The format of the output shown in line \lstref{ex-06.101} is typical of Cadabra's output for
a call to \verb|evaluate|. It displays the non-zero components as a table using a $\square$
as a place holder for the underlying tensor. Note that the expression on the far left of
line \lstref{ex-06.101} has been added here to aid in reading the output -- this term was not
generated by Cadabra but was included by hand. Cadabra's output begins with $\square_{ab}$
in that same line of output. The ${ab}$ subscripts on $\square_{ab}$ are matched to the
indices of the source on the left and the components on the right. Thus with $a=\theta$ and
$b=\varphi$ the above output reads as
\begin{align*}
   \partial_{\varphi}{V_{\theta}} - \partial_{\theta}{V_{\varphi}}
   = \square_{\theta\varphi}
   = -\cos\theta + 1
\end{align*}

Cadabra does place some restrictions on the scalar expressions that can be used when
describing a component of a tensor (given as \verb|bah| in the above paragraph). It is easier
to show by example what Cadabra will or will not accept for a scalar expression rather than
spelling out Cadabra's rules in detail. Here are set of definitions for $V_{a}$ that are
allowed by Cadabra.
\begin{cadabra}[numbers=none]
   V := { V_{\theta} = \varphi, V_{\varphi} = \sin(\theta) }.
   V := { V_{\theta} = \varphi, V_{\varphi} = \partial_{\theta}{\sin(\theta)} }.
   V := { V_{\theta} = f(\theta,\varphi), V_{\varphi} = g(\theta,\varphi)}.
\end{cadabra}
In contrast, each of the following definitions will be rejected by Cadabra.
\begin{cadabra}[numbers=none]
   V := { V_{a} = W_{a} }.                                 # don't use free indices
   V := { V_{\theta} = \theta, V_{\varphi} = V_{\theta} }. # don't use sub/super-scripts
   V := { V_{\theta} = \varphi^2, V_{\varphi} = \theta }.  # use ** for powers
\end{cadabra}

One other notable exception is that Cadabra\footnote{\CdbVersion} does not (yet) support the
use of derivative operators on the left hand side of the component rules. The following code
fragment will raise a Cadabra run time error.
\begin{cadabra}[numbers=none]
   \partial{#}::PartialDerivative.
   V_{a}::Depends(\theta,\varphi,\partial{#}).
   V := { \partial{\theta}{V_{\varphi}} = \cos(\theta) }.  # partial derivs not supported
\end{cadabra}
A workaround for problems like this is given later in Exercise \exref{ex:6.08}.

% --------------------------------------------------------------------------------------------
\subsection{Riemann curvature of a 2-sphere}

This approach can be easily extended to a somewhat more realistic example -- computing the
Riemann tensor for a 2-sphere\footnote{This example is adapted from the Cadabra web page
\url{https://cadabra.science/notebooks/sphere.html}}. The starting point in this case is the
metric of a 2-sphere, which in polar coordinates $(\theta,\varphi)$ can be written as
\begin{align*}
   ds^2 = r^2\left(d\theta^2 + \sin^2\theta\, d\varphi^2\right)
\end{align*}
The metric components are described in Cadabra by a list of non-zero components
\begin{cadabra}[numbers=none]
   gab := { g_{\theta\theta}   = r**2,
            g_{\varphi\varphi} = r**2 \sin(\theta)**2 }.
\end{cadabra}
Since the Riemann tensor also depends on the inverse metric $g^{a b}$ the $g^{a b}$ must also
be known to Cadabra before computing the Riemann tensor. In this simple example it is easy
compute the inverse by hand and then provide a list such as
\begin{cadabra}[numbers=none]
   iab := { g^{\theta\theta}   = 1/r**2,
            g^{\varphi\varphi} = 1/(r**2 \sin(\theta)**2) }.
\end{cadabra}
A second method is to use Cadabra's \verb|complete| algorithm (see the source code
of this example for a modified version that uses \verb|complete|).

Note that the lists for \verb|gab| and \verb|iab| contain \emph{all} of their non-zero
components. These are just lists of simple expressions and Cadabra knows nothing about
any symmetries that might be associated with these lists. In our case the underlying tensors
are symmetric so it is essential that \emph{all} non-zero components be included in the
list including those that could be inferred from the symmetries. Thus for a metric of the
form $ds^2 = du^2 + 2uv dudv + dv^2$ the components must be specified using
\begin{cadabra}[numbers=none]
   gab := { g_{u u} = 1,  g_{u v} = uv,
            g_{v u} = uv, g_{v v} = 1 }.
\end{cadabra}

Here is the complete code for the Riemann tensor for the 2-sphere.

\begin{cadabra}
   {\theta, \varphi}::Coordinate.
   {a,b,c,d,e,f,g,h#}::Indices(values={\theta, \varphi}, position=independent).

   \partial{#}::PartialDerivative.

   Gamma := \Gamma^{a}_{b c} -> 1/2 g^{a d} (   \partial_{b}{g_{d c}}
                                              + \partial_{c}{g_{b d}}
                                              - \partial_{d}{g_{b c}}).

   Rabcd := R^{a}_{b c d} ->   \partial_{c}{\Gamma^{a}_{b d}}
                             - \partial_{d}{\Gamma^{a}_{b c}}
                             + \Gamma^{e}_{b d} \Gamma^{a}_{c e}
                             - \Gamma^{e}_{b c} \Gamma^{a}_{d e}.

   gab := { g_{\theta\theta}   = r**2,                       |\lstlabel{ex-06.201}|
            g_{\varphi\varphi} = r**2 \sin(\theta)**2 }.

   iab := { g^{\theta\theta}   = 1/r**2,                     |\lstlabel{ex-06.202}|
            g^{\varphi\varphi} = 1/(r**2 \sin(\theta)**2) }.

   substitute (Rabcd, Gamma)                                 |\lstlabel{ex-06.203}|

   evaluate   (Gamma, gab+iab, rhsonly=True)                 |\lstlabel{ex-06.204}|
   evaluate   (Rabcd, gab+iab, rhsonly=True)                 |\lstlabel{ex-06.205}|
\end{cadabra}
There are two minor aspects of the above code that should be noted. First, each call to
\verb|evaluate| acts on a Cadabra rule and thus the argument \verb|rhsonly=True| is used
to restrict the action of \verb|evaluate| to just the right hand side of the rule.
Second, the construction \verb|gab+iab| results in a single list built from \verb|gab| and
\verb|iab| (this is standard Python syntax).

The output from the above code is as follows
\begin{align*}
   &\cdb{ex-06.201}\tag*{\lstref{ex-06.201}}\\[10pt]
   &\cdb{ex-06.202}\tag*{\lstref{ex-06.202}}
\end{align*}
\begin{align*}
   &\cdb{ex-06.204}\tag*{\lstref{ex-06.204}}\\[10pt]
   &\cdb{ex-06.205}\tag*{\lstref{ex-06.205}}
\end{align*}
The above results are mostly self-explanatory. However, the notation used in displaying the
Riemann components does require a brief explanation. Note that the order of the indices on
the left and right hand sides do not match. Despite this fact, the indices do maintain a
strict one-to-one correspondence. For example, the component
$R^\theta{}_{\varphi\theta\varphi}$ has indices $a=\theta,b=\varphi,c=\theta$ and
$d=\varphi$ which, on the right hand side, corresponds to
$\square{}_{\varphi\varphi}{}^{\theta}{}_{\theta}$. Thus $R^\theta{}_{\varphi\theta\varphi}
= \square{}_{\varphi\varphi}{}^{\theta}{}_{\theta} = \sin^2\theta$.

There is one small variation on the above code that is worth noting. Suppose that
\verb|Rabcd| had been declared as
\lstset{firstnumber=10}
\begin{cadabra}
   Rabcd :=   \partial_{c}{\Gamma^{a}_{b d}}
            - \partial_{d}{\Gamma^{a}_{b c}}
            + \Gamma^{e}_{b d} \Gamma^{a}_{c e}
            - \Gamma^{e}_{b c} \Gamma^{a}_{d e}.
\end{cadabra}
The object \verb|Rabcd| is no longer a substitution rule but rather a simple Cadabra
expression. Thus there is no need in this case for the \verb|rhsonly=True| argument in the
call to \verb|evaluate|. The output from the above code is
\begin{align*}
   &\cdb{ex-06.206}
\end{align*}
which no longer contains the informative left hand side, that is, $R^{a}{}_{bcd}\rightarrow$.
So when writing code it may be useful to apply the \verb|evaluate| algorithm to a rule (if
convenient) rather than a simple expression. The advantage in doing so is that the left hand
side of the rule retains a useful reminder of how the indices map to the components on the
right hand side.

% --------------------------------------------------------------------------------------------
\subsection{Selecting components}

Calls to \verb|evaluate| will return a Cadabra object that contains all of the non-zero
components. This raises a simple question -- How can individual components of a tensor be
found? The simplest answer is to call the Cadabra function \verb|get_component|. The
$R^{\theta}{}_{\varphi\theta\varphi}$ component of the Riemann tensor $R^{a}{}_{b c d}$ can
be obtained using the following code (appended to the above code)
\lstset{firstnumber=25}
\begin{cadabra}
   from cdb.core.component import *

   Riem := R^{a}_{b c d}.
   substitute (Riem, Rabcd)    # convert from a rule to a simple expression

   RiemCompt = get_component (Riem, $\theta, \varphi, \theta, \varphi$) |\lstlabel{ex-06.207}|
\end{cadabra}
This will return
\begin{align*}
   R^\theta{}_{\varphi\theta\varphi} &= \cdb{ex-06.207} \tag*{\lstref{ex-06.207}}
\end{align*}

The same results can also be obtained by projecting the tensor onto a suitable combination
of basis elements. Thus $R^{\theta}{}_{\varphi\theta\varphi}$ can be computed using
$R^{a}{}_{bcd} e^\theta_{a} e^{b}_{\varphi} e^{c}_{\theta} e^{d}_{\varphi} e$
where $e_{\theta} = \partial_{\theta}$ and $e_{\varphi}=\partial_{\varphi}$ are the standard
basis for the coordinates $(\theta,\varphi)$. The following code fragment will do the job.

\begin{cadabra}[numbers=none]
   theta{#}::LaTeXForm{"\theta"}.
   varphi{#}::LaTeXForm{"\varphi"}.

   gab := { g_{\theta \theta}   = r**2,
            g_{\varphi \varphi} = r**2 \sin(\theta)**2 }.

   iab := { g^{\theta\theta}   = 1/r**2,
            g^{\varphi\varphi} = 1/(r**2 \sin(\theta)**2) }.

   # define the basis for vectors and dual vectors

   basis := {theta^{\theta} = 1, varphi^{\varphi} = 1}.
   dual  := {theta_{\theta} = 1, varphi_{\varphi} = 1}.

   # obtain components by contracting with basis

   compt := R^{a}_{b c d} theta_{a} varphi^{b} theta^{c} varphi^{d}.
   substitute (compt, Rabcd)

   evaluate (compt, gab+iab+basis+dual)
\end{cadabra}
This fragment will require all the usual property definitions but more importantly it
requires a definition of the tensor \verb|Rabcd|. This can be taken from either line
\lstref*{ex-06.203} or \lstref*{ex-06.205} of the 2-sphere example given above.

% --------------------------------------------------------------------------------------------
\subsection{Components in pure Python/SymPy}

It is quite likely that one of the reasons for extracting one or more components of a tensor
is that some numerical values are sought (e.g., for plotting or for use in a separate
numerical simulation).

How can a Cadabra expression be evaluated numerically? This clearly sounds like a job for
Python/SymPy (or NumPy). But first the Cadabra expression, which may contain LaTeX markup,
needs to be reformatted for use by Python. This is rather easy -- just apply the
\verb|._sympy_()| method to convert the Cadabra expression to a SymPy expression. Note that
the \verb|._sympy()| method is the counterpart to Cadabra's \verb|Ex| function (which
converts strings to Cadabra expressions).

Thus a Python expression for the $\varphi\varphi$ component of the 2-metric could be created
using
\begin{cadabra}[numbers=none]
   r, theta, varphi = symbols('r theta varphi')
   gphiphi = compt._sympy_()
\end{cadabra}
where \verb|compt| is the result obtained above using the projection method. The first line
is needed only when \verb|gphiphi| will be subsequently processed by SymPy operations. Note
that the usual Python line
\goodbreak
\begin{cadabra}[numbers=none]
   from sympy import *
\end{cadabra}
is not needed as this is always included by Cadabra as part of its initialisation.

You can verify that the before and after expressions have the expected types by using the
following code fragment
\begin{cadabra}[numbers=none]
   print ('type compt   = ' + str(type(compt)))
   print ('type gphiphi = ' + str(type(gphiphi)))
\end{cadabra}
This produces the following output
\bgroup
\lstset{numbers=none}
\begin{lstlisting}
   type compt   = <class 'cadabra2.Ex'>
   type gphiphi = <class 'sympy.core.mul.Mul'>
\end{lstlisting}
\egroup

% \clearpage

% --------------------------------------------------------------------------------------------
\begin{Exercises}

   \begin{Exercise}
      \exlabel{ex:6.01}%
      Modify the original example by replacing line 7 with
      \begin{cadabra}[numbers=left,firstnumber=7]
         dV := dV_{a b} -> \partial_{b}{V_{a}} - \partial_{a}{V_{b}}.
      \end{cadabra}
      Observe the output then repeat using
      \begin{cadabra}[numbers=left,firstnumber=7]
         dV := dV_{a b} -> \partial_{b}{V_{a}} - \partial_{a}{V_{b}}.
         evaluate (dV, V, rhsonly=True)
      \end{cadabra}
   \end{Exercise}

   \begin{Exercise}
      \exlabel{ex:6.02}%
      Modify the original example by replacing lines 6 and 7 with
      \begin{cadabra}[numbers=left,firstnumber=6]
         V  := { V_{\theta} = f(\theta,\varphi), V_{\varphi} = g(\theta,\varphi) }.
         dV := \partial_{b}{V_{a}} + \partial_{a}{V_{b}}.
      \end{cadabra}
      Run the new code and observe the output. Not much to say here other than to admire
      your handiwork.
   \end{Exercise}

   \begin{Exercise}
      \exlabel{ex:6.03}%
      When processing a statement like \verb|evaluate(foo,bah)| Cadabra will use \verb|bah|
      as a pool of expressions to fulfil any requests while evaluating each component of
      \verb|bah|. What happens if the pool does not contain the requested component? If the
      pool contains some but not all entries for a tensor then the remaining entries are
      taken to be zero. Now run a code built on this fragment
      \begin{cadabra}
         bah := {V_{\theta} = \varphi, V_{\varphi} = \sin(\theta)}.
         foo := U_{a} V_{b}.
         evaluate (foo, bah)
      \end{cadabra}
      The output should show that Cadabra has assumed that all entries of $U_{a}$ are
      non-zero despite there being no entries for $U_{a}$ in the pool \verb|bah|.
   \end{Exercise}

   \begin{Exercise}
      \exlabel{ex:6.04}%
      Extend the sample code for the 2-sphere to also compute the scalar curvature.
      The result should be $2/r^2$ (as expected).
   \end{Exercise}

   \begin{Exercise}
      \exlabel{ex:6.05}%
      Verify that the Schwarzschild metric in isotropic coordinates
      \begin{align*}
         ds^2 = - \left(\frac{2r-m}{2r+m}\right)^2 dt^2
                + \left(1+\frac{m}{2r}\right)^4
                  \Big( dr^2 + r^2\left(d\theta^2 + \sin^2\theta\, d\phi^2\right)\Big)
      \end{align*}
      is a solution of the vacuum Einstein equations $0=R_{ab}$.
   \end{Exercise}

   \begin{Exercise}
      \exlabel{ex:6.06}%
      Compute the Ricci tensor $R_{ab}$ for the Kasner metric
      \begin{align*}
         ds^2 = -dt^2 + t^{2p_1} dx^2 + t^{2p_2} dy^2 + t^{2p_3} dz^2
      \end{align*}
      Hence verify that the Ricci tensor vanishes provided
      \begin{align*}
         p_1 + p_2 + p_3 = p^2_1 + p^2_2 +p^2_3 = 1
      \end{align*}
   \end{Exercise}

   \begin{Exercise}
      \exlabel{ex:6.07}%
      Consider the Schwarzschild metric in Schwarzschild coordinates
      \begin{align*}
         ds^2 = - f(r) dt^2
                + \frac{1}{f(r)} dr^2
                + r^2\left(d\theta^2 + \sin^2\theta\, d\phi^2\right)
      \end{align*}
      where $f(r) = (1-2m/r)$. Show that each of the following vectors
      \begin{align*}
         (i)   \quad & \xi = \xi^{a}\partial_{a} = \partial_{t}\\
         (ii)  \quad & \xi = \xi^{a}\partial_{a} = \partial_{\varphi}\\
         (iii) \quad & \xi = \xi^{a}\partial_{a}
                           = \sin\varphi\, \partial_{\theta}
                             + \cot\theta \cos\varphi\, \partial_{\varphi}\\
         (iv)  \quad & \xi = \xi^{a}\partial_{a}
                           = \cos\varphi\, \partial_{\theta}
                             - \cot\theta \sin\varphi\, \partial_{\varphi}
      \end{align*}
      is a solution of Killing's equation
      \begin{align*}
         0 = \xi_{a;b} + \xi_{b;a}
      \end{align*}

      {\bf Hint:} You will need to provide two lists of components,
                  one for $g_{ab}$ and one for $\xi^{a}$.
      {\bf Note.} To avoid a runtime error you will need to write
                  $\cot\theta$ as $\cos\theta/\sin\theta$.
   \end{Exercise}

   \begin{Exercise}
      \exlabel{ex:6.08}%
      The current version of Cadabra\footnote{\CdbVersion} does not support component rules
      that include derivative operators in the targets of the component definitions. Thus
      code like the following will raise a run time error.
      \begin{cadabra}
         {\theta, \varphi}::Coordinate.
         {a,b,c,d,e,f,g,h#}::Indices(values={\theta, \varphi}, position=independent).

         \partial{#}::PartialDerivative.

         V_{a}::Depends(\theta,\varphi,\partial{#}).

         dVrule := { \partial_{\theta}{V_{\varphi}} = \sin(\theta),
                     \partial_{\varphi}{V_{\theta}} = \cos(\theta)}.
         dV := \partial_{b}{V_{a}} - \partial_{a}{V_{b}}.

         evaluate (dV, dVrule)
      \end{cadabra}
      Though the intention is clear, Cadabra (at present) does not allow the rule
      \verb|dVrule| to be used in the call to \verb|evaluate|. One solution to this
      impasse is to hide the derivative from \verb|evaluate| by making a substitution
      \verb|\partial_{a}{V_{b}} -> dV_{a b}| then applying \verb|evaluate| to
      \verb|dV_{a b}|. Test this idea by modifying the above code to include this hack.
   \end{Exercise}

\end{Exercises}
% --------------------------------------------------------------------------------------------

\clearpage

% ============================================================================================
\section{Escape to C}
\label{sec:ex-07}
\ResetCounters

\input{./cadabra/example-07.cdbtex}

An increasingly popular approach in computational physics is to harness the power of
programs like Mathematica and Maple to convert differential equations into computational
procedures written in a language like C or Fortran (see \cite{husa:2006-01} and
\cite{ruchlin:2018-01}). Cadabra can take this one step further by first processing the
tensor equations before handing the results over to Mathematica, Maple or even Cadabra's own
internal version of SymPy. This opens up the possibility of using Cadabra, from beginning to
end, by starting with a complex tensor equation, such as Einstein's equations, and then
doing all the work required to produce a stand alone C program.

All but the last stage of this workflow can be easily handled using techniques described in
the previous examples. The final stage of this workflow, where the C-code is created, can be
easily implemented using the Codegen package from Python/Sympy. The basic idea is to iterate
over a list of expressions, passing each expression to Codegen and then saving the results
to a file. Here is a short Python code that writes raw C-code for a single
tensor\footnote{An extended version of the function, suitable for use with tensors and scalars, can be found in {\tt\footnotesize hybrid-latex/python/writecode.py}}.

\begin{cadabra}
   def write_code (obj,name,filename,rank):

      import os

      from sympy.printing.ccode import C99CodePrinter as printer
      from sympy.printing.codeprinter import Assignment

      idx=[]  # indices in the form [{x, x}, {x, y} ...]
      lst=[]  # corresponding terms [termxx, termxy, ...]

      for i in range( len(obj[rank]) ):                 # rank = num. of free indices
          idx.append( str(obj[rank][i][0]._sympy_()) )  # indices for this term
          lst.append( str(obj[rank][i][1]._sympy_()) )  # the matching term

      mat = sympy.Matrix([lst])                         # row vector of terms
      sub_exprs, simplified_rhs = sympy.cse(mat)        # optimise code

      with open(os.getcwd() + '/' + filename, 'w') as out:

         for lhs, rhs in sub_exprs:
            out.write(printer().doprint(Assignment(lhs, rhs))+'\n')

         for index, rhs in enumerate (simplified_rhs[0]):
            lhs = sympy.Symbol(name+' '+(idx[index]).replace(', ',']['))
            out.write(printer().doprint(Assignment(lhs, rhs))+'\n')
\end{cadabra}

The function \verb|write_code| takes four arguments. The first, \verb|obj|, is a list of
components of the tensor created in a prior call to \verb|evaluate|. The second,
\verb|name|, is a string representing the C-array name. Entries in this array will be the
C-code for the corresponding tensor component with indices matching exactly those of the
tensor represented by \verb|obj|. The third argument is simply the filename while the final
argument, \verb|rank|, equals the number of free indices on the tensor.

The function \verb|write_code| also applies basic optimisation of the C-code by looking for
common subexpressions and writes these (as assignments to variables like \verb|x0|,
\verb|x1|, \verb|x2| ...) to the file ahead of the tensor components.

The argument \verb|obj| is assumed to be the result of a call to \verb|evaluate|. As noted
in Example \ref{sec:ex-06} (on the 2-sphere) there are two ways to call \verb|evaluate|, the
first uses a simple expression as in
\begin{cadabra}[numbers=none]
   Rab := R_{a b};
   evaluate (Rab, ...)
\end{cadabra}
while the second uses a substitution rule as in
\begin{cadabra}[numbers=none]
   Rab := R_{a b} -> R^{c}_{a c b};
   evaluate (Rab, ...)
\end{cadabra}

The function \verb|write_code| is designed for the first case and expects a call like
\verb|write_code(Rab,...)|. However, if the components were created using the second method,
then the correct call would be \verb|write_code(Rab[1],...)|. Using \verb|Rab[1]| steps over
the leading \verb|Rab ->| part of \verb|Rab|.

The connection, Riemann and Ricci components for the 2-sphere (using the Cadabra code from
Example \ref{sec:ex-06}) could be converted to C using
\begin{cadabra}[numbers=none]
   write_code (Gamma[1], 'myGamma', 'example-07-gamma.c', 3)
   write_code (Rabcd[1], 'myRabcd', 'example-07-rabcd.c', 4)
   write_code (Rab[1],   'myRab',   'example-07-rab.c',   2)
\end{cadabra}
This creates the following C code for the connection
\bgroup
\lstset{numbers=none,style=myC}
\begin{lstlisting}
   x0 = 1.0/tan(theta);
   myGamma [varphi][theta][varphi] = x0;
   myGamma [theta][varphi][varphi] = x0;
   myGamma [varphi][varphi][theta] = -1.0/2.0*sin(2*theta);
\end{lstlisting}
\egroup
Clearly this C-code would not compile (as it stands) for it lacks some basic declarations
(e.g., array declarations for \verb|myGamma| and access to the \verb|math| library). One
solution could be to modify the function \verb|write_code| to fill in the missing pieces but
a better solution is treat the above code as a fragment to be included (either by hand or by
a \verb|#include|) into a separate C-program.

% \clearpage

% --------------------------------------------------------------------------------------------
\begin{Exercises}

   \begin{Exercise}
      \exlabel{ex:7.01}%
      Using the result of Exercise \exref{ex:3.09} (i.e., $R_{ab}$ in terms of $g_{ab}$)
      write a Cadabra code that creates C-code that could be the used to compute each of the
      components of $R_{ab}$. Assume a generic 3d-metric and assume the coordinates are
      labelled \verb|x,y,z|.

      {\bf Hint:} You may need to refer back to Exercise \exref{ex:6.08} to hide the first
                  and second partial derivatives of $g_{ab}$. You could also need to add
                  \verb|simplify=False| to the argument list in the call to \verb|evaluate|.
                  The function \verb|write_code| will optimise the C-code so there is not a
                  great deal to be gained by asking Sympy to also optimise its output.
   \end{Exercise}

\end{Exercises}
% --------------------------------------------------------------------------------------------

\clearpage

% ============================================================================================
\section{Expressions of interest}
\label{sec:ex-08}
\ResetCounters

\input{./cadabra/example-08.cdbtex}

A common paradigm in computational science is to break a given problem into smaller parts
with each part allocated to a single computer code. This obviously requires some cooperation
between the various programs, usually in the form of sharing results -- the programs exchange
information by exporting and importing data in some suitable format.

Can such a paradigm be applied in the context of Cadabra? Put another way, Is it possible to
share Cadabra content between different Cadabra programs? Though this might sound like a
simple question it does raise some serious issues. Recall that a Cadabra expression is more
than just an object with a list of indices -- it may also be subject to a set of properties
such as index sets, symmetries, commutation rules etc. Thus when an object is exported the
question arises -- how much information about its properties should also be exported? And
when that object is imported into another program might the inherited properties clash with
those declared in the host code? (e.g., an object declared as symmetric in \verb|foo.cdb|
might be imported by \verb|bah.cdb| where it is (incorrectly) declared as
anti-symmetric).

% --------------------------------------------------------------------------------------------
\subsection{Importing Cadabra code}
\label{sec:ImportNotebooks}

Cadabra supports the usual Python constructs for importing code from other sources. This make
it rather easy for sibling codes to share content. For example, suppose \verb|foo.cdb| is a
plain text file with the following content
\begin{cadabra}[numbers=none]
   {a,b,c,d,e,f,g,h}::Indices(position=independent).
   R_{a b c d}::RiemannTensor.
\end{cadabra}
Then another Cadabra program can import the above code as in this example
\begin{cadabra}[numbers=none]
   from foo import *
   expr := R_{a b c d} + R_{a b d c}.
\end{cadabra}
The result of this simple example is zero (since $R_{abcd} = - R_{abdc}$).

A good use of this method would be to include all of the common properties from a collection
of codes in one file. Each code in the collection would then import this shared library.
This saves the programmer time and also ensures consistent definitions across the
collection. However, this still leaves the problem of sharing results between one or more
programs. One solution, as described in the following section, is to use basic Python I/O to
read and write the data as required.

Note that the assumption that \verb|foo.cdb| is a plain text file is not essential as this
method works equally well with Cadabra notebooks such as \verb|foo.cnb|.

Be aware that there is a little trap that might catch the unwary. Whenever Cadabra is asked
to import a library (either from an explict call in your code or during Cadabra's startup) it
will go on a merry hunt looking for the first matching library -- and that may not be the
library you had expected. For example, during startup Cadabra will import various standard
libraries including \verb|sympy|. If you happen to have your own version of \verb|sympy.cdb|
then Cadabra will use your version rather than the system version -- and that would not play
well down the track. The simple solution is to ensure that all of your library names do not
clash with any Cadabra or Python libraries.

% --------------------------------------------------------------------------------------------
\subsection{Basic data I/O}
\label{sec:DataIO}

The basic idea is to store Cadabra objects as strings in a Python dictionary which in turn
is stored as a file in \verb|JSON| format. The Python code
\verb|hybrid-latex/python/cdblib.py| consists of three simple Python functions with the
following headers
\begin{cadabra}[numbers=none]
   def create (file_name):                 # create a library or clear an exiting library
   def put    (key_name,object,file_name): # add an object to the library
   def get    (key_name,file_name):        # retrieve an object from the library
\end{cadabra}

The implementation of these functions is not all that important here (see the source code in
\verb|hybrid-latex/python| for full details). Note that there are no explicit functions to
open or close the library as such actions are handled internally in the \verb|put| and
\verb|get| functions.

Here is a simple example that demonstrates the use of these functions. It creates two
expressions, writes them to a library, reads them back in but with new names and finally
checks that the new objects agree with the originals.

\begin{cadabra}[numbers=none]
   lib_name = 'example-08.json'

   create (lib_name)

   \nabla{#}::Derivative.

   gab := g_{a b} - 1/3 x^{c} x^{d} R_{a c b d}
                  - 1/6 x^{c} x^{d} x^{e} \nabla_{c}{R_{a d b e}}.

   iab := g^{a b} + 1/3 x^{c} x^{d} g^{a e} g^{b f} R_{c e d f}
                  + 1/6 x^{c} x^{d} x^{e} g^{a f} g^{b g} \nabla_{c}{R_{d f e g}}.

   put ('g_ab',gab,lib_name)
   put ('g^ab',iab,lib_name)

   gBar = get ('g_ab',lib_name)
   iBar = get ('g^ab',lib_name)

   tmp := @(gab) - @(gBar).
   tmp := @(iab) - @(iBar).
\end{cadabra}

The output from the above code is as follows. First, the two original objects, $g_{ab}$ and
$g^{ab}$, exported to the file \verb|example-08.json|
\begin{align*}
   g_{ab}(x) &= \cdb{ex-08-02.101}\\[10pt]
   g^{ab}(x) &= \cdb{ex-08-02.102}
\end{align*}
Second, the new objects, $\bar{g}_{ab}$ and $\bar{g}_{ab}$, imported by reading the file
\verb|example-08.json|
\begin{align*}
   {\bar{g}}_{ab}(x) &= \cdb{ex-08-02.103}\\[10pt]
   {\bar{g}}^{ab}(x) &= \cdb{ex-08-02.104}
\end{align*}
Finally, here is the proof that the new and old objects agree
\begin{align*}
   g_{ab}(x) - {\bar{g}}_{ab}(x) &= \cdb{ex-08-02.105}\\[10pt]
   g^{ab}(x) - {\bar{g}}^{ab}(x) &= \cdb{ex-08-02.106}
\end{align*}

\clearpage

% ============================================================================================
% Part 2. Concrete examples. With serious section headings.
% ============================================================================================
\hrule height 0pt
\vskip 4cm
{\Huge\bf Part 2 Applications}
\vskip 2cm

The examples in the first part of this tutorial were chosen to be sufficiently simple so as
to allow the reader to easily gain a good understanding of Cadabra. The risk in using such
simple examples is that they might convey the (incorrect) notion that Cadabra is suitable
only for such simple calculations. The examples in this second part were chosen to dispel
such notions -- they are non-trivial calculations and demonstrate that Cadabra is more than
capable of handling serious computations in general relativity.

The discussion in each of the following examples will not be as detailed as that given in
Part 1. Also, as there are no exercises in this part the reader is encouraged to experiment
with the source code (in \verb|source/cadabra/|).

\clearpage

% ============================================================================================
\section{The Gauss equation}
\label{sec:ex-09}
\ResetCounters

\input{./cadabra/example-09.cdbtex}

\hypertarget{GaussEqtn}{In this example} Cadabra will be used to derive the Gauss equation
which relates the induced and ambient curvatures of a hypersurface in an $n-$dimensional
Riemannian manifold.

The basics of the underlying mathematics are as follows. Suppose $\Sigma$ is an
$(n-1)-$dimensional subspace of an $n-$dimensional space $M$. Suppose $M$ is equipped with
Riemannian metric $g$ and a metric compatible derivative operator $\nabla$. The subspace
$\Sigma$ will, by way of its embedding in $M$, inherit a metric and a derivative operator
which will be denoted by $h$ and $D$ respectively. Let $n^a$ be the oriented unit normal to
$\Sigma$. Then the metrics of $\Sigma$ and $M$ are related by
\begin{align*}
   g_{a b} = h_{a b} + n_{a} n_{b}
\end{align*}
while, for any dual-vector $v_a$ lying in $\Sigma$ (i.e., $v_a n^a=0$),
\begin{align*}
   D_{b} v_{a} = h^{d}{}_{b} h^{c}{}_{a} \> \nabla_{d} v_{c}
\end{align*}
where $h^{a}{}_{b} = g^{a}{}_{b} - n^{a} n_{b}$ is the projection operator. The curvature
tensor for $(\Sigma,h,D)$ can then be obtained by computing $\left(D_{c} D_{b}-D_{b}
D_{c}\right) v_{a}$. This is all very standard and can be found in most textbooks on
differential geometry (see \cite{chavel:2006-01}).

Translating these equations into Cadabra code is straightforward and follows a now
familiar pattern. Unlike the previous examples, the discussion will begin by considering the
fragments of code needed to express the basic mathematical relations as just given. These
code fragments will later be glued together to form a complete Cadabra program.

Consider the definition of the projection operator $h^{a}{}_{b} = g^{a}{}_{b}-n^{a}n_{b}$
and its use in defining $D$ in terms of $\nabla$. The symbol \verb|hab| will be used to
record the projection operator and \verb|vpq| to record the covariant derivative $D_{q}v_p$.
Thus the code will contain the lines

\begin{cadabra}[numbers=none]
   hab:=h^{a}_{b} -> g^{a}_{b} - n^{a} n_{b}:
   vpq:=v_{p q} -> h^{a}_{p} h^{b}_{q} \nabla_{b}{v_{a}}:
\end{cadabra}
The code will also need an expression for the commutation of the covariant
derivatives,\Break
$\left(D_{r}D_{q} - D_{q}D_{r} \right)v_{p}$ which will be written as \verb|vpqr|
\begin{cadabra}[numbers=none]
   vpqr:=h^{a}_{p} h^{b}_{q} h^{c}_{r} ( \nabla_{c}{v_{a b}} - \nabla_{b}{v_{a c}} ).

   substitute (vpq,hab)
   substitute (vpqr,vpq)
\end{cadabra}

Some standard substitutions will also be needed to simplify and tidy the result. These
substitutions (and all of the previous definitions) are exactly what would normally be used
if these calculations were done by hand. For example, the lines
\begin{cadabra}[numbers=none]
   substitute (vpqr,$h^{a}_{b} n^{b} -> 0$)
   substitute (vpqr,$h^{a}_{b} n_{a} -> 0$)
\end{cadabra}
expresses the condition that $n^a$ is normal to the subspace, $0 = n^b h^{a}{}_{b}$ and
$0 = n^{b}h_{b}{}^{a}$. The line
\begin{cadabra}[numbers=none]
   substitute (vpqr,$\nabla_{a}{g^{b}_{c}} -> 0$)
\end{cadabra}
states that the covariant derivative of $g$ is zero while the line
\begin{cadabra}[numbers=none]
   substitute (vpqr,$n^{a} \nabla_{b}{v_{a}} -> -v_{a} \nabla_{b}{n^{a}}$)
\end{cadabra}
is a simple re-working of $0=\nabla\left(n^a v_a\right)=\left(\nabla n^a \right)v_a + n^a
\left(\nabla v_a\right)$ to eliminate first derivatives of $v^a$ from the expression
\verb|vpqr|. The next line
\begin{cadabra}[numbers=none]
   substitute (vpqr,$v_{a} \nabla_{b}{n^{a}} -> v_{p} h^{p}_{a} \nabla_{b}{n^{a}}$)
\end{cadabra}
squeezes a projection operator between $v_{a}$ and $\nabla n^a$. This is allowed because
$v^a$ has zero normal component. Finally, lines like
\begin{cadabra}[numbers=none]
   substitute (vpqr,$h^{p}_{a} h^{q}_{b} \nabla_{p}{n_{q}} -> K_{a b}$)
   substitute (vpqr,$h^{p}_{a} h^{q}_{b} \nabla_{p}{n^{b}} -> K_{a}^{q}$)
\end{cadabra}
can be used to introduce the extrinsic curvature tensor $K_{ab}$.

The above code fragments will need to be supplemented with extra statements, such as
an index set, substitution and simplification rules etc., before Cadabra can do its job.
Such pieces of code are very similar to those given in the previous examples and thus
require no further explanation here. Here then is the final code.

\begin{cadabra}
   {a,b,c,d,e,f,g,h,i,j,k,l,m,n,o,p,q,r,s,t,u#}::Indices(position=independent).

   \nabla{#}::Derivative.

   K_{a b}::Symmetric.
   g^{a}_{b}::KroneckerDelta.

   # define the projection operator

   hab:=h^{a}_{b} -> g^{a}_{b} - n^{a} n_{b}.

   # 3-covariant derivative obtained by projection on 4-covariant derivative

   vpq:=v_{p q} -> h^{a}_{p} h^{b}_{q} \nabla_{b}{v_{a}}.

   # compute 3-curvature by commutation of covariant derivatives

   vpqr:= h^{a}_{p} h^{b}_{q} h^{c}_{r} (  \nabla_{c}{v_{a b}}
                                         - \nabla_{b}{v_{a c}} ).

   substitute (vpq,hab)
   substitute (vpqr,vpq)

   distribute   (vpqr)
   product_rule (vpqr)
   distribute   (vpqr)
   eliminate_kronecker (vpqr)

   # standard substitutions

   substitute (vpqr,$h^{a}_{b} n^{b} -> 0$)
   substitute (vpqr,$h^{a}_{b} n_{a} -> 0$)
   substitute (vpqr,$\nabla_{a}{g^{b}_{c}} -> 0$)
   substitute (vpqr,$n^{a} \nabla_{b}{v_{a}} -> -v_{a} \nabla_{b}{n^{a}}$)
   substitute (vpqr,$v_{a} \nabla_{b}{n^{a}} -> v_{p} h^{p}_{a}\nabla_{b}{n^{a}}$)
   substitute (vpqr,$h^{p}_{a} h^{q}_{b} \nabla_{p}{n_{q}} -> K_{a b}$)
   substitute (vpqr,$h^{p}_{a} h^{q}_{b} \nabla_{p}{n^{b}} -> K_{a}^{q}$)

   # tidy up

   {h^{a}_{b},\nabla_{a}{v_{b}}}::SortOrder.

   sort_product   (vpqr)
   rename_dummies (vpqr)
   canonicalise   (vpqr)
   factor_out     (vpqr,$h^{a?}_{b?}$)
   factor_out     (vpqr,$v_{a?}$)               |\lstlabel{ex-09.101}|
\end{cadabra}

At this stage Cadabra's output is
\begin{align*}
( D_{r}D_{q} - D_{q}D_{r} ) v_{p}
          = \cdb{ex-09.101}\tag*{\lstref{ex-09.101}}
\end{align*}
which, although correct, is not in the familiar textbook form. This minor quibble is easily
addressed by making good use of the results from the previous example. Thus, the respective
Riemann tensors for the metrics $g$ and $h$ can be written as
\begin{align*}
   ( D_{r}D_{q} - D_{q}D_{r} ) v_{p} &= {\strut}^{h} R^{a}{}_{p q r}v_{a}\\[5pt]
   ( \nabla_{r}\nabla_{q} - \nabla_{q}\nabla_{r} ) v_{p} &= {\strut}^{g} R^{a}{}_{p q r}v_{a}
\end{align*}
These relations, along with the simple observations that $v_{a} = h^{b}{}_{a} v_{b}$ and
$K_{a}{}^{b} = h^{b}{}_{c} K_{a}{}^{c}$, can be used to massage Cadabra's output into the
familiar textbook form. The Cadabra code for this final stage is
\bgroup
\lstset{firstnumber=last}
\begin{cadabra}
   R{#}::LaTeXForm("{{\strut}^g R}").

   gRabcd := \nabla_{c}{\nabla_{b}{v_{a}}}
            -\nabla_{b}{\nabla_{c}{v_{a}}} -> R^{d}_{a b c} v_{d}.

   substitute     (vpqr,gRabcd)
   distribute     (vpqr)
   substitute     (vpqr,$v_{a} -> h^{b}_{a} v_{b}$)
   substitute     (vpqr,$h^{b}_{a} K_{c}^{a} -> K_{c}^{b}$)
   sort_product   (vpqr)
   rename_dummies (vpqr)
   canonicalise   (vpqr)
   factor_out     (vpqr,$v_{a?}$)
   substitute     (vpqr,$v_{a}->1$)
   sort_product   (vpqr)               |\lstlabel{ex-09.111}|
\end{cadabra}
\egroup
The final output is now
\begin{align*}
   {\strut}^h R^{a}{}_{pqr} = \cdb{ex-09.111}\tag*{\lstref{ex-09.111}}
\end{align*}
which is the standard textbook form for the Gauss equation.

\clearpage

% ============================================================================================
\section{The metric determinant in Riemann normal coordinates}
\label{sec:ex-10}
\ResetCounters

\input{./cadabra/example-10.cdbtex}

This and the following two examples are based on the standard leading order
expansion of a metric in Riemann normal coordinates, namely
\begin{align}
   g_{ab}(x)
   &= g_{a b}
   - \frac{1}{3} x^{c} x^{d} R_{a c b d}
   - \frac{1}{6} x^{c} x^{d} x^{e} \nabla_{c}{R_{a d b e}}\notag\\
   &\qquad + \frac{1}{180} x^{c} x^{d} x^{e} x^{f}
                           \left( 8g^{g h} R_{a c d g} R_{b e f h}
                                 -9\nabla_{c d}{R_{a e b f}}\right)
   + \cdots
   \label{ex-10:rncgab}\\
% \end{align}
\intertext{and}
% \begin{align}
   g^{ab}(x)
   &= g^{a b}
   + \frac{1}{3} x^{c} x^{d} g^{a e} g^{b f} R_{c e d f}
   + \frac{1}{6} x^{c} x^{d} x^{e} g^{a f} g^{b g} \nabla_{c}{R_{d f e g}}\notag\\
   &\qquad + \frac{1}{60} x^{c} x^{d} x^{e} x^{f} g^{a g} g^{b h}
                          \left( 4 g^{i j} R_{c g d i} R_{e h f j}
                                +3 \nabla_{c d}{R_{e g f h}}\right)
   + \cdots
   \label{ex-10:rnciab}
\end{align}
where $g_{ab}$, $g^{ab}$, $R_{abcd}$ and its derivatives are evaluated at the origin. The
expansions are valid inside a suitably chosen neighbourhood of the origin. Note that it is
customary to choose $g_{ab}={\rm diag}(-1,1,1,1)$ for Lorentzian spacetimes. For more
details on Riemann normal coordinates, their derivation and use, see
\cite{chavel:2006-01},
\cite{chern-chen-lam:2000-01},
\cite{eisenhart:1926-01},
\cite{gray:1973-01}, \cite{willmore:1996-01}.
See also the website \url{https://github.com/leo-brewin/riemann-normal-coords/} for an
extensive set of Cadabra programs for developing Riemann normal expansions (this includes
the code used to generate the above expressions for the metric and its inverse).

The above expansions includes terms up-to fourth order in the coordinates. Thus when forming
other quantities based on this metric, such as the metric determinant (in this example) and
the connection (the next two examples), care must be taken to truncate the computed
expression to fourth order (at most).

The metric determinant can be easily computed using a basic result from linear algebra. For
the $4\times 4$ matrices $N$ and $M$ built from $N^{a b}$ and $M_{a b}$ then
\begin{align*}
   \det N \det M = \frac{1}{4!} \epsilon^{abcd}_{pqrs} M_{ia} M_{jb} M_{kc} M_{ld}
                                                       N^{ip} N^{jq} N^{kr} N^{ls}
\end{align*}
where $\det N = \det(N^{a b})$ and $\det M = \det(M_{i j})$.

Choosing $M_{ab} = g_{ab}(x)$ and $N^{ab} = g^{ab} = {\rm diag}(-1,1,1,1)$ leads to the
following simple expression for the metric determinant
\begin{align*}
   \det g(x) = - \frac{1}{4!}\,\epsilon^{abcd}_{pqrs}\,
                 g_{ia}(x)\, g_{jb}(x)\, g_{kc}(x)\, g_{ld}(x)
                 g^{ip}      g^{jq}      g^{kr}      g^{ls}
\end{align*}

Implementing the above in Cadabra is straightforward though there a two minor points worth
noting. First, Cadabra provides an algorithm, \verb|asym|, that can be used to impose
antisymmetry on chosen objects. This can be used to create $\epsilon^{abcd}_{pqrs}$ using
code similar to
\begin{cadabra}
   d{#}::KroneckerDelta.
   eps := d^{a}_{p} d^{b}_{q} d^{c}_{r} d^{d}_{s}.
   asym (eps, $^{a}, ^{b}, ^{c}, ^{d}$)
\end{cadabra}
The first line declares $d$ to be a Kronecker delta while the second creates a seed for
$\eps$. The call to \verb|asym| uses this seed to create a new object that is antisymmetric
in the nominated indices (i.e., the upper indices $(abcd)$). The bonus in using \verb|asym|
is that it will include the $1/4!$ factor, that is, \verb|asym| returns $(1/4!)
\epsilon^{abcd}_{pqrs}$.

The second point concerns the truncation of $\det g$ to be consistent with that of the
metric (in this case to fourth order). The question is -- at what stage in the computation
should the truncation be imposed? The answer will have a significant impact on the
computational cost (particularly for higher order expansions). The lazy approach is to defer
the truncation until the very end. For a fourth order expansion in four dimensions this
would produce a polynomial of order $4^4 = 256$ and yet only the first four terms would be
retained. This is a huge waste of resources. A better approach would be to compute the terms
in $\det g$ in successive orders, starting from zeroth order. One way to do so would be to
first decompose $g_{ab}(x)$ into successive orders,
\begin{align*}
     g_{a b}(x) =
     \ngab{0}_{a b}
   + \ngab{2}_{a b}
   + \ngab{3}_{a b}
   + \ngab{4}_{a b}+\cdots
\end{align*}
where $\ngab{n}$ denotes the $n$-th order term of $g_{ab}(x)$. A similar expansion can be
proposed for $\det g$, namely,
\begin{align*}
     \det g =
     \ndetg{0}
   + \ndetg{2}
   + \ndetg{3}
   + \ndetg{4}+\cdots
\end{align*}
A standard procedure of substitution, expansion and matching can then be applied. The result
would be a series of equations that would allow, for example, $\ndetg{0}$ to be computed
from $\ngab{0}_{a b}$, $\ndetg{1}$ to be computed form $\ngab{0}_{a b}$ and $\ngab{1}_{a b}$
etc.

Despite the clear advantage of this second scheme the code given in the
\verb|source/cadabra/example-10| uses the previous lazy method. Why? For the simple
reason that it was easy to write and it gave results in a reasonable time (similar to most
of the other codes in this tutorial). It is certain that this lazy code will be too
expensive for higher order expansions (or in higher dimensions).

The actual computation of $\det g$ requires only a few lines of Cadabra code
\begin{cadabra}
   # compute Ndetg = negative det g
   Ndetg := @(eps) gx_{p a} gx_{q b} gx_{r c} g^{i p} g^{j q} g^{k r}.
   substitute      (Ndetg,gxab)
   distribute      (Ndetg)
   Ndetg = truncate (Ndetg,4)
   substitute      (Ndetg,$g^{a b} g_{b c} -> d^{a}_{c}$,repeat=True)
   eliminate_kronecker (Ndetg)
\end{cadabra}
where \verb|gx_{a b}| represents the metric $g_{ab}(x)$ given above and \verb|gxab| is
a rule that substitutes $g_{ab}(x)$ for \verb|gx_{a b}|.

The remainder of the code is just housekeeping in particular  the introduction of the Ricci
tensor
\begin{cadabra}
   substitute (Ndetg,$R_{a b c d} g^{a c}               -> R_{b d}$,repeat=True)
   substitute (Ndetg,$\nabla_{a}{R_{b c d e}} g^{b d}   -> \nabla_{a}{R_{c e}}$,repeat=True)
   substitute (Ndetg,$\nabla_{a b}{R_{c d e f}} g^{c e} -> \nabla_{a b}{R_{d f}}$,repeat=True)
\end{cadabra}
The final result for $\det g$, to fourth order in $x^{a}$, is
\begin{dgroup*}
   \Dmath*{-\det g(x) = \cdb{ex-10.Ndetg.701}+\cdots}
\end{dgroup*}

\clearpage

% ============================================================================================
\section{The metric connection in Riemann normal coordinates}
\label{sec:ex-11}
\ResetCounters

\input{./cadabra/example-11.cdbtex}

Though the subject of this example will be the computation of the connection for the metric
in Riemann normal form most of the discussion will concern the computational costs. These
costs will increase with higher order expansions. The surprising thing is just how easy it
is to hit the computational wall. Fortunately (for this example) there are simple ways to
manage this problem.

The starting point is the familiar equation for the connection
\begin{align*}
   \Gamma^{a}{}_{b c}(x) = \frac{1}{2} g^{a d} \left( \partial_{b}{g_{d c}}
                                                     +\partial_{c}{g_{b d}}
                                                     -\partial_{d}{g_{b c}}\right)
\end{align*}
where $g_{ab}(x)$ and $g^{ab}(x)$ are given by (\ref{ex-10:rncgab}) and (\ref{ex-10:rnciab})
respectively. A proliferation of terms will arise first through the product rule acting on
the individual terms and second through the expansion of $g^{ab}(x)$ and its coupling with
the derivative terms. It is also clear that at some point the result will need to be
truncated to an order consistent with that of the metric.

An obvious strategy (to minimise computational cost) is to avoid introducing unnecessary
terms. One clear case of this occurs when computing the derivatives of terms such as
$\partial_{a}(R_{bcde} x^{x} x^{e})$. Since the $R_{abcd}$ are constants (evaluated at the
origin of the RNC) it follows that
\begin{align}
   \partial_{a}(R_{bcde} x^{x} x^{e}) = R_{bcde} \partial_{a}(x^{x} x^{e})
   \label{ex-11:simple}
\end{align}
How is this simple step implemented in Cadabra? One (naive) option is to invoke a product
rule then set the derivatives of $R_{abcd}$ to zero. A better option is to inform Cadabra
ahead of time which objects are non-constant. Here is a short fragment that does the job.
\begin{cadabra}
   \partial{#}::PartialDerivative.                 |\lstlabel{ex-11.00}|
   x{#}::Depends{\partial{#}}.                     |\lstlabel{ex-11.01}|

   term := \partial_{a}{R_{bcde} x^{x} x^{e}}.     |\lstlabel{ex-11.02}|

   unwrap (term)                                   |\lstlabel{ex-11.03}|
\end{cadabra}
The idea is to identify the objects that will survive under the action of a derivative
operator. This can be seen in line \lstref*{ex-11.01} where $x^{a}$ is explicitly declared
to depend on \verb|\partial|. This information is used by \verb|unwrap| to pull out factors
that do \emph{not} depend on the derivative operators. Thus in line \lstref*{ex-11.03} the
$R_{abcd}$ will be pulled out as a common factor since they were not declared to depend upon
\verb|\partial|. Consequently, the value of \verb|term|, after the call to \verb|unwrap|,
will equal that of the right hand side of equation (\ref{ex-11:simple}). At this point the
computation can proceed by invoking a product rule to reduce $\partial_{a}(x^{x} x^{e})$ to
Kronecker deltas.

The other bits of Cadabra required to complete the computation include rules to define the
metric, the connection and, after the main body of the code, some basic housekeeping. The
main body of the code (excluding the rules for the metric and its inverse) is
\begin{cadabra}[numbers=none]
   ChrSym := \Gamma^{a}_{b c} -> 1/2 g^{a d} (  D_{b}{g_{d c}}
                                              + D_{c}{g_{b d}}
                                              - D_{d}{g_{b c}} ).

   Gamma := \Gamma^{a}_{b c}.

   substitute     (Gamma,ChrSym)   # the connection
   substitute     (Gamma,gab)      # the metric
   substitute     (Gamma,iab)      # the metric inverse
   distribute     (Gamma)
   unwrap         (Gamma)          # pull out constants
   product_rule   (Gamma)
   distribute     (Gamma)
   substitute     (Gamma,Dx)       # rule for partial derivs of x
   eliminate_kronecker (Gamma)
\end{cadabra}
At this point the expression for $\Gamma^{a}{}_{b c}(x)$ will contain terms beyond
3rd-order\footnote{Why focus on 3rd order? Because the metric and its inverse are known only
to 4th-order and the $\Gamma^{a}{}_{b c}(x)$ requires one derivative in $x^{a}$.} in
$x^{a}$. This is a good time to revisit the question of truncation. A good choice is to
truncate \emph{before} applying the housekeeping as in the following code
\begin{cadabra}[numbers=none]
   Gamma = truncate (Gamma,3)

   sort_product   (Gamma)
   rename_dummies (Gamma)
   canonicalise   (Gamma)
\end{cadabra}
The code for \verb|truncate| is similar to that used in Example \ref{sec:ex-04}. This works
well and produces the following result
\begin{dgroup*}
   \Dmath*{\cdb{ex-11.100}(x) = \cdb{ex-11.230}}
\end{dgroup*}

It was previously noted that it is rather easy to hit the computational wall. Here is a
slightly changes that does just that -- truncate the result \emph{after} the housekeeping,
that is
\begin{cadabra}[numbers=none]
   sort_product   (Gamma)
   rename_dummies (Gamma)
   canonicalise   (Gamma)

   Gamma = truncate (Gamma,3)
\end{cadabra}
This code was terminated (by hand) with no results after running for over 20 minutes and
using over 500 Mbytes of memory. In contrast the previous code completed in around 33
seconds and required 60 Mbytes of memory.

By conducting a few experiments it was found that the slow code stalled on the call to
\verb|canonicalise|. The problem here is that \verb|canonicalise| is being asked to do its
magic across all of the terms in \verb|Gamma| which for a 4th-order metric is approximately
200 terms. And since \verb|Gamma| is a polynomial in $x^{a}$ the housekeeping (sort, rename,
canonicalise) will naturally target each power of $x^{a}$. Thus a better approach would be
to decompose \verb|Gamma| into separate powers of $x^{a}$, do the housekeeping on each power
then rebuild \verb|Gamma|. That will work but it is a silly solution as there is no point in
doing any housekeeping on the higher order terms as they will be discarded in the later call
to \verb|truncate|. The main point in this variation is to show how simple changes (without
thought) can dramatically blow out the computational cost. The advice given earlier, to keep
as few terms as needed, is well worth remembering.

\clearpage

% ============================================================================================
\section{The third order terms of Calzetta etal.}
\label{sec:ex-12}
\ResetCounters

\input{./cadabra/example-12.cdbtex}

The metric connection $\Gamma^{a}{}_{bc}(x)$ is symmetric in its lower indices. Thus there is
no loss of information in forming a product like $z^{b} z^{c} \Gamma^{a}{}_{bc}(x)$. The
$z^{a}$ have no real meaning, they are just to help with the bookkeeping. Now define
$\Gamma^{a}$ by \begin{align*}
   \Gamma^{a} := z^{b} z^{c} \Gamma^{a}{}_{bc}(x)
\end{align*}
then using the results from the previous example it is easy to show that
\begin{dgroup*}[spread=5pt]
   \Dmath*{ \Gamma^{a} = z^{b} z^{c} \Gamma^{a}{}_{bc}(x) = \cdb{ex-12.305} }
\end{dgroup*}

Calzetta etal.\cite{calzetta:1988-01} have also computed an expression for the connection in
Riemann normal coordinates. Their result, denoted by $\GammaBar$,
\begin{equation*}
   \begin{split}
      \GammaBar^{\mu}
      &= z^{\nu} z^{\rho} \GammaBar^{\mu}{}_{\nu\rho}(x)\\
      &= z^{\nu} z^{\rho} \Bigl\lbrace
           \frac{2}{3} R^{\mu}{}_{\nu\rho\sigma} x^{\sigma}
         + \frac{1}{12} \left(5 \nabla_{\lambda}{R^{\mu}{}_{\nu\rho\sigma}}
                              + \nabla_{\rho}{R^{\mu}{}_{\sigma\nu\lambda}}
                        \right) x^{\sigma} x^{\lambda}\\[5pt]
      &\qquad\qquad
         + \frac{1}{6}
            \Big[
               \frac{9}{10} \nabla_{\tau\lambda}{R^{\mu}{}_{\rho\nu\sigma}}
             + \frac{3}{20} \left(  \nabla_{\tau\rho}{R^{\mu}{}_{\sigma\nu\lambda}}
                                  + \nabla_{\rho\tau}{R^{\mu}{}_{\sigma\nu\lambda}}
                            \right)\\[5pt]
      &\qquad\qquad\qquad
             + \frac{1}{60}
               \bigl(  21 R^{\mu}{}_{\lambda\xi\rho} R^{\xi}{}_{\sigma\nu\tau}
                     + 48 R^{\mu}{}_{\xi\rho\lambda} R^{\xi}{}_{\sigma\nu\tau}
                     - 37 R^{\mu}{}_{\sigma\xi\lambda} R^{\xi}{}_{\nu\rho\tau}
               \bigr)
            \Big] x^{\sigma} x^{\lambda} x^{\tau} \Bigr\rbrace
   \end{split}
\end{equation*}
appears (at first sight) to differ significantly from $\Gamma$. The purpose of this example
is to show that both expressions agree. The basic approach will be to compute the difference
$\Delta\Gamma:=\Gamma - \GammaBar$ and to then use the known symmetries of the Riemann
tensor to show that all terms cancel.

Note that the convention for the Riemann tensor used by Calzetta etal. is opposite to that
used in this tutorial. This is easily accounted for (in the Cadabra code) by replacing their
$R_{abcd}$ with $-R_{abcd}$.

Other simple changes will also be made to $\Gamma$ and $\GammaBar$ before attempting to show
that $\Gamma -\GammaBar = 0$. One obvious change is that a single index set should be used
for both $\Gamma$ and $\GammaBar$. Further changes include lowering indices so that each
Riemann term is of the form $R_{abcd}$ and sorting all products into a consistent order
$g,x,z,R,\nabla R,\nabla\nabla R$. The goal in making these changes is simply to maximise
the opportunity to use known Riemann symmetries and to spot the terms that cancel. These
basic preconditioning steps are implemented in Cadabra as follows.

Converting the Greek indices on $\GammaBar$ to Latin indices can be done in Cadabra by
first declaring a named pair of index sets
\begin{cadabra}[numbers=none]
   {a,b,c,d,e,f,g,h,i,j,k,l,m,n,o,p,u,v#}::Indices("latin",position=independent).
   {\mu,\nu,\rho,\sigma,\tau,\lambda,\xi#}::Indices("greek",position=independent).
\end{cadabra}
and then passing this pair to \verb|rename_dummies| as in
\begin{cadabra}[numbers=none]
   rename_dummies (\GammaBar,"greek","latin")
\end{cadabra}

The call to \verb|rename_dummies| only renames the dummy indices (now that was a surprise).
This leaves the free index $\mu$ on $\GammaBar^{\mu}$ unchanged. This little problem can be
dealt with by lowering the index using $\delta_{a\mu} \GammaBar^{\mu}$ and then eliminating
the Kronecker deltas
\begin{cadabra}[numbers=none]
   \delta{#}::KroneckerDelta.
   GammaBar := \delta_{a \mu} @(GammaBar).
   distribute (GammaBar)
   eliminate_kronecker (GammaBar)
\end{cadabra}
Note that a small liberty has been take here -- the index lowering should be done using
$g_{a \mu}$ rather than $\delta_{a\mu}$. But the nett outcome is the same and it saves
having to include extra code to implement the action of $g_{a\mu}$ on each term (the call to
\verb|eliminate_kronecker| does the same for $\delta_{a\mu}$).

Lowering the upper index on $\Gamma^{a}$ is slightly more involved as shown is this Cadabra
fragment
\begin{cadabra}[numbers=none]
   # lower free index ^{a} to _{v}

   Gamma := g_{v a} @(Gamma).

   distribute (Gamma)
   substitute (Gamma, $g_{a d} g^{d b} -> \delta_{a}^{b}$)
   eliminate_kronecker (Gamma)

   # change free index _{v} to _{a}

   foo := tmp_{v} -> @(Gamma).
   bah := tmp_{a}.
   substitute (bah, foo)

   Gamma := @(bah).
\end{cadabra}
This involves two steps. First, lower the index $a$ and covert it to $v$. Second, convert the
index $v$ back to $a$. Note that this second step could also be implemented in Cadabra using
\begin{cadabra}[numbers=none]
   Gamma := \delta^{v}_{a} @(Gamma).
   distribute (Gamma)
   eliminate_kronecker (Gamma)
\end{cadabra}
Recall that Cadabra will do the necessary index juggling to avoid any clash that might
arise in the above computation (see the results of Exercise \exref{ex:1.09}).

At this point the difference $\Delta\Gamma_{a}:=\Gamma_{a}-\GammaBar_{a}$ is given by
\Dmath*{\Delta\Gamma_{a} = \cdb{ex-12.diff.305} }

Now the fun begins (cancelling terms). First notice that $\Delta\Gamma$ consists of second
and third order terms in $x$. It is easy to see that the second order terms will vanish when
the second Bianchi identity is applied. The third order terms require a little more work.
The first step is to commute the order of the second covariant derivatives on the
$\nabla_{eb}$ and $\nabla_{ab}$ terms. This of course will introduce new $RR$ terms which
couple with the existing $RR$ terms.

Each of these steps can be implemented in Cadabra by applying suitable substitution rules on
a \verb|zoom|'ed and tagged expression (along the lines shown in Example \ref{sec:ex-05}).
For example, the following code applies the second Bianchi identity to the second order terms
\begin{cadabra}
   diff2 = get_xterm (diff,2)            |\lstlabel{ex-12.01}|
   diff3 = get_xterm (diff,3)            |\lstlabel{ex-12.02}|

   diff2 = add_tags (diff2,'\\mu')       |\lstlabel{ex-12.03}|

   # swap indices on middle term, then apply 2nd Bianchi identity

   zoom       (diff2, $\mu_{1} Q??$)                                           |\lstlabel{ex-12.04}|
   substitute (diff2, $\nabla_{b}{R_{a d c e}} -> - \nabla_{b}{R_{d a c e}}$)  |\lstlabel{ex-12.05}|
   unzoom     (diff2)                                                          |\lstlabel{ex-12.06}|

   substitute (diff2, $\mu_{1} -> \mu_{0}, \mu_{2} -> \mu_{0}$)    |\lstlabel{ex-12.07}|
   substitute (diff2, $\mu_{0} -> 0$)                              |\lstlabel{ex-12.08}|

   diff2 = clear_tags (diff2,'\\mu')                               |\lstlabel{ex-12.09}|

   diff := @(diff2) + @(diff3).                                    |\lstlabel{ex-12.10}|
\end{cadabra}
The code is rather easy to understand. The first pair of lines extracts the second and third
order terms. The second order terms are then tagged in line \lstref*{ex-12.03}. Lines
\lstref*{ex-12.04} to \lstref*{ex-12.06} isolates the target (the middle term) and applies
the substitution (swapping indices ${ad}$ on $\nabla_{b}R_{adce}$). The three terms are
united (in line \lstref*{ex-12.07}) by setting $\mu_{0}=\mu_{1}=\mu_{2}$ and then eliminated
(in line \lstref*{ex-12.08}) by setting $\mu_0 = 0$. Finally, the tags are cleared in line
\lstref*{ex-12.09}. This last step is not really needed since \verb|diff2| is zero. The last
line of the code rebuilds \verb|diff| for later processing of the third order terms.

Similar code can be used to commute the second covariant derivatives leading to
\Dmath*{\Delta\Gamma_{a} = \cdb{ex-12.diff.307} } The final steps are now rather obvious --
apply the second Bianchi identity to the first term and the first Bianchi identity to the
second term. These steps are (once again) implemented using code very similar to that given
above. The result is that $\Delta\Gamma_{a}=0$, that is $\Gamma=\GammaBar$.

\clearpage

% ============================================================================================
\section{The Weyl tensor vanishes in 3-dimensions}
\label{sec:ex-13}
\ResetCounters

\input{./cadabra/example-13.cdbtex}

The Weyl tensor in an $N$-dimensional space is given by
\begin{align*}
   C_{a b c d} = R_{a b c d} + \frac{1}{N-2} (  R_{a d} g_{b c} - R_{a c} g_{b d}
                                              + g_{a d} R_{b c} - g_{a c} R_{b d})
                             + \frac{R}{(N-1)(N-2)} (g_{a c} g_{b d} - g_{a d} g_{b c})
\end{align*}
The $C^{a}{}_{bcd}$ shares not only all of the symmetries of the Riemann tensor
but it also satisfies $C^{a}{}_{bad}=0$. Thus the number $M(N)$ of algebraically independent
components of $C^{a}{}_{bcd}$ at a point in an $N$ dimensional space is given by
\begin{align*}
   M(N) = \frac{N^2(N^2-1)}{12} - \frac{N(N+1)}{2}
\end{align*}
A common argument is that since $M(3)=0$ it follows that the Weyl tensor has zero
algebraically independent components and thus must vanish in 3 dimensions. It should also be
possible to establish the same result by direct computation, that is, show that
$C^{a}{}_{bad}=0$ for any Riemannian metric in a 3 dimensional space. That is the aim of
this example. Two methods will be presented. The first uses a brute force method where
the Weyl tensor is evaluated on a generic 3-metric. The second method uses only the known
symmetries of the Riemann tensor to show that all frame components of the Weyl tensor are
zero (and thus that the Weyl tensor must also be zero).

% --------------------------------------------------------------------------------------------
\subsection{Proof by brute force}

The computational steps are straightforward. First start with basic definitions for the
connection and the Riemann, Ricci and Weyl tensors.
\begin{cadabra}
   GammaU := \Gamma^{a}_{b c} ->  1/2 g^{a d} (   \partial_{b}{g_{d c}}
                                                + \partial_{c}{g_{b d}}
                                                - \partial_{d}{g_{b c}}).

   GammaD := \Gamma_{a b c} ->  1/2 (   \partial_{b}{g_{a c}}
                                      + \partial_{c}{g_{b a}}
                                      - \partial_{a}{g_{b c}}).

   Rabcd := R_{a b c d} ->   \partial_{c}{\Gamma_{a b d}}
                           - \partial_{d}{\Gamma_{a b c}}
                           + \Gamma_{e a d} \Gamma^{e}_{b c}
                           - \Gamma_{e a c} \Gamma^{e}_{b d}.

   Rab := R_{a b} -> g^{c d} R_{a c b d}.

   Rscalar := R -> g^{a b} R_{a b}.

   # Weyl in 3-dimensions

   Cabcd := R_{a b c d} - (R_{a c} g_{b d} - R_{a d} g_{b c})
                        - (g_{a c} R_{b d} - g_{a d} R_{b c})
                  + 1/2 R (g_{a c} g_{b d} - g_{a d} g_{b c}).   |\lstlabel{ex-13a.110}|
\end{cadabra}
Then combine these into a single expression for the Weyl tensor expressed solely in terms of
the metric.
\begin{cadabra}[firstnumber=last]
   substitute     (Cabcd,Rscalar)
   substitute     (Cabcd,Rab)
   substitute     (Cabcd,Rabcd)
   substitute     (Cabcd,GammaU)
   substitute     (Cabcd,GammaD)

   distribute     (Cabcd)

   sort_product   (Cabcd)
   rename_dummies (Cabcd)
   canonicalise   (Cabcd)     |\lstlabel{ex-13a.111}|
\end{cadabra}
The final step is to evaluate this expression on a generic metric in 3-dimensions.
\begin{cadabra}[firstnumber=last]
   gab := {g_{x x} = gxx, g_{x y} = gxy, g_{x z} = gxz,
           g_{y x} = gxy, g_{y y} = gyy, g_{y z} = gyz,
           g_{z x} = gxz, g_{z y} = gyz, g_{z z} = gzz}.

   complete (gab, $g^{a b}$)
   evaluate (Cabcd,gab)    |\lstlabel{ex-13a.112}|
\end{cadabra}

The result is that the Weyl tensor is zero (as expected).
\begin{dgroup*}
   \Dmath*{ 8C_{abcd} = \cdb{ex-13a.110}\hfill\lstref{ex-13a.110}
                      = \cdb{ex-13a.111}\hfill\lstref{ex-13a.111}
                      = \cdb{ex-13a.112}\hfill\lstref{ex-13a.112}}
\end{dgroup*}

% --------------------------------------------------------------------------------------------
\subsection{Proof using an orthonormal basis}

\def\CHat{{\hat{C}}}

This method is entirely local, that is, it only requires values of the metric and Riemann
tensors at some arbitrarily chosen point. In contrast, the previous method required
knowledge of the metric in a neighbourhood of a point (in order to compute its various
derivatives).

One of the keys steps in this method is to use the basic definitions $R_{ab} = g^{cd}
R_{acbd}$ and $R = g^{ab} R_{ab}$ to express the Weyl tensor entirely in terms of
$R_{abcd}$, $g_{ab}$ and $g^{ab}$. This leads to
\Dmath*{C_{abcd} = \cdb{ex-13b.102}}
Consider now three vectors $e^{a}_i$, $i=x,y,z$, that form an orthonormal basis at the
chosen point. Then the metric $g_{ab}$ and its inverse $g^{ab}$ can be written as
\begin{align}
   g_{ab} & = e_{a}^{x} e_{b}^{x} + e_{a}^{y} e_{b}^{y} + e_{a}^{z} e_{b}^{z}\\
   g^{ab} & = e^{a}_{x} e^{b}_{x} + e^{a}_{y} e^{b}_{y} + e^{a}_{z} e^{b}_{z}
\end{align}
where $e_{a}^{i}$ are dual to $e^{a}_{i}$, that is
\begin{align}
   e^{a}_{i} e_{a}^{j} &= \delta_{i}{}^{j}\label{ex-13:orthoA}\\
   e^{a}_{i} e_{b}^{i} &= \delta^{a}{}_{b}\label{ex-13:orthoB}
\end{align}
The main part of the calculation is to show that the frame components
\begin{align}
   \CHat_{ijkl} = C_{abde} e^{a}_{i} e^{b}_{j} e^{c}_{k} e^{d}_{l}
\end{align}
of the Weyl tensor vanish (and hence, using (\ref{ex-13:orthoB}), that the Weyl tensor must
also vanish). It is sufficient to compute just two frame components, $\CHat_{xyxy}$ and
$\CHat_{xyxz}$ as all other frame components can be found by simply permuting $x,y$ and $z$.

The scene is now set for Cadabra. The key elements in the Cadabra code include a declaration
of a Riemann tensor
\begin{cadabra}[numbers=none]
   R_{a b c d}::RiemannTensor.
\end{cadabra}
together with rules for the Weyl tensor and friends
\begin{cadabra}[numbers=none]
   Rab     := R_{a b} -> g^{c d} R_{a c b d}.
   Rscalar := R -> g^{a b} R_{a b}.

   # Weyl tensor in 3-dimensions

   Cabcd := R_{a b c d} - (R_{a c} g_{b d} - R_{a d} g_{b c})
                        - (g_{a c} R_{b d} - g_{a d} R_{b c})
                  + 1/2 R (g_{a c} g_{b d} - g_{a d} g_{b c}).

   substitute (Cabcd, Rscalar)
   substitute (Cabcd, Rab)

   Cabcd := C_{a b c d} -> @(Cabcd).
\end{cadabra}
and rules that define the metric inverse and the orthonormal basis
\begin{cadabra}[numbers=none]
   gab := g^{a b} -> ex^{a} ex^{b} + ey^{a} ey^{b} + ez^{a} ez^{b}.

   ortho := {ex^{a} ex^{b} g_{a b} -> 1,
             ey^{a} ey^{b} g_{a b} -> 1,
             ez^{a} ez^{b} g_{a b} -> 1,
             ex^{a} ey^{b} g_{a b} -> 0, ex^{a} ez^{b} g_{a b} -> 0,
             ey^{a} ex^{b} g_{a b} -> 0, ey^{a} ez^{b} g_{a b} -> 0,
             ez^{a} ex^{b} g_{a b} -> 0, ez^{a} ey^{b} g_{a b} -> 0}.
\end{cadabra}

The last part of the calculation is to apply these rules to the frame components
$\CHat_{xyxy}$ and $\CHat_{xyxz}$. For $\CHat_{xyxy}$ the code is
\begin{cadabra}
   Cxyxy := C_{a b c d} ex^{a} ey^{b} ex^{c} ey^{d}.       |\lstlabel{ex-13b.104}|

   substitute     (Cxyxy,Cabcd)                            |\lstlabel{ex-13b.105}|
   distribute     (Cxyxy)                                  |\lstlabel{ex-13b.106}|
   substitute     (Cxyxy, ortho, repeat=True)              |\lstlabel{ex-13b.107}|
   substitute     (Cxyxy, gab)                             |\lstlabel{ex-13b.108}|
   distribute     (Cxyxy)                                  |\lstlabel{ex-13b.109}|

   sort_product   (Cxyxy)                                  |\lstlabel{ex-13b.110}|
   rename_dummies (Cxyxy)                                  |\lstlabel{ex-13b.111}|
   canonicalise   (Cxyxy)                                  |\lstlabel{ex-13b.112}|
\end{cadabra}
which leads to the following output
\begin{dgroup*}
   \Dmath*{\cdb{ex-13b.104} = \cdb{ex-13b.105}\hfill\lstref{ex-13b.105}
                            = \cdb{ex-13b.106}\hfill\lstref{ex-13b.106}
                            = \cdb{ex-13b.107}\hfill\lstref{ex-13b.107}
                            = \cdb{ex-13b.108}\hfill\lstref{ex-13b.108}
                            = \cdb{ex-13b.111}\hfill\lstref{ex-13b.111}
                            = \cdb{ex-13b.112}\hfill\lstref{ex-13b.112}}
\end{dgroup*}
showing clearly (in the last line) that $\CHat_{xyxz}=0$. Similar code can be used to show
that $\CHat_{xyxz}=0$.

\clearpage

% ============================================================================================
\section{Conformal invariance of the Weyl tensor}
\label{sec:ex-14}
\ResetCounters

\input{./cadabra/example-14.cdbtex}

The Weyl tensor has the property that it is conformally invariant under conformal
transformations of the metric. That is, if two metrics, $g$ and $\overline{g}$, are related
by a conformal transformation, $\overline{g} = \phi g$ for some scalar function $\phi$, then
their corresponding Weyl tensors are equal, $\overline{C}^{a}{}_{bcd} = C^{a}{}_{bcd}$. This
simple result can be shown by direct computation. Though this result is true in any number
of dimensions, the specific case of four dimensions (in this example) is sufficient to
demonstrate the ideas behind a general proof. Note that the following computation is based
on the related expression $\overline{C}_{abcd} = \phi C_{abcd}$.

The computation is similar to the previous example and begins by first computing a general
expression for $C_{abcd}$ by forming an appropriate combination of rules. Then a copy of the
result is made (this is the Weyl tensor on the base metric $g$),
\begin{cadabra}[numbers=none]
   baseC := @(Cabcd).
\end{cadabra}
followed by a rule defining the conformal transformation
\begin{cadabra}[numbers=none]
   conformal := {g_{a b} -> \phi g_{a b}, g^{a b} -> (1/phi) g^{a b}}.
\end{cadabra}
The rule is applied to the current version of \verb|C_{a b c d}| followed by some basic
housekeeping
\begin{cadabra}[numbers=none]
   substitute   (Cabcd, conformal)
   product_rule (Cabcd)
   distribute   (Cabcd)
   product_rule (Cabcd)
   distribute   (Cabcd)
   map_sympy    (Cabcd, "simplify")
\end{cadabra}
Note that two rounds of the product rule are required because the conformal factor is
buried inside a set of second order partial derivatives. The result is the Weyl tensor for
the conformal metric. This is copied to a new object and then the difference between the two
Weyl tensors is computed
\begin{cadabra}[numbers=none]
   confC := @(Cabcd).
   diff  := @(confC) - \phi @(baseC).
\end{cadabra}
The game now is to show that \verb|diff| is zero. The code then make uses of the basic
identities (in 4 dimensions)
\begin{align*}
   g_{a b} g^{a b} = 4,\quad g_{a c} g^{c b} = \delta^{b}{}_{a},\quad \delta^{a}{}_{a} = 4
\end{align*}
to simplify the expression \verb|diff|. The result is zero (as expected).

\clearpage

% ============================================================================================
\section{The BSSN equations}
\label{sec:ex-15}
\ResetCounters

\input{./cadabra/example-15.cdbtex}

Einstein's equation of General Relativity are most simply described in the full 4-dimensional
form as
\begin{align}
   R_{ab} - \frac{1}{2} g_{ab} R = \kappa T_{ab}
\end{align}
They can also be recast in the form of a Cauchy initial value problem in which a
3-dimensional metric is evolved forward in time from a given set of initial conditions. One
such formulation is due to Arnowitt, Deser and Misner, widely known as the ADM 3+1
formulation (see \cite{mtw:1973-01}). In one of its simplest forms\footnote{That is, for
vacuum spacetimes using coordinates with a zero shift vector.}, the ADM 3+1 evolution
equations can be written as
\begin{align}
   \ddt{g_{i j}} &= -2 N K_{i j}\\[5pt]
   \ddt{K_{i j}} &= -D_{i j}{N} + N (R_{i j} + \trK K_{i j} - 2 K_{i m} K_{j n} g^{m n})
\end{align}
where $g_{ij}$ is the 3-metric, $K_{ij}$ is the extrinsic curvature, $R_{ij}$ is the Ricci
tensor, $D$ is the metric compatible covariant derivative and finally, $N$ is the lapse
function (which can be freely specified though subject to $N>0$).

For many years the ADM 3+1 equations were the cornerstone of computational general
relativity. Unfortunately, they proved to be less than ideal for long term evolutions of one
or more black holes (the evolutions were highly unstable). In recent times an alternative
set of equations, first proposed by Shibata and Nakamura \cite{shibata:1995-01} and later
popularised by Baumgarte and Shapiro \cite{baumgarte:1998-01}, now known as the BSSN
equations, have come to dominate the field (as they do allow stable long term evolutions of
black hole systems).

The BSSN evolution equations, for vacuum spacetimes and a zero shift vector, are given by
\begin{align}
   \ddt{\phi} &= -\frac{1}{6} N \trK\label{eqn:bssn01}\\[5pt]       % eqtn. 10
   \ddt{\gBar_{ij}} &= -2 N \ABar_{i j}\label{eqn:bssn02}\\[5pt]    % eqtn. 09
   \ddt{\trK} &= - g^{i j} D_{i j}{N}
                 + N (\ABar_{i j} \ABar^{i j} + \frac{1}{3} \trK^2)\label{eqn:bssn03}\\[5pt]     % eqtn. 11
   \ddt{\ABar_{ij}} &= N (\trK \ABar_{i j} - 2 \ABar_{i k} \ABar^{k}{}_{j})
                       + \exp(-4\phi) (N R_{i j} - D_{i j}{N}
                                       - \frac{1}{3} g_{i j} (N R_{k l}
                                                              - D_{k l}{N}) g^{k l})
                                                                   \label{eqn:bssn04}\\[5pt]
                                                                               % eqtn. 12
   \ddt{\GammaBar^i} &= - 2 \partial_{j}\left(N \ABar^{i j}\right)\label{eqn:bssn05}\\[5pt]      % eqtn. 19
                     &= - 2 \ABar^{i j} \partial_{j}{N}
                        + 2 N (  \GammaBar^{i}_{j k} \ABar^{k j}
                               - \frac{2}{3} \gBar^{i j} \partial_{j}{\trK}
                               + 6 \ABar^{i j} \partial_{j}{\phi})\label{eqn:bssn06}             % eqtn. 20
\end{align}
The Ricci tensor could be computed directly from the 3-metric but part of the black magic of
the BSSN formulation is to use
\begin{align}
   R_{ij} &= - 2 \DBar_{i j}{\phi}
             - 2 \gBar_{i j} \gBar^{m n} \DBar_{m n}{\phi}
             + 4 \DBar_{i}{\phi} \DBar_{j}{\phi}
             - 4 \gBar_{i j} \gBar^{m n} \DBar_{m}{\phi} \DBar_{n}{\phi}\notag\\
             &\quad
             - \frac{1}{2} \gBar^{l m} \partial_{l m}{\gBar_{i j}}
             + \gBar_{k (i} \partial_{j)}{\GammaBar^{k}}
             + \GammaBar^{k} \GammaBar_{(i j) k}
             + \gBar^{l m} \gBar^{k p} (  \GammaBar_{p l (i} \GammaBar_{j) k m}
                                        + \GammaBar_{k i m}  \GammaBar_{p l j})
                                                                           \label{eqn:bssn07}
\end{align}

The dynamical variables of the ADM formulation, $g_{ij}, K_{ij}$ are related to the those of
the BSSN formulation, $\trK,\phi,\gBar_{ij},\ABar_{ij},\GammaBar^{i}$ by the equations
\begin{align}
   \trK & = g^{ij} K_{ij}\label{eqn:bssn08}\\
   e^{4\phi} &= g^{1/3} = \left(\det(g_{ij})\right)^{1/3}\label{eqn:bssn09}\\
   \gBar_{ij} &= e^{-4\phi} g_{ij}\label{eqn:bssn10}\\
   \ABar_{ij} &= e^{-4\phi} \left(K_{ij} - \frac{1}{3} g_{ij} \trK\right)\label{eqn:bssn11}\\
   \GammaBar^{i} &= \gBar^{jk} \GammaBar^{i}_{jk} = - \gBar^{ij}{}_{,j}\label{eqn:bssn12}
\end{align}
These equations can be used to derive the BSSN equations from the ADM equations.

After a drawn out preamble here is the point of this example -- to show how Cadabra can be
used to derive the BSSN equations from the ADM equations. The calculations are non-trivial
so only the first two of the five evolution equations will be derived here. The derivation
of the full set of equations (including the constraint equations) can be found on the
website \url{https://github.com/leo-brewin/adm-bssn-equations}

It should be noted that the equations given above (\ref{eqn:bssn01}-\ref{eqn:bssn12}) are a
subset of the full set of BSSN equations. For full details see
\cite{alcubierre:2000-03,alcubierre:2003-01}

% --------------------------------------------------------------------------------------------
\subsection{Evolution equation for $\phi$}

The key elements in the Cadabra code for the first BSSN equation (\ref{eqn:bssn01}) are the
four rules
\begin{cadabra}[numbers=none]
   phi     := \phi                    -> (1/12) \log(detg).
   gdotK   := g^{i j} K_{i j}         -> trK.
   DdetgDt := \partial_{t}{detg}      -> detg g^{i j} \partial_{t}{g_{i j}}.
   DgijDt  := \partial_{t}{g_{i j}}   -> -2 N K_{i j}.
\end{cadabra}
The first three rules follow from the definitions of $\phi$, $\det g$ and $\trK$ while the
final rule is the original ADM equation for $\partial g_{ij}/\partial t$. Two other rules
are also included as they help train Cadabra to do basic calculus
\begin{cadabra}[numbers=none]
   dlog    := \partial_{a?}{\log(A?)} -> (1/A?)\partial_{a?}{A?}.
   dexp    := \partial_{a?}{\exp(A?)} -> \exp(A?)\partial_{a?}{A?}.
\end{cadabra}
The main body of the code begins with a single line
\begin{cadabra}
   dotphi  := \partial_{t}{\phi}.
\end{cadabra}
followed be a series of substitutions
\begin{cadabra}[firstnumber=2]
   substitute (dotphi, phi)          |\lstlabel{ex-15-02.101}|
   substitute (dotphi, dlog)         |\lstlabel{ex-15-02.102}|
   substitute (dotphi, DdetgDt)      |\lstlabel{ex-15-02.103}|
   substitute (dotphi, DgijDt)       |\lstlabel{ex-15-02.104}|
   substitute (dotphi, gdotK)        |\lstlabel{ex-15-02.105}|
   map_sympy  (dotphi, "simplify")   |\lstlabel{ex-15-02.106}|
\end{cadabra}
The step-by-step results are as follows
\begin{align*}
  \ddt{\phi} &=\cdb{ex-15-02.101}\tag*{\lstref{ex-15-02.101}}\\[5pt]
             &=\cdb{ex-15-02.102}\tag*{\lstref{ex-15-02.102}}\\[5pt]
             &=\cdb{ex-15-02.103}\tag*{\lstref{ex-15-02.103}}\\[5pt]
             &=\cdb{ex-15-02.104}\tag*{\lstref{ex-15-02.104}}\\[5pt]
             &=\cdb{ex-15-02.105}\tag*{\lstref{ex-15-02.105}}\\[5pt]
             &=\cdb{ex-15-02.106}\tag*{\lstref{ex-15-02.106}}
\end{align*}

% --------------------------------------------------------------------------------------------
\subsection{Evolution equation for $\gBar_{ij}$}

A similar set of rules and substitutions can be used to obtain the second BSSN equation
(\ref{eqn:bssn02}). In this case the essential rules are
\begin{cadabra}[numbers=none]
   DphiDt := \partial_{t}{\phi}   -> @(dotphi).
   gBarij := gBar_{i j}           -> \exp(-4\phi) g_{i j}.
   Kij    := K_{i j}              -> A_{i j} + (1/3) g_{i j} trK.
   A2ABar := \exp(-4\phi) A_{i j} -> ABar_{i j}.
\end{cadabra}
Note that the first rule is built using the result of the computation for
$\partial\phi/\partial t$. This construction (of building rules to record key results) is
used frequently in the full BSSN code\footnote{On the website
\url{https://github.com/leo-brewin/adm-bssn-equations}}. It avoids having to copy-paste
results for later use and thus also avoids any transcription errors. The other rules are
again built directly from the basic definitions of the BSSN variables.

The starting point for the main calculation is
\begin{cadabra}
   dotgBarij := \partial_{t}{gBar_{i j}}.
\end{cadabra}
followed by some substitutions, a product rule and dab of housekeeping
\begin{cadabra}[firstnumber=last]
   substitute   (dotgBarij, gBarij)     |\lstlabel{ex-15-03.101}|
   product_rule (dotgBarij)             |\lstlabel{ex-15-03.102}|
   substitute   (dotgBarij, dexp)       |\lstlabel{ex-15-03.103}|
   substitute   (dotgBarij, DgijDt)     |\lstlabel{ex-15-03.104}|
   substitute   (dotgBarij, DphiDt)     |\lstlabel{ex-15-03.105}|
   substitute   (dotgBarij, Kij)        |\lstlabel{ex-15-03.106}|
   distribute   (dotgBarij)             |\lstlabel{ex-15-03.107}|
   map_sympy    (dotgBarij, "simplify") |\lstlabel{ex-15-03.108}|
   substitute   (dotgBarij, A2ABar)     |\lstlabel{ex-15-03.109}|
\end{cadabra}
The corresponding output is
\begin{align*}
  \ddt{\gBar_{ij}} &=\cdb{ex-15-03.101}\tag*{\lstref{ex-15-03.101}}\\[5pt]
                   &=\cdb{ex-15-03.102}\tag*{\lstref{ex-15-03.102}}\\[5pt]
                   &=\cdb{ex-15-03.103}\tag*{\lstref{ex-15-03.103}}\\[5pt]
                   &=\cdb{ex-15-03.104}\tag*{\lstref{ex-15-03.104}}\\[5pt]
                   &=\cdb{ex-15-03.105}\tag*{\lstref{ex-15-03.105}}\\[5pt]
                   &=\cdb{ex-15-03.106}\tag*{\lstref{ex-15-03.106}}\\[5pt]
                   &=\cdb{ex-15-03.107}\tag*{\lstref{ex-15-03.107}}\\[5pt]
                   &=\cdb{ex-15-03.108}\tag*{\lstref{ex-15-03.108}}\\[5pt]
                   &=\cdb{ex-15-03.109}\tag*{\lstref{ex-15-03.109}}
\end{align*}

% --------------------------------------------------------------------------------------------
\subsection{A numerical code}

The website \url{https://github.com/leo-brewin/adm-bssn-equations} contains all of the
Cadabra code for a complete derivation of the BSSN equations (with zero shift) from the ADM
equations. A companion website, \url{https://github.com/leo-brewin/adm-bssn-numerical},
contains further Cadabra code that converts the BSSN equations into a working numerical code.

All of the tools in that second website are based on material already covered in this
tutorial (in particular, Example \ref{sec:ex-07} for exporting tensor expressions as C-code).

Using a symbolic package (in this case Cadabra) to turn a set of partial differential
equations (the BSSN equations) into a numerical code has great advantages. It frees the
researcher from the tedium of writing extensive code (try writing a code for $R_{ab}$ from
the metric by hand), it minimises the risk of coding errors and it allows for much quicker
development of new codes as changes are made in the underlying mathematics (e.g., shifting
from the ADM to the BSSN equations). This approach is quite common in the computational
general relativity community, see for example the papers by Husa etal. \cite{husa:2006-01}
and Ruchlin etal. \cite{ruchlin:2018-01}.

\clearpage

% ============================================================================================
% Part 3.
% ============================================================================================
\hrule height 0pt
\vskip 4cm
{\Huge\bf Part 3 Common traps and errors.}
\vskip 2cm
Despite our best efforts, bugs do creep in from time to time. Often the errors are
immediately obvious but on other occasions a great deal of head scratching and scouring of
web pages fills in the time before the light-bulb moment arrives. Here are some examples of
what can go wrong, how to spot the errors and tips on how to avoid them in the first place.

\clearpage

% ============================================================================================
\section*{Problems with indices}
If you have never encountered the Cadabra runtime error

\begin{lstlisting}[numbers=none,backgroundcolor=\color{white}]
   RuntimeError: Free indices in different terms in a sum do not match.
\end{lstlisting}

consider yourself lucky. For those who have (author included) here are some examples
demonstrating various ways to encounter this error.

\bgroup
\lstset{numbers=none,gobble=5}

\begin{enumerate}

   % -----------------------------------------------------------------------------------------
   \Item{\bf Inconsistent free indices}\\[5pt]
   This is trivial -- the free indices on \verb|A| and \verb|B| do not match.
   \begin{cadabra}
      foo := A_{a} + B_{b};
   \end{cadabra}

   \vskip 10pt

   % -----------------------------------------------------------------------------------------
   \Item{\bf Incorrect number of free indices}\\[5pt]
   Another trivial example. Maybe a typo (one too many indices) on \verb|B_{a b}|?
   \begin{cadabra}
      foo := A_{a} + B_{a b};
   \end{cadabra}

   \vskip 10pt

   % -----------------------------------------------------------------------------------------
   \Item{\bf Missing spaces}\\[5pt]
   The intention in the following line is to create an expression with two free
   indices
   \begin{cadabra}
      foo := A_{ab} + B_{a b};
   \end{cadabra}
   The problem here is that Cadabra will take \verb|_{ab}| to be a
   single index. Always include a space between indices (unless the indices have
   a natural separator like the slash in LaTeX names, e.g., \verb|_{\alpha\beta}|
   would be accepted as a pair of indices).

   \vskip 10pt

   % -----------------------------------------------------------------------------------------
   \Item{\bf Forgetting to declare the derivative operator}\\[5pt]
   Here is a simple and apparently correct use of indices.
   \begin{cadabra}
      foo := A_{a b} + \partial_{a}{A_{b}};
   \end{cadabra}
   So why would
   Cadabra complain? The answer is that by forgetting to declare \verb|\partial| as a
   derivative operator, Cadabra will interpret \verb|\partial_{a}{A_{b}}|
   as a function call with argument \verb|A_{b}|. Thus it thinks that this
   term has just one free index, namely, \verb|_{a}|. Hence the error.

   \vskip 10pt

   % -----------------------------------------------------------------------------------------
   \Item{\bf Cavalier use of {\tt @(...)}}\\[5pt]
   The \verb|@(...)| construct is extremely useful but it also requires the user
   to take great care less the dreaded index problem pops up. Here is a simple
   example
   \begin{cadabra}
      foo := A_{a};
      bah := B_{b};
      meh := @(A) + @(B);
   \end{cadabra}
   The problem here is obvious -- the free indices on \verb|@(foo)| and \verb|@(bah)|
   clearly do not match. Though this error is startlingly obvious in this example
   it may be much harder to detect in codes where the computation of \verb|A|
   and \verb|B| are buried deep in parts of the code far removed from each other
   and their use in \verb|@(A)+@(B)|. One way to avoid this problem is to use
   rules that define \verb|A| and \verb|B|. Here is a short example.
   \begin{cadabra}
      foo := A_{a};
      ruleA := TmpA_{a} -> @(foo);
      ...
      bah := B_{b};
      ruleB := TmpB_{b} -> @(bah);
      ...
      meh := TmpA_{c} + TmpB_{c};
      substitute (meh, ruleA)
      substitute (meh, ruleB)
   \end{cadabra}
   The ellipses in the above denote some intervening code. The rules are created
   as soon as the expressions \verb|foo| and \verb|bah| have been created.
   In most cases the right hand side of \verb|foo| and \verb|bah| will be
   substantially more complicated than that given above.

   \vskip 10pt

   % -----------------------------------------------------------------------------------------
   \Item{\bf Upstairs/downstairs index clash}\\[5pt]
   The free indices in an expression must be consistent with regards to being upstairs or
   downstairs when using either \verb|Indices(position=fixed)| or
   \verb|Indices(position=independent)|. Thus the following code snippet will cause grief
   for Cadabra.
   \begin{cadabra}
      foo := A_{a} + B^{a}.
   \end{cadabra}
\end{enumerate}

% ============================================================================================
\section*{Problems with derivatives}

\ItemNum=6

\begin{enumerate}

   % -----------------------------------------------------------------------------------------
   \Item{\bf Forgetting to declare the derivative operator}\\[5pt]
   This has already been noted (see see item 4 above) -- but a reminder can not hurt.

   \vskip 10pt

   % -----------------------------------------------------------------------------------------
   \Item{\bf Forgetting to enclose the derivative argument in {\tt \{...\}}}\\[5pt]
   The printed output for the following
   \begin{cadabra}
      foo := A_{a b} + \partial_{a} A_{b};
   \end{cadabra}
   will look like
   \begin{align*}
      A_{ab} + \partial_{a} A_{b}
   \end{align*}
   which seems fine. But without the \verb|{...}| enclosing the \verb|A_{b}| term,
   Cadabra will interpret \verb|\partial_{a} A_{b}| as a product of
   \verb|\partial_{a}| with \verb|A_{b}|. You can see that this is so by
   calling \verb|sort_product| on \verb|foo|. The output will be
   \begin{align*}
      A_{ab} +  A_{b} \partial_{a}
   \end{align*}

   \vskip 10pt

   % -----------------------------------------------------------------------------------------
   \Item{\bf Avoid applying {\tt canonicalise} to partial derivatives}\\[5pt]
   Calling \verb|canonicalise| on an expression like
   \begin{cadabra}
      foo := A_{a} \partial_{b}{B^{a}};
   \end{cadabra}
   might result in index raising/lowering of the dummy index \verb|a|. In general relativity
   this would not be allowed (except for the trivial case where the metric components are
   constants). One way to avoid this problem is to use \verb|Indices(position=independent)|.
   This will force \verb|canonicalise| to leave the indices as is. Another option (if
   possible) is to only use metric compatible derivative operators.

\end{enumerate}

% ============================================================================================
\section*{Problems with substitution rules}

\ItemNum=9

\begin{enumerate}

   % -----------------------------------------------------------------------------------------
   \Item{\bf Rules using \verb|$...$| must be confined to a single line}\\[5pt]
   Rules built using \verb|$...$| must be defined on a single line. The following
   example
   \begin{cadabra}
      substitute (foo, $A_{a}->B_{a},
                        C_{a}->D_{a}$);
   \end{cadabra}
   will raise a syntax error
   \begin{lstlisting}[backgroundcolor=\color{white}]
      SyntaxError: invalid syntax.
   \end{lstlisting}
   You can fix this either by collapsing the \verb|$...$| to a single line
   or by creating a named substitution rule (these can be split over multiple
   lines) as in the following code
   \begin{cadabra}
      myRule := {A_{a}->B_{a},
                 C_{a}->D_{a}}.
      substitute (foo, myRule);
   \end{cadabra}

   \vskip 10pt

   % -----------------------------------------------------------------------------------------
   \Item{\bf Only use \verb|->| to change index structure}\\[5pt]
   \itlabel{it:SubRuleChoices}
   There are occasions where indices need to added or deleted from expressions.
   Doing so using an equality rule like \verb|A_{a b} = A_{a}| will raise a runtime error.
   For example, the following code
   \begin{cadabra}
      foo := A_{a b};
      substitute (foo, $A_{a b} = A_{a}$)
   \end{cadabra}
   causes Cadabra great grief, reporting that
   \begin{lstlisting}[backgroundcolor=\color{white}]
      RuntimeError: Free indices on lhs and rhs do not match.
   \end{lstlisting}
   The correct code is
   \begin{cadabra}
      foo := A_{a b};
      substitute (foo, $A_{a b} -> A_{a}$);
   \end{cadabra}
   with output $A_{a}$ as expected.

   \vskip 10pt

   % -----------------------------------------------------------------------------------------
   \Item{\bf Use care when using \verb|->| to change index structure}\\[5pt]
   \itlabel{it:SubRuleDanger}
   Changing the index structure of an expression can cause runtime errors. Here is a simple
   example.
   \begin{cadabra}
      foo := A_{a} x^{a} + B_{b} x^{b}.
      substitute (foo, $x^{a} -> 1$)
   \end{cadabra}
   The problem here is that the result for \verb|foo| is \verb|A_{a} + B_{b}| and though
   Cadabra does not report an index mis-match error at this point it will do so later (most
   likely at the point when \verb|foo| is coupled to some other expression). The solution to
   this problem is to ensure that the each term in the expression uses the same index on
   \verb|x|. Here is a corrected version of the code.
   \begin{cadabra}
      foo := A_{a} x^{a} + B_{b} x^{b}.
      rename_dummies (foo)
      substitute     (foo, $x^{a} -> 1$);
   \end{cadabra}
   This works because the result after renaming the dummy indices is
   \begin{cadabra}
      foo := A_{a} x^{a} + B_{a} x^{a}
   \end{cadabra}

   But had the initial expression for \verb|foo| been
   \begin{cadabra}
      foo := A_{b} B^{b} C_{a} x^{a} + D_{d} x^{d};
   \end{cadabra}
   then this simple trick of renaming the dummies would not be sufficient to avoid the later
   problem when applying \verb|x^{a} -> 1|. In this case the call to \verb|rename_dummies|
   will return
   \begin{cadabra}
      foo := A_{a} B^{a} C_{b} x^{b} + D_{a} x^{a};
   \end{cadabra}
   The problem here is that once again the \verb|x| terms do not share a common index. This
   occurs because the renaming of dummy indices occurs left to right. As the first term
   requires two dummy indices while the second requires one the first \verb|x| will be given
   a different dummy to that assigned to the second \verb|x|.

   This minor problem can solved by first using \verb|sort_product| to bring the \verb|x|
   factors to the left of all other terms. Here is a code that does the job.
   \begin{cadabra}
      {x^{a},A_{a},B^{a},D_{a}}::SortOrder.
      foo := A_{a} B^{a} C_{b} x^{b} + D_{a} x^{a};
      sort_product   (foo)
      rename_dummies (foo)
      canonicalise   (foo)
      substitute     (foo, $x^{a} -> 1$);
   \end{cadabra}
   The corresponding output is
   \begin{align*}
      A_{b} B^{b} C_{a} + D_{a}
   \end{align*}
   This is one of the reasons why numerous exercises on sorting were included in the
   collection at the end of Example \ref{sec:ex-01}. Note that in this case
   \verb|rename_dummies| did not align the \verb|x| indices. That job fell to
   \verb|canonicalise|. The combination of \verb|sort_product|, \verb|rename_dummies| and
   \verb|canoniclaise| appears throughout this tutorial in the examples and exercises. It is
   a very standard combination.

   The take home point here is that careful inspection of the expression is required before
   operations that alter the index structure are applied.

\end{enumerate}

% ============================================================================================
\section*{Miscellaneous}

\ItemNum=12

\begin{enumerate}

   % -----------------------------------------------------------------------------------------
   \Item{\bf Syntax error}\\[5pt]
   This covers a whole raft of errors and in most cases the fix will be obvious. Here are a
   few things to look for.
   \begin{itemize}
      \item Check the termination character -- a dot, a semi-colon or the closing parenthesis
            of a function call.
      \item Check the assignment operator, use \verb|:=| for Cadabra and \verb|=| for Python.
      \item Do not use underscores in symbol names.
      \item Use standard LaTeX names such as \verb|\alpha,\beta,\mu| etc.
      \item Do not use \verb|return @(foo)|. The correct return is \verb|return foo|.
   \end{itemize}

   % -----------------------------------------------------------------------------------------
   \Item{\bf Problems with LaTeXForm}\\[5pt]
   If you want to specify the LaTeX form for an object that carries indices you
   must use \verb|{#}|, otherwise do not use \verb|{#}|. Here are two simple example.
   \begin{cadabra}
      foo{#}::LaTeXForm{"{\bar\alpha}"}.    # matches objects foo with indices
      bah::LaTeXForm{"{\hat\beta}"}.        # matches objects bah without indices
   \end{cadabra}

   % -----------------------------------------------------------------------------------------
   \Item{\bf Correct form of exponential function}\\[5pt]
   Be aware that Cadabra will treat \verb|e^{a}| as a tensor with one upstairs index. If you
   wanted the exponential function then you should write \verb|\exp{a}| or \verb|e**{a}|.

   % -----------------------------------------------------------------------------------------
   \Item{\bf Do not use underscore in expression names}\\[5pt]
   This has been mentioned before -- underscores denote subscripts and thus should
   not be used as part of an expression name (though their use in function names is
   perfectly okay).

   % -----------------------------------------------------------------------------------------
   \Item{\bf Horizontal alignment of indices}\\[5pt]
   Using braces \verb|{}| around indices, even single indices, ensures that the
   printed version of the tensor will have its indices in sequential columns.
   Thus \verb|R^{a}_{b c d}| will be printed as $R^{a}{}_{bcd}$ while
   \verb|R^a_{b c d}| will be printed as $R^a_{bcd}$.

   % -----------------------------------------------------------------------------------------
   \Item{\bf Excessively long lines}\\[5pt]
   \hypertarget{link2part3}{Each} statement in the following fragment will raise a Cadabra
   syntax error.
   \begin{cadabra}
      {\alpha,\beta,\gamma,\delta,
       \mu,\nu,\sigma,\rho,\tau,\theta}::Indices.

      {R_{\alpha\beta\gamma\delta},
       \partial_{\mu}{R_{\alpha\beta\gamma\delta}}}::SortOrder.

       substitute (foo, $R -> R_{\mu\nu} g^{\mu\nu},
                         R_{\mu\nu} -> R_{\alpha\mu\beta\nu} g^{\alpha\beta}$)
   \end{cadabra}
   One solution is to condense each statement to a single line (one for each statement). That
   will work but may lead to excessively long lines. There is an alternative -- convert the
   (single-line) statements into pure Python (using the command line tool
   \verb|cadabra2python|) and then add suitable line breaks. Suppose that the (single-line)
   statements are in the file \verb|foo.cdb|. You can then create the Python equivalent
   \verb|foo.py| using (on the command line)
   \bgroup
   \lstset{numbers=none}
   \begin{lstlisting}
      cadabra2python foo.cdb foo.py
   \end{lstlisting}
   \egroup
   The file \verb|foo.py| will contain the following lines
   \begin{cadabra}
      __cdbtmp__ = Indices(Ex(r'''{\alpha,\beta,\gamma,\delta,\mu,\nu,\sigma,\rho,\tau,\theta}'''), Ex(r''))

      __cdbtmp__ = SortOrder(Ex(r'''{R_{\alpha\beta\gamma\delta},\partial_{\mu}{R_{\alpha\beta\gamma\delta}}}'''), Ex(r''))

      substitute (foo, Ex(r'''R -> R_{\mu\nu} g^{\mu\nu}, R_{\mu\nu} -> R_{\alpha\mu\beta\nu} g^{\alpha\beta}''', False))
   \end{cadabra}
   This does not seem like much of an improvement but the good news is that as most of the
   text is written as strings the issue of line breaking is now trivial -- strings are
   easily split across lines. This is also a good time to do a bit tidying up (replacing
   triple quotes with single quotes and \verb|__cdbtmp__| with \verb|tmp|). The tidied
   version of
   \verb|foo.py| is now
   \begin{cadabra}
      tmp = Indices(Ex(r'{\alpha,\beta,\gamma,\delta,'+
                       r'\mu,\nu,\sigma,\rho,\tau,\theta}'), Ex(r'') )

      tmp = SortOrder(Ex(r'{R_{\alpha\beta\gamma\delta},'+
                         r'\partial_{\mu}{R_{\alpha\beta\gamma\delta}}}'), Ex(r'') )

      substitute (foo, Ex(r'R -> R_{\mu\nu} g^{\mu\nu},'+
                          r'R_{\mu\nu} -> R_{\alpha\mu\beta\nu} g^{\alpha\beta}', False))
   \end{cadabra}
   This code fragment can be cut-and-pasted into an existing Cadabra code (or to replace the
   original code in \verb|foo.cdb|). The new code will be happily accepted by Cadabra
   (though it is a matter of opinion whether the aesthetics of this version are an
   improvement over the original single-line statements).

   After running a few experiments you should be able to infer the basic actions of
   \verb|cadabra2python|. For a typical property like
   \begin{lstlisting}
      {list of things}::PropertyName(arguments).
   \end{lstlisting}
   the conversion will produce (after a bit of tidying up)
   \begin{lstlisting}
      tmp = PropertyName ( Ex(r'list of things'), Ex(r'arguments') )
   \end{lstlisting}
   while for a typical algorithm like
   \begin{lstlisting}
      Algorithm (foo, $a substitution rule$)
   \end{lstlisting}
   the result will be (again after manual tidying up)
   \begin{lstlisting}
      Algorithm (foo, Ex(r'a substitution rule', False))
   \end{lstlisting}

   Note that the \verb|False| argument in the \verb|substitute (foo,Ex(...,False))| call
   appears to serve no purpose and can be deleted (though, if in doubt, just leave it
   as it stands).

\end{enumerate}

\egroup

\clearpage

% ============================================================================================
% Part 4.
% ============================================================================================
\hrule height 0pt
\vskip 4cm
{\Huge\bf Part 4 Further reading.}
\vskip 2cm
The content of this tutorial reflects mainly the author's own interests. It is thus a highly
selective sample of topics in Cadabra. There is vastly more to Cadabra than has been
conveyed in this tutorial. The following few pages contain links to a wide variety of
(mostly) on-line resources for Cadabra.

\lstset{numbers=none}

\clearpage

% ============================================================================================
\section*{Webpages}

\url{https://github.com/kpeeters/cadabra2.git}\Break
This is the GitHub repository for Cadabra2. You can clone the site using
\begin{lstlisting}
   git clone https://github.com/kpeeters/cadabra2.git
\end{lstlisting}
This will create a \verb|cadabra2| directory containing the complete source code. It also
contains full instructions on how to compile and install the code (see
\verb|cadabra2/README.rst|).

\url{https://cadabra.science/help.html}\Break
This is the main online reference for Cadabra. It is written as series of short tutorials
each covering a key aspect of Cadabra. Topics covered include basic syntax for writing
expressions, properties and algorithms, basic input/output, how to manipulate expressions and
an introduction to programming in Cadabra.

\url{https://cadabra.science/man.html}\Break
This site describes every property and algorithm supported by Cadabra. It is the
first place to go when looking for information about a property or an algorithm.

\url{https://cadabra.science/qa/questions}\Break
This is a popular site for posting Cadabra questions and answers. Anyone can read the
questions and answers but to post to the site you will need to register.

\url{https://cadabra.science/notebooks/ref_patterns.html}\Break
This topic in the reference guide provides full details on how use the \verb|?| and
\verb|??| patterns. It also discusses more powerful pattern matching using conditional
patterns and regular expressions (neither of which are described in this tutorial).

\url{https://cadabra.science/notebooks/ref_programming.html}\Break
This topic contains a good discussion on how expressions are stored in Cadabra. It also
describes how you can access and manipulate the elements of an expression (such as its
indices and the individual terms). The topic contains a nice function that will return the
covariant derivative for \emph{any} tensor.

\url{https://cadabra.science/tutorials.html}\Break
This is a collection of tutorials showcasing the main features of Cadabra. The tutorials can
be viewed online or they can be downloaded as Cadabra notebooks (and thus allowing
experiments to be run in the Cadabra gui).

\url{https://cadabra.science/user_notebooks.html}\Break
This is set of user contributed notebooks. One notebook (by Mattia Scomparin) shows how
Cadabra can be used to derive the non-vacuum Einstein equations from Hilbert action
integral. Another notebook (by Oscar Castillo-Felisola) uses differential forms to derive
the second Bianchi identities.

\url{https://github.com/leo-brewin/cadabra-tutorial}\Break
This is the GitHub repository for this tutorial. It contains all of the Cadabra and LaTeX
sources as well the pdf files generated from those sources. The source files are written in
a hybrid syntax that has the Cadabra code embedded within the LaTeX source. These hybrid
sources can be compiled using a small set of tools (Python scripts and LaTeX style files)
all of which can be obtained from the authors GitHub site (i.e., the very next web page).

\url{https://github.com/leo-brewin/hybrid-latex}\Break
This site contains all the tools needed to process the hybrid LaTeX/Cadabra files used
in this tutorial. It also contains similar tools for LaTeX sources with embedded Maple,
Mathematica, Matlab and Python code. Some readers may find these tools useful beyond their
use in this tutorial.

\url{https://github.com/leo-brewin/riemann-normal-coords}\Break
This site contains all of the Cadabra code used in the authors' paper on Riemann Normal
Coordinates \cite{brewin:2009-02}.

\url{https://github.com/leo-brewin/adm-bssn-equations}\Break
This site contains the full derivation of the BSSN equations from the ADM equations (for
vacuum spacetimes and a zero shift vector). This extends the limited discussion given
in Example \ref{sec:ex-15} (where only two equations were derived).

\url{https://github.com/leo-brewin/adm-bssn-numerical}\Break
This site contains a full 3+1 evolution code (written in Ada) for a Kasner $T^3$ cosmology.
This includes the Cadabra code used to convert the BSSN equations into computer code
suitable for use in the numerical integrators. Two codes are provided, one for the ADM
system and another for the BSSN system.

\url{https://docs.python.org/3/}\Break
\url{https://docs.python.org/3/reference/index.html}\Break
\url{https://docs.python.org/3/tutorial/index.html}\Break
These are the official sites for Python. They provide excellent information on all matters
Python.

\url{https://www.sympy.org/en/index.html}\Break
\url{https://docs.sympy.org/latest/tutorial/index.html}\Break
This is a great place to start when learning how to use SymPy.

% ============================================================================================
\section*{Notebooks}

The Cadabra2 source code includes a wealth of sample notebook in the directory
\verb|cadabra2/examples|.

% ============================================================================================
\section*{Acknowledgement}

I am very grateful to Kasper Peeters for his careful reading of various drafts of this
tutorial and for his many helpful suggesstions.

% ============================================================================================
% \printbibliography  % create the bibliography using biblatex & biber

% ============================================================================================
\bibliographystyle{bibdata}  % create the bibliography using bibtex
\bibliography{bibdata}

\end{document}